\documentclass[11pt,letterpaper]{article}
\usepackage[margin=1in]{geometry}
\usepackage{fullpage}
\usepackage{float}
\usepackage{amsfonts,amsmath,amssymb,amsthm,graphicx}

\usepackage{thmtools}
\usepackage{thm-restate}

\usepackage{amstext}
\usepackage{url}
\usepackage{subfig}
\usepackage{hyperref}
\usepackage{multirow}
\usepackage{caption}
\usepackage{comment}
\usepackage[capitalize]{cleveref}
\usepackage[authoryear]{natbib}
\usepackage{enumerate}

\usepackage{algpseudocode}

\usepackage{setspace}
\allowdisplaybreaks[1]

\usepackage[linesnumbered,ruled]{algorithm2e}
\SetKwComment{Comment}{/* }{ */}

\newtheorem{theorem}{Theorem}[section]
\newtheorem{corollary}[theorem]{Corollary}
\newtheorem{lemma}[theorem]{Lemma}
\newtheorem{proposition}[theorem]{Proposition}
\newtheorem{claim}[theorem]{Claim}
\newtheorem{definition}[theorem]{Definition}

\newtheorem{example}[theorem]{Example}

\newenvironment{proofof}[1]{\begin{proof}[Proof of #1]}{\end{proof}}

\newcommand{\floor}[1]{
{\lfloor {#1} \rfloor}
}

\newcommand{\given}{\,|\,}

\newcommand{\prob}[2][]{\text{\bf Pr}\ifthenelse{\not\equal{}{#1}}{_{#1}}{}\!\left[{\def\givenn{\middle|}#2}\right]}
\newcommand{\expect}[2][]{\text{\bf E}\ifthenelse{\not\equal{}{#1}}{_{#1}}{}\!\left[{\def\givenn{\middle|}#2}\right]}

\newcommand{\tparen}{\big}
\newcommand{\tprob}[2][]{\text{\bf Pr}\ifthenelse{\not\equal{}{#1}}{_{#1}}{}\tparen[{\def\given{\tparen|}#2}\tparen]}
\newcommand{\texpect}[2][]{\text{\bf E}\ifthenelse{\not\equal{}{#1}}{_{#1}}{}\tparen[{\def\given{\tparen|}#2}\tparen]}

\newcommand{\sprob}[2][]{\text{\bf Pr}\ifthenelse{\not\equal{}{#1}}{_{#1}}{}[#2]}
\newcommand{\sexpect}[2][]{\text{\bf E}\ifthenelse{\not\equal{}{#1}}{_{#1}}{}[#2]}

\newcommand{\rbr}[1]{\left(\,#1\,\right)}

\newcommand{\cbr}[1]{\left\{\,#1\,\right\}}

\newcommand{\indicate}[1]{\mathbf{1}\left[\,#1\,\right]}

\newcommand{\eps}{\epsilon}

\newcommand{\rev}{{\rm Rev}}
\newcommand{\poly}{{\rm poly}}
\newcommand{\opt}{{\rm OPT}}
\newcommand{\np}{{\rm NP}}

\newcommand{\OPT}{\mathrm{OPT}}

\newcommand{\ps}{\widetilde\psi}
\usepackage{xcolor}

\usepackage{color-edits}
\addauthor{yl}{red}
\DeclareMathOperator*{\argmax}{arg\,max}

\usepackage{array}
\usepackage{booktabs}

\usepackage{thm-restate}

\begin{document}

\title{Fair Team Contracts}

%\author{
%Junjie Chen\thanks{School of Computer Science and Engineering, Southeast %University. Email: \texttt{junjie.chen@seu.edu.cn}}
%\and Matteo Castiglioni\thanks{DEIB, Politecnico di Milano. Email: \texttt{matteo.castiglioni@polimi.it}}
%\and Yingkai Li\thanks{Department of Economics, National University of Singapore. Email: \texttt{yk.li@nus.edu.sg} }
%}

\author{
Junjie Chen\thanks{
School of Computer Science and Engineering, Southeast University.
Email: \texttt{junjie.chen@seu.edu.cn}
}
\and
Matteo Castiglioni\textsuperscript{$\diamond$}\thanks{
DEIB, Politecnico di Milano.
Email: \texttt{matteo.castiglioni@polimi.it}
}
\and
Yingkai Li\textsuperscript{$\diamond$}\thanks{
Department of Economics, National University of Singapore.
Email: \texttt{yk.li@nus.edu.sg}
}
}

%\author{
%anonymous author(s)
%}

\date{}

\begin{titlepage}
	\clearpage\maketitle
	\thispagestyle{empty}

    \begingroup
\renewcommand{\thefootnote}{\ensuremath{\diamond}}
\footnotetext[0]{M. Castiglioni and Y. Li are listed
in alphabetical order.}
\endgroup

\begin{abstract}
A principal hires a team for a joint project and pays success-contingent shares to induce costly effort. We require that agents do not envy one another's contracts: for any pair of team members, neither would gain if they exchanged shares and all team members reoptimized their effort choices. For submodular rewards, revenue-optimal fair contracts can be chosen to pay each hired agent the maximum of her incentive threshold and a common endogenous share floor. We use this characterization to obtain an FPTAS for additive rewards and a deterministic value-oracle algorithm achieving a $0.249$ fraction of optimal fair revenue for submodular rewards. We also show that optimal fair contracts can outperform optimal non-discriminatory contracts by a factor of approximately $1.445$. This bound is tight for both additive and submodular rewards.

\end{abstract}

\end{titlepage}

\section{Introduction}

Contract theory is one of the fundamental fields in microeconomics, and it has started to attract attention from the computer science community since \citep{babaioff2006combinatorial,dutting2019simple}. 
It tackles the moral hazard problem in principal-agent models, where a principal can only observe noisy signals about the actions privately chosen by the agents, 
and designs contracts to incentivize the agents to choose costly actions through monetary compensation. 
The primary focus of the computer science literature is on the complexity of computing optimal contracts~\citep[e.g.,][]{dutting2021complexity,castiglioni2022designing,castiglioni2025reduction}, approximations~\citep[e.g.,][]{dutting2022combinatorial,dutting2023multi,deo2024supermodular,alon2025multi}, and learning in contracts~\citep[e.g.,][]{bacchiocchi2023learning,chen2024bounded,han2024learning}.

In this paper, we focus on the design of optimal team contracts for revenue maximization. Specifically, a principal hires multiple agents as a team to collaborate on a single project. 
In addition to moral hazard, a central challenge in this setting is the externality. 
That is, the principal can only monitor the success of the project, not the individual efforts of the agents in the team, 
and the success probability is determined by the joint effort of the agents in the team. 
This externality leads to novel computational challenges, which have been investigated in \citet{dutting2023multi} and subsequently generalized by \citet{dutting2025multi}.

However, a crucial aspect neglected by \citep{dutting2023multi} and subsequent papers is the fairness concerns among agents in the team. 
The following example provides a simple illustration of the fact that purely maximizing the principal's revenue may lead to {\it unfairness} among agents. 
\begin{example}\label{unfairexample_intro}
    Consider an example with $2$ agents collaborating on a single project, and the success of the projects gives the principal a value of $1$. The costs of the agents for exerting efforts are $c_1=c_2 =0.05$. If none of them exerts effort, the project fails. Moreover, agent $1$ (agent $2$) can increase the probability of success by $p_1=0.5$ ($p_2=0.25$) by exerting costly effort. 
    By \citep{dutting2023multi}, it is optimal to recruit both agents in the team and offer agent $1$ a linear contract of $\alpha_1=\frac{c_1}{p_1} = 0.1$, and agent $2$ a linear contract of $\alpha_2=\frac{c_2}{p_2} = 0.2$. 
    In this example, it is unfair for agent $1$ since agent $1$ has a higher contribution to the optimal, but he gets a lower utility (also lower payments) compared to agent $2$. 
\end{example}

Beyond this simple illustrative example, fairness in team contracts is a central concern in various practical applications. 
A local survey in Singapore shows that junior employees are getting paid less than new hires~\citep{CNA_GraduateSalaries2022}, which raises a reflection on fair remuneration. 
Indeed, the issue of fair remuneration has long been a central topic in economics, labor studies, and organizational design. 
Moreover, fairness also plays a critical role in shaping incentives, productivity, and long-term organizational performance. For example,  \cite{akerlof1990fair} demonstrates that ``workers proportionately withdraw effort as their actual wage falls short of their fair wage'', underscoring how fairness perceptions translate into behavioral responses.
The Ministry of Manpower in Singapore promotes the Progressive Wage Model (PWM)~\citep{SG_MOM_PWM} to link wage growth to skills upgrading and productivity gains.

These observations highlight an important gap: the fairness dimension, which is central to organizational practice and labor policies, has been largely overlooked by theoretical studies in the community.\footnote{A prominent exception is \citet{castiglioni2025fair} who consider the fair allocation of tasks in contracting settings.} In many real-world environments, principals must balance revenue considerations with equitable treatment of workers, both to maintain morale and to comply with institutional or regulatory guidelines. 
In light of this tension, our paper formulates and studies the \emph{Fair Team Contracts} problem: how to design optimal team contracts that not only incentivize effort and maximize expected revenue, but also satisfy principled fairness constraints across agents.

\subsection{Model}\label{intro_model}

We first briefly introduce our team contract model and a justifiable {\it fairness} notion proposed in this paper. 
Our model consists of one principal who hires a team of agents $S \subseteq \mathcal{N}$ to accomplish a single project. 
Each agent can choose whether to exert {\it costly} efforts for the project or not. 
If the set of agents $S \subseteq \mathcal{N}$ chooses to exert efforts for the project, a reward $f(S)$ will be generated, where $f: 2^\mathcal{N} \to \mathbb{R}_{\ge 0}$ is a non-negative set function. 
Following common assumptions in the literature \cite{dutting2022combinatorial,duetting2025multi,dutting2023multi}, the principal assigns a {\it linear} contract $\alpha_i  \ge 0$ to agent $i$ and hence the agent receives a payment $\alpha_i f(S)$ from the principal.

In team contracts, some agents contribute more, while others make less important contributions. To ensure fairness, one naive approach is to employ envy-freeness \citep{foley1966resource}: agent $i$ does not envy agent $j$ if agent $i$ gains more utility under his own contract $\alpha_i$ than agent $j$'s contract $\alpha_j$, fixing the choice of other agents. This leads to offering the same contract to all incentivized agents.~While simple, one issue with this notion is that it fails to account for the {\it externality effects}, where agents could affect the payoffs of others through their actions in team contracts \citep[c.f.,][]{dasaratha2023equity}. 
Moreover, intuitively, an agent who is more crucial to the success of the project, such as the team leader or core contributor, is naturally entitled to larger compensation compared to others. 
\vspace{-3pt}
\begin{example}\label{fairness_def_example}
    Consider an example with $2$ agents collaborating on a single project, and the success of the projects gives the principal a value of $1$. If none of them exerts effort, the project fails. Moreover, agent $1$ (agent $2$) can increase the probability of success by $p_1=0.3$ ($p_2=0.2$) by exerting effort with cost $c_1 = 0.1$ ($c_2=0.05$). 
    A feasible team contract is to recruit both agents in the team and offer agent~$1$ a linear contract of $\alpha_1=\frac{c_1}{p_1} = \frac{1}{3}$, and agent $2$ a linear contract of $\alpha_2=\frac{c_2}{p_2} = \frac{1}{4}$. 
\end{example}
\vspace{-3pt}
In this example, the contract that offers different compensations to different agents is not fair based on the naively defined notion of fairness. 
However, due to the existence of externalities, agent $1$ also contributes significantly to the reward of agent $2$, even more than the contribution of agent $2$ himself in this example.  
Thus, it would not be unreasonable to accommodate at least a mild degree of compensation differences due to the heterogeneity of the agents and the externality. 
Indeed, most legislation only requires equal pay for similar work, as documented in \citet{gentile2024equal}, and the compensation is allowed to differ for workers with drastically different performance.

In our paper, we propose a new fairness notion that builds upon \citet{velez2016fairness}, which utilizes envy-freeness under a {\it swap} operation to address the additional externality effects.
Specifically, to examine whether there is envy between any two agents $i, j\in S$ under contract $\alpha$, a new contract $\hat{\alpha}$ is obtained by {\it swapping} $\alpha_i$ and $\alpha_j$. 
Then, a new equilibrium will be reached, producing a new set of incentivized agents $\hat{S}$. If both agents $i, j$ obtain weakly higher utilities under the equilibrium set $S$ than $\hat{S}$, there is no envy between agent $i$ and $j$. 
If there is no envy between any pair of agents in $S$, contract $\alpha$ is fair. 
Note that this definition is a generalization of the naive notion, since the contract in which all agents in $S$ are paid the same trivially satisfies the fairness requirement under the swap operation.
Moreover, under this new definition of fairness, the contract defined in \cref{fairness_def_example} is fair, as agent $1$ would not exert effort after the swap, leading to low utilities for both agents under the new equilibrium. 
The main goal of the principal in our paper is to find the revenue maximizing contract under this new fairness constraint.

\subsection{Our Results}

In this section, we present three sets of results that range from characterizing the optimal contracts to computing the (approximately) optimal fair contracts.

Our results focus on settings with submodular reward functions. We first show that, even with the additional complexity introduced by the fairness concerns compared to the team contracts design of \citep{dutting2023multi}, the optimal fair contract has a rather simple and intuitive structure. 

\vspace{1mm}
\noindent\textbf{Theorem 1. (\cref{theorem_optimal_share})} {\it In optimal fair contracts, there exists a value $\mathcal{L} \ge 0$ such that the agents exerting efforts are paid the maximum of their respective cut-off wages and $\mathcal{L}$.}
\vspace{1mm}

We call $\mathcal{L}$ the \emph{minimum share}. The cut-off wage for an agent is defined as the minimally sufficient incentive contract that induces the agent to exert effort in equilibrium. Our result, which implies that the agent must be paid at least contract $\mathcal{L}$, has important practical implications. In particular, it suggests that setting a wage floor can be an effective mechanism to achieve fairness. Accordingly, a regulatory policymaker may consider imposing a minimum wage for laborers. Such policies are already widely adopted in practice. For instance, Hong Kong has implemented the Minimum Wage Ordinance since 2011 to protect grassroots employees and support sustained economic growth.

One interesting observation from the above simple structure is that in the optimal fair contracts, the contracts of any two agents (except for at most one agent) differ by at most a factor of 2 (\cref{lm:1/2}).
This allows us to further investigate the revenue loss relative to the optimal fair contract for a particular class of fair contracts, namely non-discriminatory contracts in which all incentivized agents are paid the same wage.
 
\vspace{1mm}
{\noindent\textbf{Theorem 2. (\cref{thm_additive_tight_gap})} {\it Under submodular reward functions, the tight approximation of a non-discriminatory contract to the optimal fair contract is $1+\rho$, where $\rho\in (0, 1)$ is the unique root of $\rho^3 - \rho^2 - 2\rho +1 = 0$. Moreover, this approximation factor is also tight for additive cases.}}
\vspace{1mm}

While non-discriminatory contracts are restrictive, they are perhaps the \emph{simplest} form of fair contracts under our swap-based notion of fairness. This simplicity may have practical appeal in many applications, and we show that the worst-case loss from imposing this restriction is tolerable, especially in the additive case, when simplicity is of paramount concern. 
In contrast, if such a limitation is not required for applications, we show that the principal can achieve strictly higher revenue by providing discriminatory treatment to agents while maintaining our notion of fairness. 
This further highlights the importance of investigating optimal fair contracts beyond the class of non-discriminatory contracts studied in the literature \citep[e.g.,][]{feldman2026equal}.

Finally, we address the computational challenges of finding the optimal contracts. 
We derive an almost optimal fair contract for additive rewards. Moreover, we show that it is $\np$-hard to compute the optimal fair contracts even in additive cases.

\vspace{1mm}
\noindent\textbf{Theorem 3. (\cref{amoreformalfptasstatement} and \cref{hardness_additive})} {\it Under additive reward functions, there exists an FPTAS that finds an almost optimal contract. Moreover, it is $\np$-hard to compute the optimal fair contract.}

Our last result concerns the canonical submodular reward functions. In this case, we show that there exists a polynomial-time algorithm for computing a constant approximation to the optimal fair contracts for submodular rewards. 

\vspace{1mm}
\noindent\textbf{Theorem 4. (\cref{thm:main})} {\it Under submodular reward functions and given access to only value oracles, there exists a polynomial-time algorithm that finds a fair contract achieving a constant approximation to the optimal fair contract.}
\vspace{1mm}

Motivated by Theorem 2, our polynomial-time algorithm for submodular rewards uses non-discriminatory candidates for large teams and enumerates optimal fair contracts for small teams, which may have unequal shares. It guarantees at least a $0.249$ fraction of optimal fair revenue directly.

\subsection{Our Techniques}\label{app_tech_applied}

{\bf Characterizing Optimal Fair Contracts.} 
A key challenge in characterizing the optimal contracts is that, under our definition of fairness, swapping contracts between two agents may lead to multiple equilibria for the resulting incentivized set $\hat{S}$. As a result, the incentivized set of agents $S$ before swapping can differ substantially from $\hat{S}$ in general, which complicates the characterization. However, a crucial observation in our setting is that when the reward function is submodular, swapping contracts yields a unique equilibrium. Moreover, under our fairness constraint, if one swaps the contracts of two agents $i,j \in S$ in a fair team contract with $\alpha_i < \alpha_j$, the resulting equilibrium incentivized set has a simple and unique structure: $\hat{S} = S \setminus {j}$. That is, only the agent with strictly higher initial compensation chooses to opt out. This structure allows us to establish tight relationships between the contracts of any two agents, which ultimately leads to our characterization featuring a minimum-share structure.

\vspace{1mm}

\noindent{\bf Tight Approximations of Non-discriminatory Contracts.} Interestingly, to derive the tight approximation bound, it suffices to focus on the additive-reward case.
Our critical observation, which may be of independent interest, is that if the optimal non-discriminatory contract can achieve at least $\frac{1}{\Gamma}$ fraction of the optimal fair contracts for every additive instances, then the same guarantee extends to submodular instances. 
To see this, consider an optimal set of agents $S^*$ in a submodular instance.
Since every submodular function is XOS, there exists an additive function $a(\cdot)$ such that $f(S^*) = \sum_{i\in S^*} a(i)$, $a(T) \le f(T)$ for every $T \subseteq S^*$, and $a(i) \ge f(i| S^* \setminus \{i\})$ for every $i \in S^*$. 
We then construct an additive instance using the additive function $a(\cdot)$, where only agents in $S^*$ can be incentivized and their cut-off wages are the same as in submodular instances. This allows us to relate the quantity $\mathcal{L}$ from Theorem 1 across the two instances, as well as their respective non-discriminatory contracts. These relationships are what is needed to transfer the $1/\Gamma$ approximation guarantee to the submodular setting.

To derive the tight approximation bound for additive cases, we divide the analysis according to the size of the optimal incentivized set of agents, onsidering separately the cases $|S^*|=2$ and $|S^*|>2$. 

The tight bound $1+\rho$ arises from the case of $|S^*|=2$. Let $S^*=\{a, b\}$ and the cut-off wages $r_a\le r_b$.  We consider three candidate non-discriminatory contracts: (1) incentivizing only agent $a$, (2) incentivizing only agent $b$, and (3) incentivizing only agents $a$ and $b$ through non-discriminatory contracts. The upper bound of approximation is obtained by comparing the revenue from the optimal non-discriminatory contract with the maximum revenue achieved by these three candidate non-discriminatory contracts. An interesting feature of the analysis is that the worst-case bound can occur when $r_a=\mathcal{L}$ and both agents are incentivized via non-discriminatory contracts. This observation also guides our construction establishing the matching lower bound of $1+\rho$. Specifically, we construct an instance of two agents, where the smaller cut-off wage is exactly $\mathcal{L}$ and the optimal non-discriminatory contract incentivizes both agents. 

Lastly, we show that when $|S^*|>2$, the approximation upper bound is strictly less than $1+\rho$. Let $h$ be the agent receiving the highest payment in the optimal fair contract and $G=S^*\setminus\{h\}$. The key intuition is as follows. If $f(h)$ is large relative to $f(S^*)$, then incentivizing only agent $h$ alone may already capture a substantial fraction of the optimal value, while additionally incentivizing agents in $G$ can be too costly under a non-discriminatory contract. On the other hand, if $f(h)$ is small, then the value $\mathcal{L}$ would be relatively large and close to the payment of agent $h$, which implies that extending the non-discriminatory contract to additional agents in $G$ requires only a modest increase from their original contracts.
Taking these two cases into account, we construct a non-discriminatory contract that incentivizes agent $h$ together with the $j$ highest-value agents in $G$.
A careful analysis shows that the approximation upper bound is at most $\frac{24}{17}$.

\vspace{1mm}

\noindent{\bf FPTAS for Additive Rewards.} 
For additive rewards $f(S)=\sum_{i\in S} f(i)$, we obtain an FPTAS via a dynamic
program that exploits two structural facts specific to our minimum-share characterization.
We begin with a dichotomy: if some agent contributes almost all of the optimal reward, then paying only
that agent (at its cut-off wage) is already near-optimal; otherwise we proceed to a fully polynomial scheme.
In the main case, the challenge is that we need a \emph{multiplicative} approximation to revenue
$(1-\sum_{i\in S}\alpha_i)\,f(S)$, so we must control both the approximation of $f(S)$ and the approximation
of the minimum share $\mathcal{L}$ that determines which agents are paid the floor versus their cut-off wages.
These are coupled: $\mathcal{L}$ is not a free variable but is endogenously determined by the chosen set $S$
through the formula $\mathcal{L}(S,f(S))=\max_{i\in S}\frac{c_i}{f(i)}(1-\frac{f(i)}{f(S)})$.

To address this coupling, the algorithm enumerates two ``pivotal'' agents that identify the correct
scales in the optimum: (i) an agent $i^*$ with the largest contribution in $S^*$, which sets the
discretization granularity, and (ii) an agent $\bar i$ that attains the maximum in $\mathcal{L}(S^*,f(S^*))$.
Given a guess of $i^*$, we discretize each $f(i)$ on a grid with a step size proportional to $f(i^*)$,
which guarantees that for any set whose largest contributor is indeed $i^*$, the total reward $f(S)$ and
its discretized proxy $\widetilde f(S)$ remain within a controlled multiplicative factor after aggregation.
Given guesses of $\bar i$ and the discretized total $\widetilde f(S^*)$, we obtain an \emph{upper} estimate
of the true minimum share $\mathcal{L}(S^*,f(S^*))$ and restrict attention to agents consistent with these
guesses. The remaining optimization becomes a knapsack-style DP: among feasible agents, find a set
$S$ achieving a target discretized total $\widetilde f(S)$ while minimizing the induced total payment
$\sum_{i\in S}\widetilde\alpha_i$ (where $\widetilde\alpha_i$ is either the estimated minimum share or the
cut-off wage).
A final technical ingredient is a reserve lemma showing that in any optimal solution the principal keeps
a non-negligible fraction of the reward (i.e., $1-\sum_{i\in S^*}\alpha_i^*$ is bounded away from $0$),
so small additive errors in the DP-induced payments translate into only a small multiplicative loss in
revenue. Combining these pieces yields a fully polynomial-time approximation scheme for computing the optimal fair contract.

\vspace{1mm}

\noindent{\bf Approximation for Submodular Rewards.}
Although our minimum-share characterization determines the optimal fair
contract for any fixed team, finding a good team remains challenging:
changing the team changes agents' marginal contributions and hence their
cut-off wages, while the common share needed to incentivize an agent
affects the payments to all selected agents. Consequently, a team with
high reward need not generate high revenue. Our approach constructs
non-discriminatory contracts whose revenue is compared directly with
the optimal fair revenue.

The first challenge is to obtain a useful set-selection surrogate objective from
the optimal fair contract, so that one can find a good candidate set by optimizing that objective. 
We enumerate the team size, a seed agent,
and a common share $\bar\alpha$, and assign each agent a price
$\pi_i=c_i/\bar\alpha$, with which we construct a surrogate problem $\max_{S\subseteq \mathcal N} f(S)-\pi(S)$.
We then apply a seeded variant of distorted
greedy~\cite{harshaw2019submodular} to the surrogate objective
$f(S)-\pi(S)$, followed by pruning agents whose incentive constraints
are violated. In the analysis, the seed is an agent with the largest
cut-off wage in the optimal team. If its marginal contribution is
large, a singleton contract already gives the desired guarantee;
otherwise, the minimum-share characterization provides a lower bound
on the optimal team's total payment. This bound guides the choice of
$\bar\alpha$ so that the greedy certificate
$G=f(U)-\pi(U)$ captures a controlled fraction of the optimal fair
revenue.

The second challenge is to convert this certificate into actual
revenue while controlling both the non-discriminatory contracts and the team size.
Given the candidate set $S$, we repeatedly delete an agent with the largest price-to-marginal
ratio $z = \frac{\pi_{i^*}}{f(i^*|S\setminus {i^*})}$, obtaining a sequence of teams feasible under decreasing common
shares $z\bar{\alpha}$. The obtained team is feasible because $z\bar{\alpha} =\frac{c_{i^*}}{f(i^*|S\setminus {i^*})} $ is indeed the minimum non-discrimatory contract to incentivize set $S$.  The key observation is an exact integral identity: if $V(z)$ (i.e., the total contribution of all agents in $S$ whose price-to-marginal ratios are less than $z$)
is the reward retained at ratio threshold $z$, then
$\int_0^1 V(z)\,dz=G$.
Starting separately from each such team, we repeatedly delete an
agent with the smallest marginal contribution. Submodularity ensures
that retaining a given fraction of its agent size preserves at least
the same fraction of its reward, while reducing the total payment
share. Combining this cardinality trade-off with the integral identity
guarantees a high-revenue candidate. Finally, optimizing the size of the candidate set and
controlling the discretization and rounding errors yields a
$(\Gamma-\varepsilon)$ fraction of the optimal fair revenue, where
$\Gamma>0.24988$, using only value queries.

\subsection{Related Work} \label{app_related_workdsics}

\paragraph{Team and combinatorial contracts.}
Our model follows the joint-project setting of \citet{dutting2023multi}, in which several agents choose hidden actions affecting a common outcome. Extensions include multiple actions per agent~\citep{cacciamani2024multi}, combinatorial actions~\citep{duetting2025multi}, and multiple projects~\citep{alon2025multi}. The single-agent combinatorial model allows an agent to select a subset of actions~\citep{dutting2022combinatorial}; related work studies combinatorial contracts beyond gross substitutes~\citep{dutting2024combinatorial}, sequential actions~\citep{ezra2024contract}, and inspections~\citep{ezra2024contracts}. Our fairness constraint concerns the utility comparisons induced by exchanging success shares within a team working on the same project.

\paragraph{Fairness in contract design.}
\citet{castiglioni2025fair} study fair contracts for projects assigned among independent agents, connecting contracting to fair allocation of goods or chores~\citep{bei2021price,caragiannis2019unreasonable}. Both that work and ours use envy-based comparisons. Our joint-project setting requires accounting for the effort response to a contract swap and its effect on both agents' payoffs through the common outcome.

Concurrent work by \citet{feldman2026equal} studies non-discriminatory (equal-pay) contracts and their revenue loss relative to unconstrained contracts. We quantify the revenue advantage of optimal fair contracts over non-discriminatory contracts, obtaining the tight factor $1+\rho\approx1.445$. For submodular rewards with binary actions, our polynomial-time algorithm uses only value queries and guarantees a $0.249$ fraction of optimal fair revenue. Its large-team candidates are non-discriminatory, while small-team enumeration can return unequal fair shares. Their value-query approximation targets optimal non-discriminatory revenue; the estimate $1/(2480e)$ applies to a subroutine handling instances without a large agent. For additive rewards, they compute an optimal non-discriminatory contract exactly in polynomial time, while we obtain an FPTAS for optimal fair contracts.

\citet{feng2024price} study equal contracts in a multi-agent setting where agents undertake their own projects and receive payments from those projects, with equal contracts imposed as a restriction. Our utility-based fairness test permits unequal shares as well as equal shares. This distinction relates to fair remuneration~\citep{gentile2024equal} and subjective fairness~\citep{yaari1984dividing}.

\paragraph{Algorithmic contract theory.}
Contract theory has a rich literature in microeconomics dating back several decades \citep[e.g.,][]{holmstrom1980theory,myerson1982optimal,carroll2015robustness,gottlieb2015simple,chade2019disentangling}. Computational approaches include combinatorial models~\citep{babaioff2006combinatorial}. In this space, computational approaches and approximation tools are usually used to understand the general problems. For example, \cite{dutting2019simple} study the approximation ability of simple linear contracts and \cite{dutting2021complexity} investigate the computational complexity of a general contract model. Further approximation results include \citet{deo2024supermodular,alon2025multi}, while learning contracts is studied by \citet{bacchiocchi2023learning,chen2024bounded,han2024learning}.

Subsequent works in computer science also introduce adverse selection into contract settings by assuming private types for agents. These include single-parameter contracts~\citep{alon2021contracts,alon2022bayesian}, multi-parameter contracts~\citep{castiglioni2021bayesian,castiglioni2022designing,guruganesh2021contracts,guruganesh2023power}, and reductions between them~\citep{castiglioni2025reduction}. Other directions include ambiguous contracts~\citep{dutting2023ambiguous} and joint information and contract design~\citep{castiglioni2025hiring}; see \citet{dutting2024algorithmic} for a survey.

\paragraph{Fairness with externalities.}
Our work is related to the literature on fair division. Specifically, our fairness notion is built upon the extensively studied notion, i.e., {\it envy-freeness}~\citep{beynier2019local,lipton2004approximately,foley1966resource}. In the context of fair division of goods, it ensures that all agents gain weakly higher utilities from their assigned set of goods than from sets assigned to other agents. Other related envy-freeness notions include envy-freeness up to one item (EF1)~\citep{budish2011combinatorial}, envy-freeness up to any item (EFX)~\citep{caragiannis2019unreasonable} and weighted-envy-freeness (WEF)~\citep{chakraborty2021weighted}.

Related work on fair division with externalities includes cake cutting~\citep{branzei2013externalities,li2015truthful} and indivisible goods~\citep{aziz2023fairness}. Most directly, \citet{velez2016fairness} studies swap-based envy comparisons in the presence of externalities. We apply this perspective to success-contingent team contracts, where the swap also changes agents' effort incentives.

\section{Model}\label{model_sec}

A principal hires a team of agents to collaborate on a joint project. 
The set of agents available for the project is $\mathcal{N} = [n] = \{1, 2, \dots, n\}$. 
Each agent can privately choose to exert effort or not, and if agent $i\in \mathcal{N} $ chooses to exert effort, he incurs {a cost $c_i \ge 0$.}
The cost of not exerting effort is {\it zero}. 
Moreover, the success of the project depends on the set of agents exerting effort. 
Specifically, let $f: 2^N \to [0, 1]$ be the success probability function, i.e., $f(S)$ is the probability that the project will be successful when the agents in $S\subseteq \mathcal{N} $ choose to exert effort. 
We assume that the success probability function $f$ is a monotone function. That is, for any set of agents $S' \subseteq S \subseteq \mathcal{N} $, we have $f(S') \leq f(S)$. 
Moreover, we assume that $f$ is normalized in the sense that $f(\emptyset) = 0$.

In our model, upon the success of the project, the principal receives a reward normalized to 1, while the failure of the project gives the principal a reward of $0$. 
Therefore, when the set of agents exerting effort is~$S$, the expected reward of the principal is also $f(S)$. 
Hence, we also refer to $f$ as the reward function in the remainder of the paper. {Throughout the paper, we define the value $\frac{0}{0} = 0$ and $\frac{0}{c}=\infty$ for any $c>0$.}

\paragraph{Team Contracts.} 
In our model, the agents' effort choices are hidden from the principal. Therefore, to incentivize the agents to exert effort, the principal needs to design contracts that provide monetary incentives to the agents. Following the literature on combinatorial contracts~\citep{dutting2023multi,dutting2022combinatorial}, we adopt {\it linear} contracts in our model. 
Specifically, the principal designs a team contract $(S,\alpha)$, 
where $S\subseteq \mathcal{N} $ is the team of agents chosen by the principal, and $\alpha = (\alpha_i)_{i\in S}$
is the set of linear contracts for incentivizing all agents in team $S$ to exert costly effort. 
That is, all agents $i\in S$ receive a reward of $\alpha_i\in [0,1]$ upon the success of the project.
All agents who are not selected for the team will receive zero payments, and they will not exert effort. 
In our paper, we focus on the widely adopted tie-breaking rule where all the agents exert effort if they are indifferent about exerting effort or not. That is, it ties-breaks in favor of the principal's utility.

Our discussions mainly focus on the pure Nash equilibrium~(\Cref{pure_nash_def}), and we leave the analysis of the mixed Nash equilibrium as potential future work.\footnote{The existence of pure Nash equilibria is guaranteed in canonical settings such as submodular reward functions (\cref{def:Submodular}). See  \citet{dutting2023multi,deo2024supermodular} for the proof of existence. For more general settings where the existence of a pure Nash equilibrium is not guaranteed, considering more general mixed Nash equilibria may be necessary.}

\begin{definition}[Pure Nash Equilibrium]
Given any team contract $(S,\alpha)$, the set of agents $E\subseteq S$ forms a pure Nash equilibrium for exerting costly effort if
\begin{align*}
\alpha_i \cdot f(E) - c_i \geq \alpha_i \cdot f(E\backslash \{i\}) &\qquad \forall i\in E,\\
\alpha_i \cdot f(E\cup \{i\}) - c_i < \alpha_i \cdot f(E) &\qquad \forall i\in S\backslash E. 
\end{align*}
Let $\mathcal{E}_{S,\alpha}$ be the set of pure Nash equilibria given the team contract $(S,\alpha)$.
\label{pure_nash_def}
\end{definition}
In our paper, we say a team contract $(S,\alpha)$ is \emph{feasible} if all agents in the team are incentivized to exert costly effort. We only focus on feasible team contracts in our paper.\footnote{If an agent in $S$ does not exert effort, the principal will exclude that agent from
the team, resulting in zero contracts for that agent by our definition of team contracts.} 
To ensure this, we must have 
$\alpha_i \cdot f(S) - c_i \geq \alpha_i \cdot f(S\backslash \{i\})$. That is, for any $i\in S$,
\begin{align*}
\alpha_i \geq \frac{c_i}{f(i\given S {\setminus \{i\}} )}, \quad\text{where }
f(i\given S {\setminus \{i\}}) \triangleq f(S) - f(S\backslash \{i\}).
\end{align*}
We assume that the above inequalities hold for all team contracts $(S,\alpha)$ considered in our paper. In the rest of the paper, we refer $\frac{c_i}{f(i\given S {\setminus \{i\}} )}$ as agent $i$'s {\it cut-off wage} in set $S$. {Note that if $f$ is an additive function, agent $i$'s cut-off wage does not depend on $S$, and it is $\frac{c_i}{f(i)}$.}
The expected revenue of the principal given team contract $(S,\alpha)$ is 
\begin{align*}
\rev(S,\alpha) = \rbr{1-\sum_{i\in S} \alpha_i} \cdot f(S).
\end{align*}

\paragraph{Fair Contracts} As illustrated in Example~\ref{unfairexample_intro}, simply maximizing the principal’s revenue in team contracts may lead to outcomes that are {\it unfair} to certain agents. In practice, such unfairness can significantly {\it disincentivize} agents from exerting effort, ultimately undermining the success of the project. This motivates the need to incorporate explicit fairness requirements into our model.

A distinguishing feature of team contracts is the presence of externalities in agents’ action choices: the action taken by one agent can influence the incentives and actions of another agent. To address this challenge, we follow the intuition of \citet{velez2016fairness} and define fairness through a {\it swap}-based comparison of agents’ contracts within the team. Our fairness notion builds on the well-established concept of envy-freeness~\citep{foley1966resource}, adapting it to settings where agents’ actions are interdependent.

\begin{definition}[Fair Team Contracts] 
\label{def:fair_contract}
A {(feasible)} team contract $(S,\alpha)$ is \emph{fair} if all agents are envy free under the swap of contracts. That is, for any pair of agents $i,j\in S$, letting $(S, \hat{\alpha})$ be the team contract {(not necessarily feasible)} where $\hat{\alpha}_i = {\alpha}_j$, $\hat{\alpha}_j = {\alpha}_i$, and $\hat{\alpha}_k = {\alpha}_k$ for all $k\in S\backslash\{i,j\}$, 
there exists an equilibrium $E \in \mathcal{E}_{S,\hat{\alpha}}$ such that 
\begin{equation}\label{fairnessconstraints_fariteacon}
\begin{aligned}
\alpha_i \cdot f(S) - c_i &\geq \alpha_j \cdot f(E) - c_i \cdot \indicate{i\in E},\\
\alpha_j \cdot f(S) - c_j &\geq \alpha_i \cdot f(E) - c_j \cdot \indicate{j\in E}.
\end{aligned}
\end{equation}
\end{definition}
Fairness compares agents within the selected team $S$. In this hypothetical swap, the hired team remains fixed while agents reoptimize their effort. An agent who stops exerting effort still receives her share under the swapped contract if the project succeeds.  Intuitively, \Cref{def:fair_contract} of fair contracts requires that both agents in each pair gain weakly higher utility from the original contract than at some common equilibrium under the new contract obtained by swapping. In Example~\ref{unfairexample_fairteamdef}, we further illustrate that the optimal solution by \cite{dutting2023multi} (also in Example~\ref{unfairexample_intro}) in team contracts is not fair either according to our definition of fairness.
\begin{example}  
  [Example \ref{unfairexample_intro} continued.]  Consider agent $1$. Swap the contracts of the two agents and obtain $\hat{\alpha} =(0.2, 0.1)$. We can see that agent $2$ will opt out in this case, and hence $E = \{1\}$. In the original contract $\alpha$, $u_{1}(\{1, 2\}, \alpha) = 0.75\times 0.1 - 0.05 = 0.025$, while $u_{1}(E, \hat{\alpha})=0.2\times 0.5 - 0.05 = 0.05$. After the swap, the agent $1$ gains a strictly higher utility, violating (\ref{fairnessconstraints_fariteacon}).
\label{unfairexample_fairteamdef}
\end{example}

The objective of the principal is to design a fair team contract that maximizes her expected revenue.
\begin{align}
\max_{(S, \alpha)} \quad & \rev(S, \alpha) \tag{Fair Team Contracts} \label{program:fair_contracts} \\
& (S, \alpha) \text{ is fair.} \nonumber
\end{align}
Let $(S^*,\alpha^*)$ be the optimal fair team contract, 
and $\opt \triangleq \rev(S^*, \alpha^*)$.

\paragraph{Submodular Rewards} In this paper, most of our results concern the settings with $f$ being submodular set functions. Note that an additive set function is a subset of submodular functions~\citep{lehmann2001combinatorial}. 

\begin{definition}[Submodularity]
\label{def:Submodular}
The reward function $f$ is a {\it submodular} function if, for any two sets $S'\subseteq S\subseteq N$ and any element $i\in [n]$, it holds that 
$f(i\given S) \leq f(i\given S')$.
\end{definition}
In \Cref{def:fair_contract} of fair team contracts, we only require the existence of an equilibrium $E$ such that the constraint (\ref{fairnessconstraints_fariteacon}) is satisfied. One crucial fact is that there may be multiple equilibria after the swap, i.e., $|\mathcal{E}_{S,\hat{\alpha}}| \ge 1$. This poses a challenge of equilibrium selection, and we defer the discussion of this problem in general environments to \cref{conclusion_section}.

However, if we restrict $f$ to be a submodular function, which is the main focus of our paper, there exists a unique equilibrium after the swap, i.e., $|\mathcal{E}_{S,\hat{\alpha}}| = 1$. This result greatly simplifies the analysis, allowing us to derive a simple structure of the optimal fair team contracts.

\begin{proposition}[Uniqueness of Equilibrium for Submodular]
Given any team contract $(S, \alpha)$ where all agents in $S$ are incentivized to exert costly effort, for any pair of agents $i,j\in S$ and the contract $\hat{\alpha}$ after swapping $i,j$, i.e., $\hat{\alpha}_i = \alpha_j$ and $\hat{\alpha}_j = \alpha_i$, there exists a unique equilibrium under contract $\hat{\alpha}$, i.e., $|\mathcal{E}_{S,\hat{\alpha}}| = 1$. 
\label{uniquen_propoistion}
\end{proposition}

\Cref{uniquen_propoistion} demonstrates the uniqueness of equilibrium in submodular settings. Since the remainder of the paper mostly focuses on submodular functions, we define $S_{i,j}$ as the unique equilibrium obtained.
 \begin{definition}
     $S_{i,j} \subseteq \mathcal{N}$ is the unique equilibrium reached under contracts $\hat{\alpha}$, which is obtained by swapping contracts $\alpha_i$ and $\alpha_j$ of agents $i, j \in S$.
 \end{definition}

%\paragraph{Primitives for Accessing $f$.} 
Finally, following the literature on combinatorial contract design \citep[e.g.,][]{dutting2023multi, dutting2022combinatorial, alon2025multi}, we assume the standard value oracles for accessing the value of function $f$: they return the value of $f(S)$ given the input set $S\subseteq [n]$.
%\begin{itemize}
%    \item {\it Value Oracle}: It returns the value of $f(S)$ given the input set $S\subseteq [n]$.
   % \item {\it Demand Oracle}: It returns a set $S\subseteq [n]$ which maximizes the value $f(S) - \sum_{i\in S}p_i$ given any price vector $p=(p_1, p_2, \dots, p_n) \in \mathbb{R}_{+}^n$.
%\end{itemize}

\paragraph{Non-discriminatory Team Contracts} A special class of contracts that is of particular interest is the non-discriminatory team contracts: all the agents $i\in S$ exerting effort receive the same payment.

\begin{definition}[Non-discriminatory Team Contracts]
\label{def:non-discriminatory}
A team contract $(S,\alpha)$ is \emph{non-discriminatory} if $\alpha_i=\alpha_j$ for all $i,j\in S$.
\end{definition}

Clearly, by swapping the contracts of any two agents $i, j \in S$, the team contract remains the same and the agents' incentives do not change, indicating the satisfaction of constraints (\ref{fairnessconstraints_fariteacon}). Hence, we have the unique equilibrium $S_{i,j} = S$. This immediately implies the following proposition.

\begin{proposition}
\label{prop:non-discriminatory_is_fair}
All non-discriminatory team contracts are fair. 
\end{proposition}

In the following, \Cref{example_not_discr} shows that the class of non-discriminatory team contracts is only one subset of fair team contracts.

\begin{example}[\Cref{unfairexample_fairteamdef} continued] By setting contract $\alpha = (0.2, 0.2)$, both two agents $S = \{1, 2\}$ will choose to exert effort. Hence, $(S, \alpha)$ is a non-discriminatory contracts and fair. There are still other fair contracts. Let the contract be $\alpha=(0.15, 0.2)$. Clearly, both agents will choose to exert efforts, $S = \{1, 2\}$. Now, we swap the contracts two agents, $\hat \alpha = (0.2, 0.15)$, and we obtain a new equilibrium $\hat{S} = \{1\}$. For agent $1$, comparing its utility obtained before and after the swap, i.e., $0.15\times 0.75-0.05 > 0.2 \times 0.5 -0.05$; for agent $2$, comparing its utility obtained before and after the swap, i.e., $0.2\times 0.75-0.05 \ge 0$. The constraints (\ref{fairnessconstraints_fariteacon}) are satisfied. Hence, $(S, \alpha)$ is fair but not non-discriminatory.
    \label{example_not_discr}
\end{example}

\section{Properties of Optimal Fair Contracts}
\label{characterizationofoptimacontracts}

In this section, we present several properties of fair contracts. Perhaps surprisingly, \cref{theorem_optimal_share} shows that the optimal fair contracts exhibit a remarkably simple structure: although fairness constraints substantially complicate the model, it suffices to impose a lower bound on agents' cut-off wages to ensure fairness. {All missing proofs are in Appendix~\ref{app_missproffsec3}.}

\begin{definition}[Minimum-share Structure]\label{minimumsharedefineiotn}
For any $\mathcal{L} \ge 0$  and any $S\subseteq [n]$, a {team contract $(S, \alpha)$} is a least $S$-incentive contract with a minimum share $\mathcal{L}$ 
if, for any agent $i\notin S$, $\alpha_i=0$,
and for any agent $i \in S$, 
\begin{align*}
\alpha_i = \max\cbr{\mathcal{L}, \frac{c_i}{f(i\given S\setminus \{i\})}}.
\end{align*}
\end{definition}

Obviously, any non-discriminatory contract has a minimum share structure where we can set the minimum share as $\mathcal{L} \ge \max_{i\in S} \frac{c_i}{f(i\given S\setminus \{i\})}$ so that all the agents $i\in S$ receive a contract $\mathcal{L}$. In this way, all agents in set $S$ will choose to exert effort, and the contract is fair by \Cref{prop:non-discriminatory_is_fair}. 
Our result will indicate that a properly chosen lower minimum share can still ensure fairness. 
Moreover, such a minimum share structure is optimal among all possible fair contracts.

\begin{theorem}[Optimality of Minimum Share]\label{theorem_optimal_share}
For the submodular reward function $f$, given any set $S\subseteq \mathcal{N}$, there exists $\mathcal{L} \geq 0$ such that a least $S$-incentive contract with a minimum share $\mathcal{L} $ provides the least expected payments to the agents among all fair team contracts that incentivize the set $S$. 
Moreover, the optimal minimum share $\mathcal{L}^*_{S} $ can be computed in polynomial time, which, given a set of incentivized agents, is defined as
\begin{equation}\label{optimal_minim_shared}
    {\mathcal{L}_{S}^* = \max_{i \in S}\Big\{ \frac{c_i}{f(i|S\setminus \{i\})} \cdot \Big( \frac{ f(S\setminus \{i\})}{f(S)} \Big)\Big\}= \max_{i \in S}\Big\{ \frac{c_i}{f(i|S\setminus \{i\})} \cdot \Big( \frac{f(S) - f(i|S\setminus \{i\})}{f(S)} \Big)\Big\}.}
\end{equation}
\end{theorem}

\noindent\textbf{Example~\ref{unfairexample_intro} continued.}
For the team $S=\{1,2\}$, \Cref{theorem_optimal_share} gives the minimum share
$\mathcal{L}_S^*=2/15$. Comparing this team with the singleton teams gives
the following optimal revenues, allowing each contract class to choose its team:
\begin{center}
\begin{tabular}{lcc}
\toprule
Contract class & Shares on $\{1,2\}$ & Optimal revenue \\
\midrule
Unrestricted & $(0.1,0.2)$ & $0.525$ \\
Fair & $(2/15,1/5)$ & $0.500$ \\
Non-discriminatory & $(0.2,0.2)$ & $0.450$ \\
\bottomrule
\end{tabular}
\end{center}
For the non-discriminatory class, hiring only agent $1$ at share $0.1$ also earns $0.450$.
Under the optimal fair contract, agent $1$ still receives a smaller expected payment
and lower utility than agent $2$: their utilities are $0.05$ and $0.10$.
After swapping, only agent $1$ exerts effort; her utility remains $0.05$,
while agent $2$'s utility falls to $1/15$.
Thus, fairness removes the gain from swapping while allowing differences in pay and utility.

Before the formal proof of \Cref{theorem_optimal_share}, we first determine the unique equilibrium (\cref{lem:S_ij}) and then develop sufficient and necessary conditions for a team contract to be fair (\cref{sufficient_necessaryforequi}).

\begin{lemma}\label{lem:S_ij}
Given any team contract $(S, \alpha)$, any pair of agents $i,j\in S$ such that $\alpha_i < \alpha_j$, 
if the team contract $(S, \alpha)$ is fair, the unique equilibrium $S_{i,j}$ after swapping the contracts for $i$ and $j$ must satisfy 
$S_{i,j} = S\setminus\{j\}$.
\end{lemma}

\begin{lemma}\label{sufficient_necessaryforequi}
For the submodular reward function $f$, a team contract $(S, \alpha)$ is fair if and only if for any pair of agents $i,j\in S$ such that $\alpha_i < \alpha_j$, the following holds:
\begin{equation}\label{upperlowerboundofcont}
{\frac{c_j}{f(j| S\setminus \{j\})} > \alpha_i \ge \alpha_j \Big(  \frac{f(S\setminus \{j\})}{f(S)} \Big).}
% ,\quad \alpha_i \ge \frac{c_i}{f(i| S\setminus\{i\})}, \quad \alpha_j \ge \frac{c_j}{f(j| S\setminus\{ j\})}
\end{equation}
\end{lemma}

Now, we are ready to present the proof of \Cref{theorem_optimal_share}.

\begin{proofof}{\Cref{theorem_optimal_share}}
Consider any given set of incentivized agents $S$ with size $|S| = h$. By \Cref{sufficient_necessaryforequi}, we know that its fair contracts $\alpha$ must satisfy (\ref{upperlowerboundofcont}). 

Without loss of generality, we sort agents in $S$ according to their cutoff wages in increasing order and relabel agents according to their rank. This implies that we must have $\alpha_i \le \alpha_j$ for $i \le j$. Otherwise, if $\alpha_i > \alpha_j$, the incentive constraints and the monotonicity in cutoff wages imply that 
\[
\alpha_j \ge \frac{c_j}{f(j| S\setminus j)} \ge \frac{c_i}{f(i| S\setminus i)}
% \frac{c_i}{f(i| S\setminus i)} > \alpha_j
\]
which violates \eqref{upperlowerboundofcont}. {Note that if \eqref{upperlowerboundofcont} is satisfied by agent $1$,  it is satisfied by all the agents in $S$. Therefore, the sufficient and necessary condition for team contract $(S, \alpha)$ to be fair are as follows: (1) By comparing all agents $i\neq 1$ with agent $1$, \Cref{sufficient_necessaryforequi} implies that 
\begin{equation}\label{inequality_fair_alaph}
\alpha_{h} \ge \alpha_{h-1}\ge \cdots \ge \alpha_1 \ge \max\Big\{\alpha_1 \Big( \frac{f(S\setminus \{1\})}{f(S)} \Big), \alpha_2 \Big( \frac{f(S\setminus \{2\})}{f(S)} \Big),  \cdots, \alpha_h \Big(  \frac{f(S\setminus \{h\})}{f(S)} \Big)\Big\}, 
\end{equation}
and (2) For any two agents $i$ and $i+1$, either $\alpha_i=\alpha_{i+1}$ or $\frac{c_{i+1}}{f(\{i+1\}|S\setminus\{i+1\})} > \alpha_{i}$ holds. 

 Recall that a team contract $(S, \alpha)$ incentivizes all the agents in $S$ to exert efforts. Hence, $\alpha_i \ge \frac{c_i}{f(i|S\setminus{i})}$ for all agents $i\in S$. This implies that the lower bound in \eqref{inequality_fair_alaph} is minimized by setting contracts as $\alpha_i=\frac{c_i}{f(i|S\setminus{i})}$ for all agents $i\in S$, which is exactly $\mathcal{L}_{S}^*$ defined in \eqref{optimal_minim_shared}.

 To maximize the principal's revenue, it is equivalent to minimize the principal's expected payment to agents. Let $k^*$ be the agent with the smallest index such that $\frac{c_{k^*}}{f(k^*|S\setminus{k^*})} \ge \mathcal{L}_{S}^*$. By the above arguments, the expected payment is minimized by setting contracts as 
\begin{equation}\label{two_sub_optimal_share}
\alpha_i =
\begin{cases} 
\frac{c_i}{f(i | S \setminus i)},  & i \ge k^* \\
\mathcal{L}_S^*, & i < k^*
\end{cases}
\end{equation}
Equivalently, the optimal contract has the minimum-share structure where for any agent $i\in S$, $\alpha_i = \max\{\mathcal{L}_S^*, \frac{c_i}{f(i|S\setminus\{i\})}\}$. The theorem is proved.}
\end{proofof}

One interesting observation from the proof of  \Cref{theorem_optimal_share} is that by increasing the minimum share to $\mathcal{L}' \ge \mathcal{L}_S^*$ and setting the contract for agent $i\in S$ as $\alpha_i' = \max\cbr{\mathcal{L}', \frac{c_i}{f(i\given S\setminus i)}}$, the fairness constraints in \eqref{upperlowerboundofcont} still hold. By \Cref{sufficient_necessaryforequi}, the team contract $(S, \alpha')$ remains fair. 
\begin{corollary}\label{lprimgretearthanlfaircontractcoro}
Consider a fair team contract $(S,  \alpha)$ with $\mathcal{L}_S^*$. 
For any $\mathcal{L}' \ge \mathcal{L}_S^*$, the team contract $S ,\alpha'$ consisting of $\alpha'_i = \max\cbr{\mathcal{L}', \frac{c_i}{f(i\given S\setminus i)}}$ for all agents $ i \in S$ is fair.
\end{corollary}

Finally, we investigate the connections among the agents' contracts. 
An interesting observation from \Cref{theorem_optimal_share} is that in the optimal fair contracts, except for at most one agent,  the difference between any two agents' contracts $\alpha_i, \alpha_j$ is indeed small.

\begin{proposition}[Difference between contracts of agents is small] \label{lm:1/2}
In any fair team contract $(S,\alpha)$ {with $f(S)>0$}, let $\hat{S} = \{\hat{i}\}$ where $\hat{i} \in S$ and $f(\{\hat i\}) > \frac{1}{2} f(S)$ if it exists; otherwise $\hat{S} = \emptyset$.
  For any pair of agents $i,j \in S \setminus \hat{S}$, we have 
       $\alpha_i \ge \frac{1}{2} \alpha_j$.
\end{proposition}
{Note that in a fair team contract $(S,\alpha)$ with $f(S)=0$, there could be infinite choices of $\alpha$. This is because $f(S)=0$ implies the cut-off wages for all the agents in $S$ are {\it zero}. Therefore, any contract with $\alpha_i\ge 0$ for $i\in S$ is feasible. Nevertheless, among all feasible contracts, the contract $\alpha_i=0, \forall i \in S$ also satisfies the property in \Cref{lm:1/2}.}

\section{Loss of Non-discrimination}\label{lossofnondiscrminatory}

In this section, we restrict our attention to one special class of simple contracts: non-discriminatory contracts.
While \Cref{lprimgretearthanlfaircontractcoro} shows that any minimum share $\mathcal{L} \ge \mathcal{L}_S^*$ immediately results in a fair contract,  one concern is that by increasing the minimum share, the principal may lose a large portion of revenue. However, the property in \Cref{lm:1/2} motivates us to conjecture that since the contracts of any two agents are close, using a non-discriminatory contract, possibly on a different team, need only lose a constant fraction of the principal's optimal fair revenue.

Given the above observation, \Cref{thm_additive_tight_gap} is interested in quantifying the worst-case approximation of the simple non-discriminatory contracts when compared with optimal contracts. In particular, 
we show a tight approximation of $1+\rho$ for additive and submodular functions $f$, where $\rho \in (0, 1)$ and it is the unique root of $\rho^3 - \rho^2 - 2\rho +1 = 0$.
\begin{theorem}[Tight loss of non-discrimination]\label{thm_additive_tight_gap}
Consider additive functions $f$. Let $\opt_{\rm nd}$ and $\opt$ be the optimal revenue of non-discriminatory contracts and fair contracts, respectively. Then, it holds that 
\[
\opt_{\rm nd}\ge \frac{1}{1+\rho}\opt.
\]
where $\rho\in (0, 1)$ is the unique root of $\rho^3 - \rho^2 - 2\rho +1 = 0$. 
Moreover, this approximation factor $1+\rho$ is tight.  The same tight approximation also holds for submodular functions $f$.
\end{theorem}

When $f$ is an additive function, there exists an algorithm that computes optimal non-discriminatory contracts. Our algorithm is depicted in Algorithm \ref{non_discriminatory_add} (in Appendix~\ref{app_missing_algo}). 

\begin{proposition}\label{them_additive_2appr}
For additive reward functions, there exists a polynomial-time algorithm that returns an optimal non-discriminatory contract. Moreover, the returned non-discriminatory contract is an $1+\rho$-approximation to the optimal fair team contract.
\end{proposition}

\subsection{A Reduction to Additive Fair  Contracts}

Before going to the proof of \cref{thm_additive_tight_gap}, we present an interesting reduction from submodular fair contracts to additive fair contracts, which will be useful in showing the approximation for submodular cases later.  Specifically, we show that if the additive non-discriminatory contracts can guarantee a $\frac{1}{\Gamma}$ fraction of revenue in additive fair contracts, then submodular non-discriminatory contracts guarantee a $\frac{1}{\Gamma}$ fraction of revenue in submodular fair contracts.

\begin{proposition}[Reduction to additive fair contracts]\label{thm_reduction_to_additive_gap}
Suppose that for every additive reward function, the optimal non-discriminatory contract achieves at least a $1/\Gamma$ fraction of the optimal fair contract. 
Then the same guarantee holds for every submodular reward function.
\end{proposition}

\begin{proof}
Fix an instance with a submodular function $f$, and let $(S^*,\alpha^*)$ be an optimal fair contract. Let $\opt^s$ be the principal's revenue from the optimal fair submodular contracts.
If $\opt^s=0$, the statement is immediate.  Hence, in the following, we assume $\opt^s>0$ and thus, $f(S^*)>0$.
Since every agent in $S^*$ is incentivized and costs are nonnegative, feasibility implies $f(i\given S^*\setminus \{i\})\ge0$ for every $i\in S^*$, and 
\[ r_i=\frac{c_i}{f(i|S^*\setminus\{i\})}\le 1.
\]
Moreover, by \Cref{theorem_optimal_share}, let $\mathcal{L}^* = \max_{i\in S^*} \frac{c_i}{f(i|S^*\setminus\{i\})} \left(1-\frac{f(i|S^*\setminus\{i\})}{f(S^*)}\right)  $ be the minimum share.

Recall that submodular functions are a subset of XOS functions. Hence, by the definition of XOS functions, there exists an additive function $a(\cdot)$ such that for the set $S^*$, $f(S^*) = \sum_{i\in S^*} a(i) = a(S^*)$, $a(T) \le f(T)$ for every $T \subseteq S^*$, and $a(i) \ge f(i| S^* \setminus \{i\})$ for every $i \in S^*$.

We construct an additive instance using the above additive function $a$. Specifically, the additive reward function is defined as $h(i) = a(i)$ for every $i \in S^*$, and $h(i) = 0$ for $i\in \mathcal{N}\setminus S^*$. Moreover, we define the costs such that $c_i^h = \frac{c_i}{f(i|S^*\setminus\{i\})} h(i)$ for every $i \in S^*$ and $c_i = 1$ for every $i \in \mathcal{N} \setminus S^*$. Hence, the cut-off wage for agent $i \in \mathcal{N} \setminus S^*$ is $\frac{c_i}{h(i)} = \frac{1}{0} = \infty$.

Let $\mathcal{L}^A$ be the optimal minimum share of the constructed additive instance on the team $S^*$. Then, we have 
\[
\mathcal{L}^A=\max_{i\in S^*} \frac{c_i^h}{h(i)}\left(1-\frac{h(i)}{h(S^*)}\right)
\le
\max_{i\in S^*} \frac{c_i}{f(i|S^*\setminus\{i\})} \left(1-\frac{f(i|S^*\setminus\{i\})}{f(S^*)}\right)
=\mathcal{L}^*.
\]
where the inequality is by $h(i) \ge f(i|S^*\setminus\{i\})$ for $i\in S^*$.  Note that the cut-off wage for $i\in S^*$ in additive instance is $\frac{c_i^h}{h(i)} = \frac{c_i}{f(i|S^*\setminus{\{i\}})} = r_i$, the same as the submodular instance. 
Therefore, the least-payment fair contract for $S^*$ in the additive instance pays each agent weakly less than $\alpha_i^*$. 
Since the total additive value of $h(S^*)=f(S^*)$, the revenue of the optimal fair additive instance, denoted by $\opt^A$, satisfies
\begin{equation}\label{eq:additive_proxy_fair_dominates}
\begin{aligned}
\opt^A
&\ge
\left(1-\sum_{i\in S^*}\max\{\mathcal{L}^A,r_i\}\right)a(S^*)\\
&\ge \left(1-\sum_{i\in S^*}\max\{ \mathcal{L}^*, r_i\}\right)f(S^*)=
\left(1-\sum_{i\in S^*}\alpha_i^*\right)f(S^*)
=\opt^s.
\end{aligned}
\end{equation}

Next, we compare non-discriminatory contracts. By our construction of the additive instance, we know that the optimal non-discriminatory contract cannot include an agent $i$ from $\mathcal{N}\setminus S^*$. Hence, 
consider any $T\subseteq S^*$. 
In the additive instance, the least non-discriminatory contract that incentivizes set $T$ is $\max_{i\in T} r_i$. 
In the original submodular instance, the corresponding non-discriminatory contract for set $T$ is
\[
\max_{i\in T}\frac{c_i}{f(i\given T\setminus\{i\})}
\le
\max_{i\in T}\frac{c_i}{f(i\given S^*\setminus\{i\})}
=
\max_{i\in T}r_i,
\]
where the inequality follows from submodularity and $T\subseteq S^*$. 
Recall that $f(T)\ge h(T)$ for every $T\subseteq S^*$. 
Hence, we have that the revenue of the optimal non-discriminatory contract for submodular reward (denoted as $\opt_{\rm nd}^s$) is at least 
Thus
\begin{equation}\label{eq:submodular_nd_dominates_proxy}
\opt_{\rm nd}^s\ge \max_{T\subseteq S^*} f(T) \left( 1 - |T| \max_{i\in T}\frac{c_i}{f(i\given T\setminus\{i\})}  \right) \ge \max_{T\subseteq S^*} h(T) \left( 1 - |T| \max_{i\in T}r_i  \right)  = \opt_{\rm nd}^A,
\end{equation}
where $\opt_{\rm nd}^A$ is the revenue of the optimal non-discriminatory contract for the constructed additive instance.

By the assumed additive guarantee, $\opt_{\rm nd}^A\ge \opt^A/\Gamma$. 
Combining this with \eqref{eq:additive_proxy_fair_dominates} and \eqref{eq:submodular_nd_dominates_proxy} gives
\[
\opt_{\rm nd}^s\ge \opt_{\rm nd}^A\ge \frac{1}{\Gamma}\opt^A\ge \frac{1}{\Gamma}\opt^s,
\]
as desired.
\end{proof}

 \subsection{Proof of \cref{thm_additive_tight_gap}} 

In this section, we present the proof of our main theorem. The road map of the proof is as follows: We first show that for additive contracts, an upper bound and lower bound of $1+\rho$ can be obtained when the size of the optimal incentivized agent set $|S^*|=2$ holds. After that, we show that when $|S^*|\ge 3$, the approximation $\opt/\opt_{\rm nd} \le  \frac{24}{17}<1+\rho$ holds. This implies that $1+\rho$ is indeed a tight approximation for cases of additive functions. Finally, by \cref{thm_reduction_to_additive_gap}, we can conclude that the same tight approximation also holds for cases of submodular functions.

\begin{proofof}{\cref{thm_additive_tight_gap}}
Let $(S^*, \alpha^*)$ be the optimal fair contract for additive instance. Moreover, let $\mathcal{L}^*$ be the optimal minimum share.
The proof proceeds by (i) first considering the case of $|S^*|=2$, where we show that the tight $1+\rho$ loss of non-discrimination is obtained; and then (ii) we show that for the case of $|S^*|>2$, the loss of non-discrimination is at most $1+\rho$.

If $\opt = 0$ or $|S^*|=1$, we immediately have the loss of non-discrimination is $1$. Hence, in the following, we consider settings of $\opt >0$ and $|S^*| >1$, and thus, $f(S^*)>0$

\begin{lemma}\label{lemma_upperbound_twoage}
    Consider settings of $|S^*|=2$. The upper bound for the loss of non-discrimination is $1+\rho$ for additive rewards. 
\end{lemma}
\begin{proof}
    Let $S^* = \{a, b\}$, and the optimal contracts are $\alpha^*_a =s \le \alpha^*_b = p$. Without loss of generality, we can assume $f(S^*) = 1$, which is obtained by scaling the function $f$ and costs. Hence, let the probabilities be $f(a) = x$ and $f(b)= 1-x$. Note that here $s, p, x \in [0, 1]$ are variables. Let $r_a, r_b$ be the cut-off wages where $r_i = \frac{c_i}{f(i)}, i\in \{a, b\}$

\begin{figure}[t]
    \centering
    \includegraphics[width=0.5\textwidth]{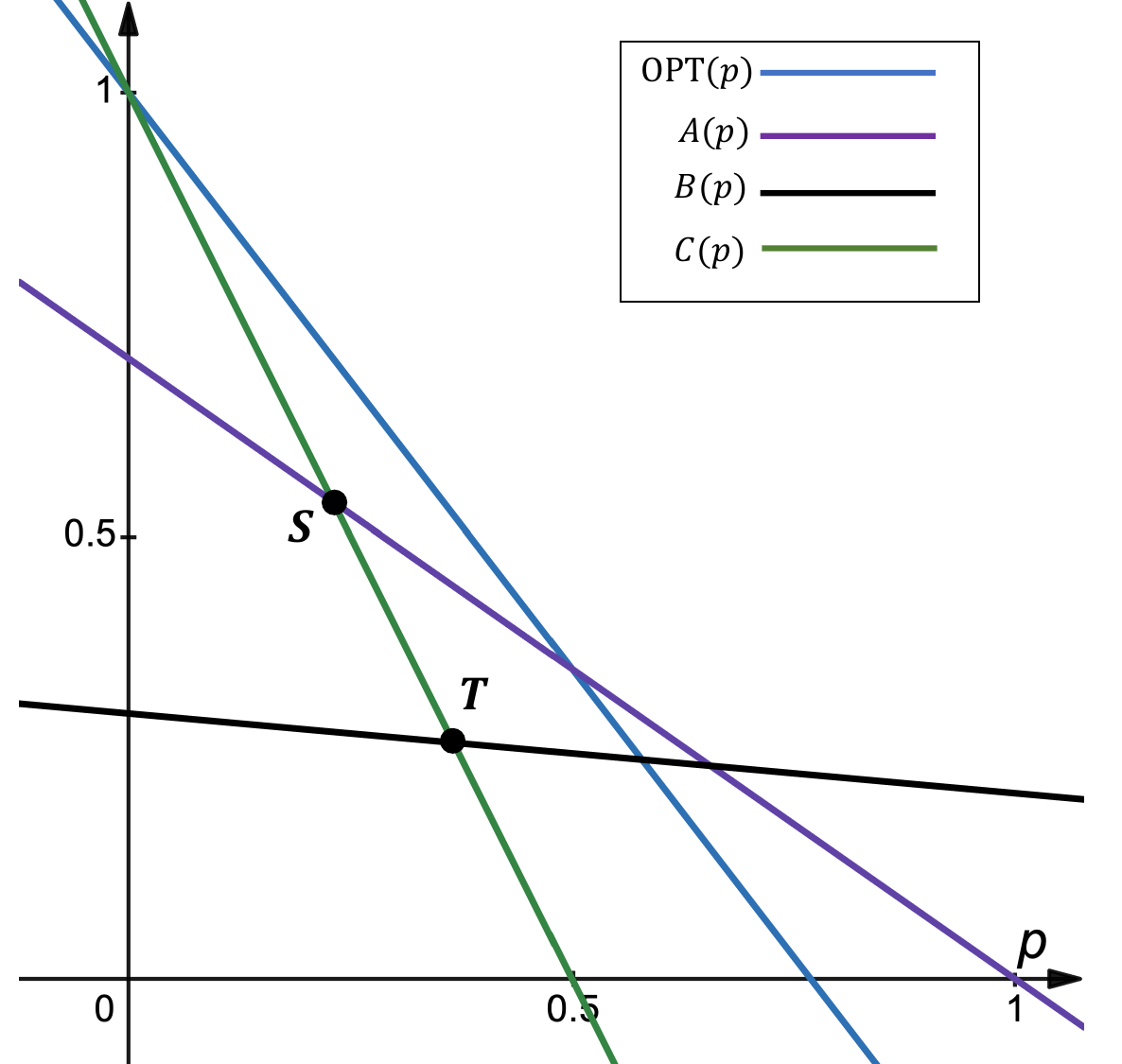}
    \caption{Illustration of $\opt(p)$, $A(p)$, $B(p)$ and $C(p)$.}
    \label{fig_my_image}
\end{figure}

    Since $|S^*| = 2$, \Cref{theorem_optimal_share} and the definition of minimum share $\mathcal{L}$ together imply $p=r_b$ and 
    \begin{equation}\label{s_lower_bound}
        s \ge p x.
    \end{equation}
    Now, consider the following non-discriminatory contracts: (i) if agent $a$ is incentivized, the optimal revenue is $(1-r_a) f(a) = (1-r_a)x$; (ii) if agent $b$ is incentivized, the optimal revenue is $(1-r_b) f(b) = (1-p)(1-x)$; and (iii) if both agents $a$ and $b$ are incentivized, the optimal revenue is $1-2p$. Hence, we have 
    
    \begin{equation}\label{opt_nd_lowerbound_tw}
        \opt_{\rm nd} \ge \max\{(1-r_a)x, (1-p)(1-x), 1-2p\}
    \end{equation}
    Notice that $s \ge r_a$. Since we aim to maximize the loss of non-discrimination, the lower bound of \eqref{opt_nd_lowerbound_tw} is minimized by $r_a = s$. This minimum can indeed be obtained in instances where the minimum share $\mathcal{L}$ is smaller than the cutoff wage $r_a$.
    
    %which is obtained by simply setting $c_a$ so that $r_a = \frac{c_a}{f(s)} = s$. Note that this is possible, as $c_a$ does not enter the calculation of revenue while only the cut-off wages and rewards matter.

    Therefore, the upper bound of loss of non-discrimination is obtained as
    \begin{equation}\label{loss_of_nond_2agent}
    \frac{1-p-s}{\max\{(1-s)x, (1-p)(1-x), 1-2p\}}
    \end{equation}
    To maximize \eqref{loss_of_nond_2agent}, observe that $\frac{1-p-s}{(1-p)(1-x)}$ and $\frac{1-p-s}{1-2p}$ increases as $s$ decreases. Moreover, $\frac{1-p-s}{(1-s)x}$ increases as $s$ decreases, since its derivative $\frac{-xp}{[(1-s)x]^2} \le 0$. Hence, we set $s = px$ by \eqref{s_lower_bound}.

Fixing $x$, consider the following function: $\opt(p) = -(1+x)p+1$, $C(p) = -2p + 1$, $A(p) = -(1-x)p + (1-x)$, and $B(p) = -x^2p + x$. They are depicted in Figure~\ref{fig_my_image}. Given that the slopes of $A(p)$ and $B(p)$ are less than $1$, we know that the maximum loss of non-discrimination \eqref{loss_of_nond_2agent} is obtained either at point $S$ or point $T$ in Figure~\ref{fig_my_image}, where $p_T = \frac{1-x}{2-x^2}$ and $p_S = \frac{x}{1+x}$. When $x =\rho$, we know that $p_T = p_S$. This is where the equation $\rho^3 - \rho^2 - 2\rho +1 = 0$ comes from.

If $x\le \rho$, we know that the maximum of  \eqref{loss_of_nond_2agent} is obtained at point $S$, which is
\[
\frac{-(1+x) \frac{x}{1+x} +1 }{-2 (\frac{x}{1+x}) + 1}  = 1+x \le 1+\rho
\]

If $x \ge \rho$, we know the the maximum of  \eqref{loss_of_nond_2agent} is obtained at point $T$, which is
\[
\frac{-(1+x) \frac{1-x}{2-x^2} +1 }{-2 (\frac{1-x}{2-x^2}) + 1} = \frac{1}{x(2-x)}\le \frac{1}{\rho(2-\rho)} = 1+\rho
\]
where the last inequality is by that $\rho^3 - \rho^2 -2\rho +1=0$ is equivalent to $1= \rho (\rho+1) (2-\rho)$.
\end{proof}

Following the same intuition of the proof for \cref{lemma_upperbound_twoage}, we can construct an instance where the lower bound of the loss of non-discrimination is at least $1+\rho$.

\begin{lemma}\label{additve_lowerbound_app}
    There exists an instance for additive rewards where the $\opt_{\rm nd} \le \frac{1}{1+\rho}\opt$.
\end{lemma}
\begin{proof}
    Following the proof of \cref{lemma_upperbound_twoage}, we construct an instance of two agents $a, b$. Let $f(a) =\rho$, $f(b) = 1-\rho$. The cutoff wages are set $r_a = \frac{\rho^2}{1+\rho}$ and $r_b = \frac{\rho}{1+\rho}$ such that $r_a = r_b (1-\frac{f(b)}{f(\{a, b\})})$, which later is shown to be the minimum share.
    Hence, we can accordingly set the costs as $c_a = r_a f(a) = \frac{\rho^3}{1+\rho}$ and $c_b = r_b f(b) = \frac{\rho(1-\rho)}{1+\rho}$.

    Consider optimal fair contracts. By \cref{theorem_optimal_share}, the revenue is as follows: (i) By hiring only agent $a$, the revenue is $(1-r_a) f(a) = \rho (1- \frac{\rho^2}{1+\rho}) = \frac{1-\rho}{1+\rho}$ where the last equality is by $\rho^3 - \rho^2 -\rho = \rho-1$; (ii) By hiring only agent $b$, the revenue is $(1-r_b) f(b) = (1-\rho) (1- \frac{\rho}{1+\rho}) =\frac{1-\rho}{1+\rho}$;  and (iii) By hiring both agents, the revenue is $(1-r_a -r_b) f(\{a, b\}) = 1-\rho$ since the minimum share $\mathcal{L} = r_a = r_b (1-\frac{f(b)}{f(\{a, b\})})$ by construction. Hence, the optimal revenue is obtained by incentivizing two agents.

    Now, consider the optimal non-discriminatory contracts. Incentivizing either agent $a$ or agent $b$ gives revenue $\frac{1-\rho}{1+\rho}$, following the above arguments. Incentivizing both agents gives a revenue of $(1-2r_b) f(\{a, b\}) = \frac{1-\rho}{1+\rho} $. Hence, the optimal non-discriminatory contract gives revenue $\frac{1-\rho}{1+\rho}$.

    Therefore, we have that
    \[
    \frac{\opt_{\rm nd}}{\opt} = \frac{(1-\rho)/(1+\rho)}{1-\rho} = \frac{1}{1+\rho}
    \]
    This concludes the proof.   
\end{proof}

    Finally, we show that for the case of $|S^*|\ge 3$, the loss of non-discrimination is less than $1+\rho$. 

\begin{lemma}\label{lem_additive_large_team_bound}
If $|S^*|\ge 3$, then $\opt_{\rm nd}\ge \frac{17}{24}\opt$ for additive rewards. 
Consequently, $\opt/\opt_{\rm nd} \le  \frac{24}{17}<1+\rho$.
\end{lemma}
\begin{proof}
    Let $(S^*, \alpha^*)$ be the optimal fair contract. Assume $|S^*| = m+1$ with $m\ge 2$. Let the optimal minimum share be $\mathcal{L}^*$. By scaling the values and costs in set $S^*$, we can, without loss of generality, assume $f(S^*)=1$.  Consider agent $h \in S^*$ with the highest payment. By \cref{theorem_optimal_share}, we know that $\alpha^*_h = r_h = \frac{c_h}{f(h)}$ and let 
    \[
    p = \alpha^*_h, \quad G=S^*\setminus \{h\}.
    \]
    If $p=0$, then the optimal non-discriminatory contract obtains the same revenue as the optimal fair contracts by paying {\it zero} to all agents. Hence, we assume $p>0$, and we define $\theta = \frac{\mathcal{L}^*}{p} \le 1$. It holds that $f(h) \ge 1-\theta$ by the definition of $\mathcal{L}^*$ such that $p(1-\frac{f(h)}{f(S^*)}) \le \mathcal{L}^*.$ Moreover, we assume $\mathcal{L}^*>0$. This assumption is without loss of generality, as it is equivalent to $f(S^*)>f(h)$; otherwise, if one has $f(S^*) = f(h)$, it is equivalent to the case $|S^*|=1$. Hence, we assume $\theta \in (0, 1]$.

By \cref{theorem_optimal_share}, we know that all the agents in $G$ are paid at least the minimum share $\mathcal{L}^*$. Hence, we have an upper bound for the optimal fair contracts as 
\begin{equation}\label{opti_rev_upp}
0<\opt \le R_0 \triangleq f(S^*)(1-p - m\mathcal{L}^*) = 1-p(1+m\theta)
\end{equation}

Notice that if the optimal fair contract pays all the agents in $G$ with the payment $\mathcal{L}^*$, then a non-discriminatory contract with a good guarantee may be obtained by incentivizing agent $h$ and as many agents in $G$ with large values as possible. Following this intuition, we consider the following non-discriminatory contracts: it incentivizes agent $h$ and the $j$ largest-value agents in $G$ where $j \in \{0, 1, \dots, m\}$; these agents are paid with contract $p$.

The total value of the incentivized $j$ largest-value agents is at least $\frac{j}{m}$ fraction of $f(G) = f(S^*)- f(h)$. Hence, we have the lower bound (denoted as $Z_j$) for the total value of incentivized agents as
\begin{equation}%\label{value_j_agent}
 Z_j \triangleq f(h) + (f(S^*)- f(h))\frac{j}{m} = f(S^*) + f(h)(1-\frac{j}{m}) \ge 1-\theta + \frac{j\theta}{m} 
\end{equation}
Therefore, we have the lower bound revenue (denoted as $L_j$) for our considered contracts as 
\begin{equation}\label{value_j_agent}
 \opt_{\rm nd} \ge L_j = Z_j (1-(j+1)p) \ge (1-\theta + \frac{j\theta}{m} )(1-(j+1)p)
\end{equation}
where we need to enusre that $1-(j+1)p \ge 0$. The first inequality in \eqref{value_j_agent} is due to that $\opt_{\rm nd}$ is the revenue of the optimal non-discriminatory contract. Therefore, by \eqref{value_j_agent} and \eqref{opti_rev_upp}, we have 
\[
\frac{\opt_{\rm nd}} {\opt}\ge \frac{L_j}{R_0}, \quad \forall j \in \{0, 1, \dots, m\}
\]

The remaining analysis is devoted to showing that the ratio $\max_{j}\frac{L_j}{R_0} \ge \frac{17}{24} > \frac{1}{1+\rho}$, concluding the proof.

\paragraph{Case: $m\theta$ is an integer.} Let $j = m\theta$. Note that we have $L_{m\theta} = (1-\theta +\theta^2)(1-(m\theta+1)p) \ge 0$ since $R_0 = (1-(m\theta+1)p) > 0$ and $1-\theta +\theta^2>0$. Hence,  we have that
\[
\max_j\frac{L_j}{R_0} \ge \frac{L_{m\theta}}{R_0}  = 1-\theta +\theta^2 \ge \frac{3}{4}>\frac{17}{24}
\]
\paragraph{Case: $m\theta$ is not an integer.} Let $\ell=\floor{m\theta}$ and $m\theta = \ell + a$ where $a \in (0, 1)$. We consider two choices: (i) incentivizing $\ell$ largest-value agents in $G$, and (ii) incentivizing $\ell+1$ largest-value agents in $G$. 

We have the ratio for these two cases as
\[
H_{\ell} = \frac{L_{\ell}}{R_0} = (A-ad)(1+a\frac{p}{R_0}), \quad H_{\ell+1} = \frac{L_{\ell+1}}{R_0} = (A+(1-a)d)(1-(1-a)\frac{p}{R_0})
\]
where $A=1-\theta+\theta^2$ and $d =\frac{\theta}{m}>0$ by $\theta\in (0, 1]$. Since $a\in (0, 1)$, it holds that $A-ad >A-d \ge 1-\frac{3}{2}\theta +\theta^2 > 0$ where the second-to-the-last inequality is by $d=\frac{\theta}{m}$ and $m\ge 2$, and $A+(1-a)d >0$. Hence, we have $H_{\ell} >0$. Clearly, the ratio is lower bounded as
\[
\max_j\frac{L_j}{R_0} \ge \max \{H_{\ell}, H_{\ell +1}\}
\]

Let $t = p/R_0$. Accordingly, define $H_{\ell}(t) = (A-ad)(1+at)$ and $H_{\ell+1}(t) = (A+(1-a)d)(1-(1-a)t)$, where $H_{\ell}(t)$ is an increasing function,  while $H_{\ell+1}(t)$ is a decreasing function. We know that 
\[H_\ell (0) = A-ad < A+(1-a)d = H_{\ell+1}(0)\]
where the strictly inequality holds due to $a \in (0, 1)$ and $d=\frac{\theta}{m} >0$. This implies that there exists some $t_0 >0$, such that $H_{\ell}(t_0) = H_{\ell+1}(t_0)$. Therefore, the minimum of $\max \{H_{\ell}(t), H_{\ell +1}(t)\}$ is attained at $t=t_0$. Let $H = H_{\ell}(t_0) = H_{\ell+1}(t_0)$. We thus have that 
\[
H = A-\frac{(1-a)ad^2}{A+(1-2a)d}, \quad t_0 = \frac{d}{A-2ad + d}
\]

Next, we bound the value of $H$. Since $(1-a)a\le \frac{1}{4}$, and $A+(1-2a)d \ge A-d >0$, we have 
\[
H \ge A - \frac{\frac{1}{4}d^2}{A-d} \ge 1-\theta+\theta^2 - \frac{\frac{1}{4}(\theta/2)^2}{1-\theta+\theta^2-\frac{\theta}{2}} \ge 1-\theta+\theta^2 - \frac{\frac{\theta^2}{16}}{\frac{7}{16}} = 1-\theta+\theta^2-\frac{\theta^2}{7}
\]
where the second inequality is by $d = \frac{\theta}{m} \le \frac{\theta}{2}$ and $A-d >0$, and the third inequality is by that $1-\frac{3\theta}{2} +\theta^2 \ge \frac{7}{16}$. Finally, we have that $H\ge \frac{17}{24}>\frac{1}{\rho+1}$. This concludes the proof.
\end{proof}

Finally, \cref{lemma_upperbound_twoage} and \cref{lem_additive_large_team_bound} together imply that for additive cases, it holds $\opt_{\rm nd}\ge \frac{1}{1+\rho}\opt$. Moreover, \cref{additve_lowerbound_app} shows that in the worst case, $\opt_{\rm nd}\le \frac{1}{1+\rho}\opt$ holds. Hence, the approximation $1+\rho$ is tight for additive settings. 
Moreover, since $\opt_{\rm nd}\ge \frac{1}{1+\rho}\opt$ holds for any instance of additive functions $f$, the same approximation holds for submodular cases by \cref{thm_reduction_to_additive_gap}. Furthermore, note that the worst-case lower bound in \cref{additve_lowerbound_app} also holds for submodular rewards cases, since additive functions are subset of submodular functions. Therefore, $1+\rho$ is also a tight approximation for submodular cases. This concludes the proof.
\end{proofof}

\section{Computations of (Approximately) Optimal Fair Contracts}

The minimum-share characterization in \cref{theorem_optimal_share} determines the optimal fair shares for a fixed team. The remaining computational challenge is selecting that team. 
Due to the inherent combinatorial structure of the team contract setting, we show that even for additive reward functions, 
computing the optimal fair team contract is \np-hard (proved in Appendix~\ref{app_nphardpproof}). 
\begin{theorem}\label{hardness_additive}
    It is \np-hard to compute an optimal fair contract with additive reward functions.
\end{theorem}

In the remaining part of the section, we utilize our  minimum-share structure in \Cref{theorem_optimal_share} to design polynomial-time algorithms for computing approximately optimal fair team contracts. In particular, \Cref{amoreformalfptasstatement} provides an FPTAS for additive 
reward functions and \Cref{thm:main} provides a constant approximation for submodular reward functions.  Missing proofs in \cref{section_additive_tpfas} are in Appendix~\ref{app_additive_fptsa_proofs}, and missing proofs in \cref{sec_submodular_approximation} are in Appendix~\ref{apx:computation},.

\subsection{An FPTAS for Additive Reward Functions}\label{section_additive_tpfas}

In this section, we show that if we restrict ourselves to additive reward functions, there is an FPTAS for computing the optimal fair team contract. The FPTAS is depicted in {Algorithm \ref{fptasforadditive}} (Appendix~\ref{app_missing_algo}). Contributions and costs are rational inputs. For this algorithm first discard agents with $f(i)=0$ or $c_i>f(i)$: the former cannot increase optimal revenue, and the latter cannot be incentivized by a share at most one. If none remain, return the empty contract; otherwise relabel the remaining agents as $[n]$, so $f(i)>0$ and $c_i/f(i)\in[0,1]$. We use the convention that the empty contract has revenue zero. The case $\gamma=1$ in the theorem is immediate from this contract; the algorithm and analysis below use $0<\gamma<1$.

\begin{theorem}[FPTAS for Additive Reward Functions]\label{amoreformalfptasstatement}
    Suppose that the success function is additive and let $\opt$ be the revenue of optimal fair contract.
    For any $\gamma \in( 0, 1]$, there exists an algorithm that returns a fair contract with revenue at least $(1-\gamma)^2 \cdot \opt$ in time polynomial in the instance size and $1/\gamma$. 
\end{theorem}

{If $\opt=0$, the approximation is immediate. Hence, in the following, we assume $\opt>0$, and thus, $f(S^*)>0$. Under such an assumption, the optimal minimum share in \cref{theorem_optimal_share} can be rewritten as
\[
\mathcal{L}_{S^*}^* = \max_{i \in S^*}\Big\{ \frac{c_i}{f(i)} \cdot \Big( 1- \frac{f(i)}{f(S^*)} \Big)\Big\}.
\]
}

Given any constant $\gamma >0$, we divide the proof for \cref{amoreformalfptasstatement} into two cases. In the first case, we show that it is sufficient to incentivize a single agent. Thereafter, we show in the second case an FPTAS where an almost optimal solution can be computed through dynamic programming approaches.
\begin{itemize}
    \item \textbf{Case 1.} There exists some agent $i\in S^*$ such that $f(i) \ge (1-\gamma) f(S^*)$.
    \item \textbf{Case 2.} All the agents $i \in S^*$ have contributions $f(i) < (1-\gamma) f(S^*)$.
\end{itemize}
If Case 1 holds, incentivizing a single agent is enough to achieve the desired approximation. 
\begin{lemma} \label{lemma:approxlarge}
    If there is an agent $i \in S^*$ such that $f(i) \ge (1-\gamma) f(S^*)$, then incentivizing only this agent gives an approximation of $1/(1-\gamma)$.
\end{lemma}

\paragraph{The FPTAS Algorithm.}
In the remainder of the proof, we will focus on Case 2. The algorithm for Case 2 is depicted in Algorithm \ref{case2algoadditive}.
To ease the exposition, we give the following definition.

\begin{definition} \label{def:minimumShare}
Given a set $S$ and a value $x\in \mathbb{R}_{\ge 0}$,  define a lower bound function as
    \begin{equation*}
\mathcal{L}(S, x) = \max_{i \in S} \Big\{ \frac{c_{i}}{f(i)}\Big( 1-\frac{f(i)}{x} \Big)\Big\}
\end{equation*}
We can see that $\mathcal{L}(S, x)$ is increasing in $x$ and $\mathcal{L}(S, x)\le \mathcal{L}(S', x)$ for any $S \subseteq S'$. Define $i^*(S)$ as the agent giving the largest contribution, i.e., 
$i^*(S) = \arg\max_{i \in S} f(i)$,
and define $\bar{i} (S, x)$ as the agent giving the largest lower bound, i.e.,
\[
\bar{i} (S, x) = \arg\max_{i \in S} \Big\{ \frac{c_{i}}{f(i)}\Big( 1-\frac{f(i)}{x} \Big)\Big\}
\]
\end{definition}

Let $(S^*,\alpha^*)$ be the optimal fair team contract. Obviously, it holds $\mathcal{L}^*_{S^*} = \mathcal{L}(S^*, f(S^*))$ by \cref{def:minimumShare}. Thus, in the following, we use $\mathcal{L}^*_{S^*}$ and $\mathcal{L}(S^*, f(S^*))$ interchangeably to represent the optimal minimum share for easy exposition.

Let $\eps = \frac{1}{2} \frac{\gamma^2}{\gamma +n(1-\gamma)} >0$. Let $\delta>0$ be the discretization step, and we discretize the success probability of each agent as $\tilde{f}(i) = \lfloor \frac{f(i)}{\delta } \rfloor \delta$. Moreover, let $\tilde{f}(S) = \sum_{i \in  S} \tilde{f}(i)$.  To ensure we find an almost optimal and fair solution, we need to address two challenges:
 
\medskip

\noindent\textbf{Challenge 1: $f(S)$ and $\tilde{f}(S)$ of the computed set $S$ should be close.}
 
\medskip

This challenge arises since we make the discretization agent by agent, 
and the discretization grid of each agent is proportional to their contribution to the success probability in order to achieve a small multiplicative approximation in polynomial time. 
However, agents are heterogeneous and their contributions to the success probability could be at different scales. 
Thus, our algorithm needs to ensure that the error is not too large after the aggregation. 

To address this challenge, in {Line \ref{line_enumeration_istar} of Algorithm \ref{case2algoadditive}}, we enumerate over all agents $i^*\in \mathcal{N}$, and for any enumerated agent $i^*$, we define the discretization step as $\delta = \frac{\eps}{2n^2} f(i^*)$. 
Intuitively, this is enumerating over all agents that have the largest contribution in the chosen set, and we fix the discretization to be proportional to this largest contribution.
This is ensured in Line \ref{every_returen_smallerethatistart} of Algorithm \ref{case2algoadditive}. 
\cref{lemma_upper_lower} shows that if the enumerated $i^*$ is indeed the agent of the largest contribution in $S$, the rewards of $f(S)$ and $\tilde{f}(S)$ will not differ too much.

\begin{algorithm}[t]
\caption{Algorithm for Case 2 (FPTAS) \\
\textbf{Input:} costs of efforts $\{c_i\}_{i \in [n]}$, an additive reward functions $f$, number of agents $n$, a constant $\gamma >0$   \\
\textbf{Output:} an incentivized set of agent $\hat{S}$.
} \label{case2algoadditive} 
$\hat{S} \gets \emptyset$, $\eps \gets \frac{1}{2} \frac{\gamma^2}{\gamma +n(1-\gamma)}$, $v_{\max} \gets 0$ \;
\For{$i^* = 1, 2, \dots, n$ \label{line_enumeration_istar}}{
   $\delta \gets \frac{\eps}{2n^2} f(i^*)$\;
   Discretize all the $f(i)$ to be $\tilde{f}(i)$  by step $\delta$ for all agent $i \in [n]$ \label{lineBuildApx}\;
   \For{$\tilde{x} = k \delta$~ in  $k =0,1, 2, \dots, \lfloor \frac{2n^3} {\eps} \rfloor $\label{line_enumeration_tildex}}{

   \For{$\bar{i} = 1, 2, \dots, n$\label{line_enumeration_bari}}{
      \If{$f(\bar{i}) \le f(i^*)  \le\tilde x/(1-\eps/(2n))$}{\label{every_returen_smallerethatistart}
         Construct domain set $S_{\bar{i}, \tilde x, i^*}$ by (\ref{domainagentsetsbaritilexistart}). \; 
         \If{$\{i^*,\bar i\}\subseteq S_{\bar i,\tilde x,i^*}$ and $\tilde f(\{i^*,\bar i\})\le\tilde x$}{
         $(S',v')\gets$ Applying dynamic programming (Algorithm~\ref{syanmicnprogram28} in \cref{app_endymica_proendom}) to Program (\ref{dynamic_pro}), returning a set $S'$ and its minimum total share $v'$\;
         \If{$S'\ne\emptyset$ and $v'\le1$}{
         $v'\gets(1-v')\tilde x$\;
         \If{ $v'> v_{\max}$}{
         $\hat S = S'$, $v_{\max} = v'$ \;
         }
         }
         }
      }
   }
   
   }
}
\Return $\hat S$.
\end{algorithm}

\begin{lemma}\label{lemma_upper_lower}
    Consider Algorithm \ref{case2algoadditive}. Let $i^*$ the agent chosen at Line \ref{line_enumeration_istar} and let $\tilde f$ be defined as per Line \ref{lineBuildApx}. For any set $S$ such that $i^* = \arg\max_{i \in S}f(i)$, we have $\tilde{f}(S) \le f(S) \le \frac{\tilde{f}(S)}{1-\eps/(2n)}$.
\end{lemma}

\medskip

 \noindent\textbf{Challenge 2: Determine the minimum share $\mathcal{L}^*_{S^*}$.}
 
 \medskip

By Theorem \ref{theorem_optimal_share}, we know that the optimal solution for the set $S^*$ consists of a minimum-share structure determined by a lower bound $\mathcal{L}^*_{S^*}$ on contracts. However, we cannot simply guess this $\mathcal{L}^*_{S^*}$ by discretization, since  $\mathcal{L}_{S^*}^*$ all determined by the set $S^*$, but finding a set $S'$ under some discrete estimate $\mathcal{L}'$ of $\mathcal{L}^*_{S^*}$ may finally lead to a minimum share $\mathcal{L}^*_{S'}$ drastically different from $\mathcal{L}_{S^*}^*$.
 
Note that given any set $S\subseteq [n]$, the optimal minimum share is computed as $\mathcal{L}(S, f(S))$.
To address the second challenge, we observe that given a set $S$, once we know $\bar{i}(S, f(S))$ (\Cref{def:minimumShare}), we can immediately calculate the minimum share $\mathcal{L}(S, f(S))$. Departure from this intuition, {Line \ref{line_enumeration_bari} in Algorithm \ref{case2algoadditive}} enumerate $\bar i$ to guess the agent $\bar{i}(S^*, f(S^*))$ of the optimal set $S^*$.

However, even if we correctly guess $\bar{i} = \bar{i}(S^*, f(S^*))$, since $f(S^*)$ is unknown, we cannot compute the minimum share. Fortunately, if we correctly guess $i^*(S^*)$ which is definitely possible by at most $n$ iterations, we should be able to have some estimated value that is close to $f(S^*)$ by  \cref{lemma_upper_lower}.

Hence, we circumvent this problem by defining $\tilde{x} = k\delta$ as an estimate to $f(S^*)$ where $k = 0,1, 2, \dots, \lfloor \frac{2n^3} {\eps} \rfloor $. We consider only guesses with $\tilde x/(1-\eps/(2n))\ge f(i^*)$. 
Finally, we define an {\it approximate minimum share} as an estimate to $\mathcal{L}^*_{S^*}$.
\begin{definition} [Approximate Minimum Share]
Given an $\bar{i}$ and  an estimate $\tilde{x}$ to $f(S^*)$, the approximate minimum share is defined as:
    \begin{equation*}
\tilde{\mathcal{L}} (\bar i, \tilde{x}/(1-\frac{\eps}{2n})) = \frac{c_{\bar{i}}}{f(\bar{i})}\Big(1-\frac{f(\bar{i})}{ \tilde{x}/(1-\eps/(2n)) } \Big).
\end{equation*}
\end{definition}
Clearly, if $\bar{i} = \bar{i}(S^*, f(S^*))$ and $\tilde{x}= \tilde{f}(S^*)$, then we have $\tilde{\mathcal{L}} (\bar i, \tilde{x}/(1-\frac{\eps}{2n})) \ge \mathcal{L}(S^*, f(S^*))$ by \cref{lemma_upper_lower}.

\medskip

 \noindent\textbf{Constructing the Dynamic Programming Problem.}
 
 \medskip

Now, we are ready to define our problem for dynamic programming. Given agents $i^*$,  $\bar i$ with $f(\bar{i}) \le f(i^*)$ and an estimate $\tilde{x}$, we define a domain set of agents aligning with the definition of $i^*$,  $\bar i$ as follows:
\begin{definition}\label{somainsetofagents}
Given an estimate $\tilde x$  and two agents $i^*, \bar i$ with $f(\bar{i}) \le f(i^*)$, we define the domain set of agents $S_{\bar i,\tilde x,i^*}$ as follows:
\begin{equation}\label{domainagentsetsbaritilexistart}
S_{\bar{i}, \tilde x, i^*}=\Big\{i \in [n] \Big| f(i) \le f(i^*) \text{\quad and \quad }   \frac{c_{i}}{f(i)}\Big(1-\frac{f(i)}{ \tilde{x}/(1-\eps/(2n)) } \Big) \le \tilde{\mathcal{L}} (\bar i, \tilde{x}/(1-\frac{\eps}{2n})) \Big\}.
\end{equation}
\end{definition}
{By \cref{somainsetofagents}, for any set $S\subseteq S_{\bar{i}, \tilde x, i^*}$ such that $i^*, \bar{i} \in S$ and $\tilde{f}(S) = \tilde{x}$, we have $i^* = i^*(S)$ and $\bar{i} = \bar{i}(S, \tilde{f}(S)/(1-\frac{\eps}{2n}))$}.  The algorithm then set the contract $\tilde{\alpha}$ as follows,

\begin{equation}\label{setcontractforaddivittildealpha}
\tilde \alpha_i =
\begin{cases} 
\tilde{\mathcal{L}} (\bar i, \tilde x/(1-\frac{\eps}{2n}))  & \text{if~} i\in S_{\bar{i}, \tilde x, i^*} \text{~and~} \frac{c_i}{f(i)} \le \tilde{\mathcal{L}} (\bar i, \tilde x/(1-\frac{\eps}{2n})), \\
\frac{c_i}{f(i)} & \text{if~} i\in S_{\bar{i}, \tilde x, i^*} \text{~and~} \frac{c_i}{f(i)} > \tilde{\mathcal{L}} (\bar i, \tilde x/(1-\frac{\eps}{2n})),  \\
0 & \text{if~} i\notin S_{\bar{i}, \tilde x, i^*}. 
\end{cases}
\end{equation}

Our algorithm then solves the exact optimum of the following problem using standard dynamic programming (Algorithm \ref{syanmicnprogram28} in  \cref{app_endymica_proendom}).
\begin{equation}\label{dynamic_pro}
\begin{aligned}
    \max_{\{i^*, \bar{i}\}\subseteq S \subseteq S_{\bar{i}, \tilde x, i^*}} & \quad (1-\sum_{i\in S}\tilde\alpha_{i}) \tilde x, \qquad\qquad
    \text{subject. to}  & \tilde{f}(S) = \tilde x
    \end{aligned}
\end{equation}
Given the enumerated agents $i^*$, $\bar i$ and the estimate $\tilde{x}$, where $\tilde{x} = k\delta$ are for $k = 0,1, 2, \dots, \lfloor \frac{2n^3} {\eps} \rfloor $, the dynamic programming approaches require a running time $O(n\cdot \frac{2n^3}{\eps})$. If the Program (\ref{dynamic_pro}) does not find a set $S$ such that $\tilde{f}(S) = \tilde x$, it returns an empty set $\emptyset$. Otherwise, 
the condition $\{i^*, \bar{i}\}\subseteq S$ ensure that $i^*, \bar{i} \in S$ holds for the returned set $S$.
We show that if $S\neq \emptyset$, the returned solution is fair.
\begin{claim}\label{dynamicprogramingreturensafaircontract}
    Given an estimate $\tilde x$  and enumerated agents $i^*, \bar i$ such that $f(\bar{i}) \le f(i^*)$, if Program (\ref{dynamic_pro}) returns a set $S\neq \emptyset$, then $(S, \{\tilde{\alpha}_i\}_{i\in S})$ is a fair contracts.
\end{claim}

\paragraph{The Returned Solution is $(1-\gamma)^2$-approximate.}
{Algorithm \ref{case2algoadditive}} returns a set $\hat S$ and its associated contract $\tilde{\alpha}(\hat{S})$ that finds the maximum approximate principal's revenue by dynamic programming iterating over all possible $i^*, \bar{i}$ and $\tilde{x}$. \cref{1minsgammaapproximate} shows that the returned solution is almost optimal.

\begin{lemma}\label{1minsgammaapproximate}
    The returned set of agents $\hat{S}$ and the least $\hat{S}$-incentive contract with a minimum share $\mathcal{L}(\hat{S}, f(\hat{S}))$ (\cref{minimumsharedefineiotn}) constitute 
     an $\frac{1}{(1-\gamma)^2}$ approximation.
\end{lemma}

\begin{proof} 
Let $\alpha(\hat S)$ be the least $\hat{S}$-incentive contract with a minimum share $\mathcal{L}(\hat{S}, f(\hat{S}))$. 
First, we  have that 
\[
(1-\sum_{i \in \hat{S}} \alpha_{i} (\hat{S}) ) f(\hat{S}) \ge (1-\sum_{i\in \hat{S} }\tilde\alpha_{i} (\hat{S})) \tilde{f}(\hat S) 
\]
where the inequality is by $f(\hat{S}) \ge \tilde{f}(\hat{S})$ and that $\big(\tilde\alpha(\hat S), \hat S\big)$ is one fair contract by \cref{dynamicprogramingreturensafaircontract} while $\big(\alpha(\hat S), \hat S\big)$ is the optimal fair contract by Theorem \ref{theorem_optimal_share}.

{Suppose in some iteration of Line \ref{line_enumeration_istar}, Line \ref{line_enumeration_tildex} and  Line \ref{line_enumeration_bari} of Algorithm \ref{case2algoadditive},} we {\it simultaneously} have  $\tilde x = \tilde{f}(S^*)$, $i^* = i^*(S^*)$ and $\bar{i} = \argmax_{i \in S^*} ~\tilde{\mathcal{L}} (\bar i, \tilde{x}/(1-\frac{\eps}{2n}))$. Note that such an iteration must exist.
Hence, $f(\bar{i}) \le f(i^*)$.
Then, by \cref{somainsetofagents}, we have $\{i^*, \bar i\}\subseteq S^* \subseteq S_{\bar{i}, \tilde{x}, i^*}$. We overload the notation and use $\tilde{\alpha}$ to denote the contract for this specific iteration, which is set as (\ref{setcontractforaddivittildealpha}).

Let $\tilde{S}$ be the set obtained by applying dynamic programming to Program (\ref{dynamic_pro}) given the above $\tilde{x}, i^*, \bar{i}$. 
We have
\[
(1-\sum_{i\in \hat{S} }\tilde\alpha_{i} (\hat{S})) \tilde{f}(\hat{S})  \ge (1-\sum_{i\in \tilde{S} }\tilde\alpha_{i} ) \tilde f(S^*) \ge (1-\sum_{i\in S^* }\tilde\alpha_{i} ) \tilde f(S^*) 
\]
where the first inequality follows since $\hat{S}$ is the optimal solution returned by the algorithm, and the second inequality follows by the guess of $\tilde x$, $i^*$ and $\bar{i}$, and from the fact that $S^* \subseteq S_{\bar{i}, \tilde{x}, i^*}$ is one feasible solution.

Additionally, we define an intermediate contract $\hat{\alpha}$ for set $S^*$. Let the set $S'$ be 
\[
S' = \Big\{i \in S^* \big| \tilde\alpha_i = \tilde{\mathcal{L}} \big(\bar i, \tilde{f}(S^*)/ (1-\frac{\eps}{2n}) \big) \Big\}
\]
We define contract $\hat{\alpha}$ as 
\begin{equation*}
\hat \alpha_i =
\begin{cases} 
 \frac{c_{\bar{i}}}{f(\bar{i})}\Big(1-\frac{f(\bar{i})}{\tilde{f}(S^*)} \Big),  & \text{if~} i \in S', \\
\tilde{\alpha}_i, & \text{if~} i \in S^*\setminus S', 
\end{cases}
\end{equation*}

Next, we show two technical lemmas. Intuitively, \cref{1minussumgreatereps} shows that in the optimal solution, the principal reserves a sufficiently large portion of reward as its own revenue. This also allows us to further derive  \cref{technicalclaimreserveparotionintermideate}. Their proofs are relegated to \cref{app_additive_fptsa_proofs}.

\begin{lemma}\label{1minussumgreatereps}
    Given the optimal set $S^*$, we have $(1-\sum_{i\in S^*} \alpha_i^*) \ge \frac{\eps}{\gamma}$.
\end{lemma}

\begin{lemma}\label{technicalclaimreserveparotionintermideate}
 It hold that $1-\sum_{i\in S^* }\tilde\alpha_{i} > 0$  and $1-\sum_{i\in S^* }\tilde\alpha_{i} \ge  \Big( 1- \sum_{i \in S^*} \hat{\alpha}_i  \Big) - \eps $.
Moreover, $ 1- \sum_{i \in S^*} \hat{\alpha}_i  \ge 1-\sum_{i \in S^*} \alpha_i^*$, where $\alpha^*$ is the optimal fair contract.
\end{lemma}

Therefore, by  \cref{lemma_upper_lower} and \cref{technicalclaimreserveparotionintermideate} that $1-\sum_{i\in S^* }\tilde\alpha_{i} > 0$,  we can further have a lower bound that 
\[
(1-\sum_{i\in S^* }\tilde\alpha_{i}) \tilde f(S^*) \ge (1-\sum_{i\in S^* }\tilde\alpha_{i} ) (1-\frac{\eps}{2n}) f(S^*)  \ge (1-\sum_{i\in S^* }\tilde\alpha_{i}) (1-\eps) f(S^*)
\]
Furthermore, by \cref{technicalclaimreserveparotionintermideate}, and \cref{1minussumgreatereps}, we have  
\[1- \sum_{i \in S^*} \hat{\alpha}_i \ge 1-\sum_{i \in S^*} \alpha_i^* \ge \frac{\eps}{\gamma},
\]
which implies that 
$
\eps < \gamma \Big( 1- \sum_{i \in S^*} \hat{\alpha}_i \Big).
$
Then, by \cref{technicalclaimreserveparotionintermideate}, we have  
\[
1-\sum_{i\in S^* }\tilde\alpha_{i}   \ge \Big( 1- \sum_{i \in S^*} \hat{\alpha}_i  \Big) - \eps\\
\ge (1-\gamma) (1-\sum_{i \in S^*} \hat\alpha_i )    
\]
Therefore, we have 
\[
(1-\eps)(1-\sum_{i\in S^* }\tilde\alpha_{i} )f(S^*)  \ge (1-\gamma)^2(1-\sum_{i\in S^* }\hat\alpha_{i} )f(S^*),
\]
where the inequality is by $\eps = \gamma \Big(\frac{1}{2} \frac{\gamma}{\gamma +n(1-\gamma)}\Big) < \gamma $. Finally, by the second part of \cref{technicalclaimreserveparotionintermideate}, we have 
\[
(1-\gamma)^2(1-\sum_{i\in S^* }\hat\alpha_{i} )f(S^*) \ge (1-\gamma)^2(1-\sum_{i\in S^* }\alpha_{i}^*)f(S^*) = (1-\gamma)^2\opt
\]
This concludes the proof.
\end{proof}

\paragraph{Complexity.} Finally, we examine the running time of Algorithm \ref{fptasforadditive}.

\begin{lemma}\label{fptas_compexity_reslut}
    The algorithm can be implemented in time $\poly(\frac{1}{\gamma} , n)$.
\end{lemma}

\begin{proofof}{Theorem \ref{amoreformalfptasstatement}}
    Finally, Algorithm \ref{fptasforadditive} selects  either a single agent by Line \ref{line_enumeraset_algo} that gives the largest revenue or a set of agents from Algorithm \ref{case2algoadditive}. The fair contracts can then be constructed as in \cref{theorem_optimal_share}.
By combining all the above results, we have an FPTAS that gives $(1-\gamma)^2$ approximation. Since $(1-\gamma)^2\ge1-2\gamma$, choosing $\gamma$ as half the requested relative error gives the usual FPTAS guarantee, with running time polynomial in the inverse of that error and the rational input length,  concluding the proof.
\end{proofof}

\subsection{A Constant Approximation for Submodular Reward Functions}\label{sec_submodular_approximation}

In this section, we present a polynomial-time algorithm that achieves a principal revenue of at least $0.249 \cdot \OPT$ for the submodular reward case, where $\OPT$ is the optimal fair revenue. In particular, we prove the following theorem:

\begin{theorem}\label{thm:main}
For every fixed $\varepsilon>0$, there exists a deterministic polynomial-time algorithm using only value oracles, and it returns fair revenue at least $(\Gamma-\varepsilon)\OPT$, such that
\begin{equation}\label{eq:constant}
 \Gamma:=\max_{t>0}(1-e^{-t})\ps\left(\frac{t^2}{1-e^{-t}}\right)
 \ge 0.24988.
\end{equation}
where for $C, P\ge 0$, the function $\ps\left(x\right)$ is defined as,
\begin{equation}\label{eq:conversion}
 H(C)=\max_{0\le r\le1}r(1-Cr)=
 \begin{cases}1-C,&C\le\frac12,\\ (4C)^{-1},&C\ge\frac12,\end{cases}
 \qquad
 \ps(P)=\left(\int_0^1\frac{dz}{H(Pz)}\right)^{-1}.
\end{equation}
\end{theorem}

Taking the fixed accuracy $\varepsilon=0.0008$ gives revenue at least $0.24908\OPT$, and hence the advertised $0.249\OPT$ guarantee.

The function $H$ is a continuous benchmark for the trade-off between reward preservation and total payment when choosing team size. The factor $\ps$ converts the greedy surrogate into revenue through the integral identity below.

Let $t_*$ denote the maximizer in \eqref{eq:constant}. The following lemma evaluates the conversion factor and locates $t_*$; its proof is in Appendix~\ref{apx:computation}. 

%Before presenting the formal algorithm for \cref{thm:main}, we show the calculation of \eqref{eq:constant} through the following lemma. 

\begin{lemma}\label{lemma_computating_psand_t}
    Let $c=\log2-1/2 >0$ be a constant. We have that 
    \begin{equation}\label{eq:psi}
 \ps(P)=\begin{cases}
 P/[-\log(1-P)],&0<P\le\frac12,\\
 P/(2P^2+c),&P\ge\frac12,
 \end{cases}\qquad \ps(0)=1.
\end{equation}
Moreover, the maximizer $t_*$ to \eqref{eq:constant} is the solution to 
\begin{equation}\label{eq:root}
 2t_*^4(1-e^{-t_*}-t_*e^{-t_*})=c(1-e^{-t_*})^3.
\end{equation}
Moreover, $t_*$ is in the range $
0.5063\le t_* \le 0.5064,$
which implies that $
 \Gamma \ge 0.24988$.
\end{lemma}

\paragraph{A Polynomial-Time Constant-Factor Approximation Algorithm.}
For a team $S$ and an agent $i\in S$, let
$r_i(S)=c_i/f(i\mid S\setminus\{i\})$ denote the cut-off wage.
When $f(S)>0$, our minimum-share characterization
(\cref{minimumsharedefineiotn,theorem_optimal_share}) gives the
optimal fair shares for this fixed team:
\[
    \alpha_i^S=\max\{r_i(S),\mathcal{L}_S^*\},
    \qquad i\in S.
\]
Thus, the main computational challenge is to select a good team.
Under submodular rewards, changing the team changes agents'
marginal contributions and hence their cut-off wages. Moreover,
under a common-share contract, including an agent with a large
cut-off wage may increase the payments to every team member.
Consequently, finding a team with high reward alone does not
guarantee high revenue. Our algorithm addresses this difficulty
by first constructing a team with a useful surrogate guarantee
and then converting this guarantee into actual revenue.

For the algorithm and its analysis, replace $\varepsilon$ by
$\min\{\varepsilon,1/16\}$ and set
$K_0=\lceil1000/\varepsilon\rceil$ and $\delta=\varepsilon/100$.
The parameter $t$ controls the number of greedy rounds. Choose rational
$t\in[0.5063,0.5064]$ satisfying $|t-t_*|\le\varepsilon/1000$;
its instance-independent computation is given immediately after the proof of
\cref{lemma_computating_psand_t} in Appendix~\ref{apx:computation}.

The algorithm, presented in Algorithm~\ref{alg:fair-team-contracts},
handles small and large teams separately. For small teams, it
enumerates every set of at most $K_0$ agents and computes its
optimal fair contract using \cref{theorem_optimal_share}.
This procedure is given in Algorithm~\ref{alg:small-teams}.
For every fixed accuracy parameter, $K_0$ is a constant, so this
enumeration takes polynomial time and includes the optimal team
whenever its size is at most $K_0$.

For larger teams, the algorithm enumerates team-size guesses $K$,
seed agents $h$, and common shares $\bar{\alpha}$. For each choice,
it defines rescaled costs $\pi_i=c_i/\bar{\alpha}$ and runs a
seeded variant of distorted greedy~\citep{harshaw2019submodular}
for $\ell=\lfloor tK\rfloor$ rounds, where $q=1-1/K$.
Starting from the seed, in round $j$ it adds an outside agent
with the largest distorted marginal gain
\[
    q^{\ell-j-1}
    \bigl[f(S\cup\{i\})-f(S)\bigr]-\pi_i,
\]
provided that this gain is positive. Since subsequent additions
can decrease the marginal contributions of previously selected
agents, the algorithm then repeatedly removes any agent satisfying
$f(i\mid S\setminus\{i\})<\pi_i$.
Each such removal increases $f(S)-\pi(S)$, and the resulting
team $U$ is feasible under the common share $\bar{\alpha}$.
For suitable guesses, the minimum-share characterization allows
us to relate the surrogate value $f(U)-\pi(U)$ directly to the
optimal fair revenue.

The remaining challenge is that a large surrogate value does not
immediately imply high contract revenue: paying the common share
to every member of $U$ may leave the principal with only a small
fraction of its reward. We address this through two deletion
procedures. First, starting from $U$, we repeatedly delete an
agent with the largest price-to-marginal ratio
$\pi_i/f(i\mid S\setminus\{i\})$. This produces a sequence of
teams that can be incentivized by progressively smaller common
shares. An integral identity relates the rewards along this
sequence to the surrogate value $f(U)-\pi(U)$.
Second, starting separately from each team in this sequence,
we repeatedly delete an agent with the smallest marginal
contribution and evaluate the contract at every remaining team
size, keeping the individual share fixed. By submodularity, repeatedly deleting a minimum-marginal agent
from a team of $k$ agents leaves, at every size $j$, a team
retaining at least a $j/k$ fraction of the original reward. Meanwhile, the
surviving agents remain incentivized, and the total payment
share decreases with the team size. Combining this trade-off
with the integral identity guarantees a candidate with high
revenue. The algorithm returns the highest-revenue candidate,
also considering the empty contract and the zero-share contract
on all zero-cost agents.

Finally, the parameter $t$ controls the number of greedy rounds
and is chosen close to the optimizer $t_*$ of our approximation
guarantee. By \cref{lemma_computating_psand_t}, $t_*$ is the
unique root of \eqref{eq:root} in $[0.5063,0.5064]$.
We apply bisection to isolate this root in an interval of length
at most $\varepsilon/1000$ and choose $t$ as its rational
midpoint, ensuring $|t-t_*|\le\varepsilon/1000$.
Under exact-real bisection,
$
    \max\left\{
        0,\left\lceil\log_2\frac{1}{10\varepsilon}\right\rceil
    \right\}
$
steps suffice, since the initial interval has length $10^{-4}$
and each step halves its length.

\begin{algorithm}[htbp]
\caption{Approximation algorithm for fair team contracts}
\label{alg:fair-team-contracts}
\footnotesize
\DontPrintSemicolon

\KwIn{Value oracle for $f$, costs $(c_i)_{i\in[n]}$,
      and accuracy $\varepsilon>0$.}
\KwOut{A feasible fair contract and its induced team.}

$\varepsilon\gets\min\{\varepsilon,1/16\}$,
$K_0\gets\lceil1000/\varepsilon\rceil$,
$\delta\gets\varepsilon/100$\;

Choose rational $t\in[0.5063,0.5064]$ satisfying
$|t-t_*|\le\varepsilon/1000$\;

$J\gets\lceil\log6/\log(1+\delta)\rceil$,
$\mathcal W\gets
\{\tfrac12(1+\delta)^j:j=0,\ldots,J\}$\;

$\mathcal C\gets
\textsc{EnumerateSmallTeams}(f,(c_i)_{i\in[n]},K_0)$\;
Add the zero-share contract on $\{i:c_i=0\}$
to $\mathcal C$\;

\BlankLine
\For{$K=K_0+1,\ldots,n$}{ \label{line_iteration_K}
    $\ell\gets\lfloor tK\rfloor$,
    $R\gets\ell+1$, $q\gets1-1/K$\;

    \ForEach{$h\in[n]$ and $w\in\mathcal W$
             with $Rw/K\le1$}{ \label{line_w_app}
        $\bar\alpha\gets w/K$,
        $\pi_i\gets c_i/\bar\alpha$ for every $i\in[n]$\;
        $S\gets\{h\}$\;

        \tcp{Seeded greedy construction}
        \For{$j=0,\ldots,\ell-1$}{ \label{line_11_algo_app}
            \If{$[n]\setminus S\ne\varnothing$}{
                Choose
                $\displaystyle
                i\in\operatorname*{arg\,max}_{a\in[n]\setminus S}
                \left\{
                    q^{\ell-j-1}
                    [f(S\cup\{a\})-f(S)]-\pi_a
                \right\}$\; \label{distortedselection_app}

                \If{$q^{\ell-j-1}
                     [f(S\cup\{i\})-f(S)]-\pi_i>0$}{
                    $S\gets S\cup\{i\}$\; 
                }
            }
        }

        \tcp{Pruning}
        \While{there exists $i\in S$ with $f(i|S\setminus i)<\pi_i$}{ \label{line19_pruning_app}
            $S\gets S\setminus\{i\}$\;
        } \label{line_21_algo_app}

        \tcp{Critical-ratio suffixes and cardinality thinning}
        \While{$S\ne\varnothing$}{ \label{pruning_s_set}
            Choose
            $\displaystyle
            i\in\operatorname*{arg\,max}_{a\in S}
            \frac{\pi_a}{f(a|S\setminus a)}$\;
            $z\gets\pi_i/f(i|S\setminus i)$, \quad $T\gets S$\; \label{line_z_and_S}

            \While{$T\ne\varnothing$}{ \label{pruning_z_start}
                Add $(T,\alpha)$ to $\mathcal C$, where
                $\alpha_a=z\bar\alpha$ for every $a\in T$\;

                Choose
                $\displaystyle
                a\in\operatorname*{arg\,min}_{b\in T}f(b|T\setminus b)$\; \label{minimum_delectionline}
                $T\gets T\setminus\{a\}$\;
            } \label{pruning_z_end}

            $S\gets S\setminus\{i\}$\;
        } \label{pruning_s_set_end}
    }
}

\BlankLine
\KwRet{$(S,\alpha)\in\mathcal C$ maximizing
       $\bigl(1-\sum_{i\in S}\alpha_i\bigr)f(S)$}\;
\end{algorithm}

% --------------------------------------------------
% Small-team enumeration
% --------------------------------------------------
\begin{algorithm}[t]
\caption{\textsc{EnumerateSmallTeams}}
\label{alg:small-teams}
\footnotesize
\DontPrintSemicolon

\KwIn{Value oracle for $f$, costs $(c_i)_{i\in[n]}$,
      and threshold $K_0$.}
\KwOut{A collection $\mathcal C$ of feasible fair contracts.}

Initialize $\mathcal C$ with the empty contract\;

\ForEach{$S\subseteq[n]$ with
         $1\le |S|\le\min\{K_0,n\}$}{
    \If{$f(S)>0$ and $r_i(S)\le1$ for every $i\in S$}{
        Set, for every $i\in S$,
        $\displaystyle
        \alpha_i^S\gets
        \max\left\{
            r_i(S),
            \max_{j\in S}
            \frac{r_j(S)f(S\setminus\{j\})}{f(S)}
        \right\}$\;

        \If{$\sum_{i\in S}\alpha_i^S\le1$}{
            Add $(S,\alpha^S)$ to $\mathcal C$\;
        }
    }
}

\KwRet{$\mathcal C$}\;
\end{algorithm}

% --------------------------------------------------
% Main algorithm
% --------------------------------------------------

\paragraph{Distorted greedy algorithms for finding candidate sets.} In this part, we analyze the distorted greedy algorithms, i.e., the Line~\ref{line_11_algo_app} to \ref{line_21_algo_app} in Algorithm~\ref{alg:fair-team-contracts}. 

Let the optimal team be $S^*$ with positive optimal revenue, i.e., $f(S^*)>0$. Suppose that in some iteration, we choose the correct size of the team (i.e., $K=|S^*|$) and the agent with the largest cut-off wage (i.e.,  $h = \arg\max_{i\in S^*} r_i(S^*)$). Define 
\[
 p=r_h(S^*),\qquad W=\frac{ f(h|S^*\setminus h)}{f(S^*)}.
\]
We first have the following technical claim:
\begin{claim}\label{technical_claim_15}
   \begin{equation}\label{eq:cost}
 \sum_{i\in S^*\setminus\{h\}}c_i\le p(1-W)f(S^*).
\end{equation}
\end{claim}

For each enumerated $w$ in Line~\ref{line_w_app}, it applies a non-discriminatory contract $\bar{\alpha} = \frac{w}{K}$ to incentivize every added agent. To do so, we reconstruct the cost $c_i$ to be $\pi_i = \frac{c_i}{\bar \alpha}$ fo every agent $i$ and then, in iteration $j$,  select the (approximately) incentivized agent $i$ for the team such that $q^{\ell-j-1}[f(S\cup\{i\})-f(S)]-\pi_i>0$. Since the agents selected are approximately incentivized (Line~\ref{distortedselection_app}), we finally perform pruning (Line~\ref{line19_pruning_app} to Line~\ref{line_21_algo_app}) so that all the selected agents are incentivized to exert effort under contract $\bar{\alpha}$. 
The following lemma shows that the distorted algorithm returns a good candidate set $U$  such that $f(U)-\pi(U)$ achieves a controlled fraction of the optimal value $f(S^*)$.
\begin{lemma}\label{lem:greedy}
Given the correct guesses $K = |S^*|$ and $h = \arg\max_{i\in S^*} r_i(S^*)$, define $A=1-q^\ell(1-W)$ where $q=1-\frac{1}{K}$. For each enumerated $w$, construct new cost $\pi_i = \frac{c_i}{\bar{\alpha}}$ for each agent $i$, where $\bar{\alpha} = \frac{w}{K}$.  The greedy output $U \subseteq \mathcal{N}$ satisfies
\begin{equation}\label{eq:greedy}
 G:=f(U)-\pi(U)
 \ge\left[A-\frac{Kp}{w}\left(W+\frac\ell K(1-W)\right)\right]f(S^*),
\end{equation}
where $\pi(S)=\sum_{i\in S}\pi_i$ and $\ell = \lfloor tK\rfloor$. Also $|U|\le R = \ell+1$, and $f(i|U\setminus i)\ge\pi_i$ for $i\in U$.
\end{lemma}

\paragraph{Further pruning the candidate sets from distorted greedy algorithms.}

In this part, we let the set returned by the greedy algorithm (\cref{lem:greedy}) be $U\subseteq \mathcal{N}$. Moreover, we know that 
 $|U|\le R$ and common share $\bar{\alpha} = \frac{w}{K}  = \frac{Rw/K}{R}$ where $Rw/K \le 1$ (by Line~\ref{line_w_app}). For general purposes, we let  $\bar\alpha=P/R$, $P = Rw/K\le1$, and $R\ge4$.

 Consider the pruning algorithm from Line~\ref{pruning_s_set} to Line~\ref{pruning_s_set_end}. Let $S_j$ be the team after $j$ deletions in this part, with $S_0=U$. Let $i_j$ be the agent deleted at step $j$, and let the record ratio be
\[
\lambda_j=\frac{\pi_{i_j}}{f(i_j|S_{j-1}\setminus i_j)}.
\]
\begin{lemma}\label{monotone_price_to_margianl}
    The deletion process (from Line~\ref{pruning_s_set} to Line~\ref{pruning_s_set_end}) generates a sequence where
    \[
 1\ge\lambda_1\ge\lambda_2\ge\cdots\ge0.
\]
\end{lemma}

Next, we show that the distorted revenue term $G:=f(U)-\pi(U)$ can be written as an integral, which will be useful for our later analysis. 
For $z\in[0,1]$, delete all records with $\lambda_j>z$, leaving the set $S(z)$, and put $V(z)=f(S(z))$. We have the following technical lemma.
\begin{lemma}\label{integration_equality_revenue}
    The distorted revenue returned can be written as
\begin{equation}\label{eq:integral}
 \int_0^1V(z)\,dz
 =\sum_j(1-\lambda_j)v_j=f(U)-\pi(U)=G.
\end{equation}
\end{lemma}

Next, we show the key lemma for our analysis of the pruning. In particular, we analyze the 
\begin{lemma}\label{lemma_z_pruning_lowerbound}
    For every ratio $z \in [0, 1]$, all the sets of agents $T$ generated by pruning (Line~\ref{pruning_z_start} to Line~\ref{pruning_z_end}) can be incentivized by the contract $z\bar{\alpha}$. Moreover, for every ratio $z \in [0, 1]$, there exists one candidate set of agents among those generated by Line~\ref{pruning_z_start} to Line~\ref{pruning_z_end} such that the principal's revenue from non-discriminatory contract $z\bar{\alpha}$ is at least 
    \[
    (H(Pz)-Pz/R^2)V(z).
    \]
    where $H(Pz)$ is defined as \eqref{eq:conversion}.
\end{lemma}
\begin{proof}
Recall that for every agent $i\in U$, $f(i|U\setminus i) \ge \frac{c_i}{\bar{\alpha}}$. For every set $S\subseteq U$ and $z=\frac{\pi_i}{ f(i|S\setminus i)}=\frac{c_i}{f(i|S\setminus i) \bar{\alpha}}$ at Line~\ref{minimum_delectionline}, we have $\bar{\alpha}z \ge \frac{c_i}{f(i|S\setminus i)}$ by submodularity. 
At every deletion, the remaining terminal marginals weakly increase, so the contract $z\bar\alpha$ remains sufficient to induce all remaining agents. This proves the first part of the lemma.

Note that for any team $S$ of size $s$, submodularity implies that its smallest terminal marginal is at most $f(S)/s$. If $i$ has that minimum marginal, then
\[
 f(S\setminus\{i\})=f(S)-f(i|S\setminus i)
 \ge\left(1-\frac1s\right)f(S).
\]

    Now, we consider a ratio $z$ and the set of agents $S$ of size $k$, as in Line~\ref{line_z_and_S}. 
Repeated deleting the minimum-marginal agent (Line~\ref{minimum_delectionline}) from size $k$ down to size $j$ therefore retain at least the fraction
\[
 \prod_{s=j+1}^{k}\left(1-\frac1s\right)
 =\prod_{s=j+1}^{k}\frac{s-1}{s}=\frac jk
\]
of $V(z)$, for $1\le j\le k$.  
Therefore, when the remaining set of agents has size $j\le k\le R$ and the non-disriminatory contract is $z\bar{\alpha} = \frac{Pz}{R}$, the principal's revenue is at least 
\begin{equation}\label{eq:subset}
 \frac jk\left(1-\frac{Cj}{R}\right)V(z)
 =\frac{y}{k/R}(1-Cy)V(z).
\end{equation}
where we let $C=Pz$ and $y=j/R$. Here $0\le C\le1$ and $j\le k\le R$, so $1-Cy\ge0$. 
Choosing the remaining cardinality balances reward preservation against total payment. A suitable admissible cardinality guarantees a revenue factor of at least $H(Pz)-Pz/R^2$, with the correction accounting for integer rounding.
The cardinality rounding and finite-candidate argument completing the proof
are in Appendix~\ref{app:ratio-thinning-completion}.
\end{proof}

Finally, we reach the main result of this part, where we show that one can find a set of agents, by incentivzing which with non-discriminatory contracts, the principal can retains some fraction of $G$ in \cref{lem:greedy} as its revenue.

\begin{proposition}\label{proposition_prunining}
    Let the set returned by the greedy algorithm (\cref{lem:greedy}) be $U\subseteq \mathcal{N}$, and the common share is $\bar\alpha=P/R$, $P = Rw/K\le1$, and $R\ge4$. By applying the pruning process (Line~\ref{pruning_s_set} to Line~\ref{pruning_s_set_end}), the best principal revenue $M$ obtained among all the candidate sets satisfies
    \begin{equation}\label{eq:revenue}
 M\ge\left(1-\frac4{R^2}\right)\ps(P)G
 \ge\left(\ps(P)-\frac4{R^2}\right)G.
\end{equation}
where the agents are incentivized by contract $z\bar{\alpha}$ for some $z\in [0, 1]$.
\end{proposition}
\begin{proof}
    By \cref{lemma_z_pruning_lowerbound}, we know that 
for every $z$, a candidate of revenue at least
\[
 \left(H(Pz)-\frac{Pz}{R^2}\right)V(z)
 \ge\left(1-\frac4{R^2}\right)H(Pz)V(z),
\]
where the second inequality follows because $Pz\le1$ and $H(Pz)\ge1/4$. 
Let $M$ be the best candidate revenue over all candidate sets in the iteration. Since the above lower bound holds for each $z$, we must have that 
\[
 M\ge\left(1-\frac4{R^2}\right)H(Pz)V(z)
 \qquad\text{for every }z\in[0,1].
\]
Since $R\ge4$, $1-4/R^2>0$, and $H(Pz)>0$. Dividing by these factors gives
\[
 V(z)\le\frac{M}{(1-4/R^2)H(Pz)}.
\]
Integrate this pointwise upper bound and use \cref{integration_equality_revenue}:
\[
 G\le\frac{M}{1-4/R^2}\int_0^1\frac{dz}{H(Pz)}.
\]
By the definition of $\ps$, the integral on the right is $1/\ps(P)$. Rearranging, and using $G\ge0$, gives
\begin{equation}
 M\ge\left(1-\frac4{R^2}\right)\ps(P)G
 \ge\left(\ps(P)-\frac4{R^2}\right)G,
\end{equation}
where the second inequality follows from $\ps(P)\le1$. $\ps(P)\le1$ is due to that $H(Pz)\le1$, which makes its reciprocal integral at least one. 
%Finally, the integral is an analysis tool, not an instruction to enumerate continuously many shares. There are only $|U|+1$ suffixes. If a nonempty suffix is $S_j$ on the ratio path, its largest current ratio is $\lambda_{j+1}$. For every $z$ that produces this suffix, $\lambda_{j+1}\le z$. The algorithm uses share $\lambda_{j+1}\bar\alpha$, which is sufficient to induce the suffix and every subset on its thinning chain. For a fixed retained team, lowering its common share weakly increases revenue. Thus each candidate compared at a continuous value of $z$ is dominated by an actual candidate in Step 3. Ties in the recorded ratios cause no problem: the suffix after all records strictly above $z$ is still on the path, and the algorithm includes it.
\end{proof}

\paragraph{Proof of \cref{thm:main}.} Here $\varepsilon$ denotes the capped accuracy, so its guarantee implies the guarantee for the original input accuracy. If $\OPT=0$, the empty contract suffices. If $|S^*|\le K_0$, Algorithm~\ref{alg:small-teams} includes an optimal fair contract on $S^*$. Otherwise, use $K,h,p,W$ defined above. If $p=0$, all agents in $S^*$ have zero cost, and the zero-share contract on $\{i:c_i=0\}$ earns at least $\OPT$. We may therefore assume $\OPT>0$, $|S^*|>K_0$, and $p>0$.

If only the agent $h$ is incentivized, then assign the agent with contract $\frac{c_h}{f(h)}$, the principal gain revenue
\begin{align*}
 (1-\frac{c_h}{f(h)}) f(h)
 &\ge 
 (1-p)Wf(S^*)\\
 &\ge W\OPT.
\end{align*}
The first inequality is by submodularity, $p = \frac{c_h}{f(h|S^*\setminus h)}\ge \frac{c_h}{f(h)}$ and $f(h) \ge f(h|S^*\setminus h) = Wf(S^*)$. The second inequality is by \cref{theorem_optimal_share} that in the optimal solution, $h$ receives a contract $p = r_h(S^*)$ and other incentivized agents receives non-negative contracts. If $W> \frac{1}{4}$, then incentivizing a single agent gives the desired approximation.  Hence, in the following, we assume that $W\le \frac{1}{4}$.

\Cref{theorem_optimal_share} gives an upper bound on the optimal fair revenue.  We know the minimum-share term corresponding to agent $h$ is
\[
 r_h(S^*)\frac{f(S^*\setminus\{h\})}{f(S^*)}
 =p\left(1-\frac{f(h|S^*\setminus h)}{f(S^*)}\right)=p(1-W).
\]
Every other member of $S^*$ therefore also receives at least $p(1-W)$. The total share is at least
\[
 p+(K-1)p(1-W)=p\bigl[K-(K-1)W\bigr].
\]
Consequently,
\begin{equation}\label{eq:benchmark}
 \OPT\le\bigl[1-p(K-(K-1)W)\bigr]f(S^*).
\end{equation}
By \cref{lem:greedy}, we know that $G \ge \left[A-\frac{Kp}{w}\left(W+\frac\ell K(1-W)\right)\right]f(S^*)$, where $A=1-q^\ell(1-W)$. Indeed, if we let 
\begin{equation}\label{eq:cap}
 w^*=\frac{W+(\ell/K)(1-W)}{A(1-W+W/K)} \ge 0,
\end{equation}
then \eqref{eq:benchmark} implies that 
\[
G \ge \left[A-\frac{Kp}{w^*}\left(W+\frac\ell K(1-W)\right)\right]f(S^*)\ge A\cdot \OPT.
\]
However, the algorithm does not know $W$, neither the $w^*$. Therefore, we find $w^*$ by grid searching, as in Line~\ref{line_w_app}. That is, the algorithm finds a $w$ from a discrete set $W$, such that $w \in [w^*, (1+\delta)w^*]$ and $P(w) \le 1$ by Line~\ref{line_w_app}. If such a $w$ exists, then we have $P(w) = Rw/K \le (1+\delta)Rw^*/K = (1+\delta) P(w^*)$, and it continues to hold that 
\[
G \ge \left[A-\frac{Kp}{w}\left(W+\frac\ell K(1-W)\right)\right]f(S^*)\ge \left[A-\frac{Kp}{w^*}\left(W+\frac\ell K(1-W)\right)\right]f(S^*) \ge A\cdot \OPT.
\]
Moreover, \cref{proposition_prunining} implies that one can find a set of agents such that the obtained revenue $M$ satisfies
$M\ge\left(\ps(P(w))-\frac4{R^2}\right)G$. If we have $\left(\ps(P(w))-\frac4{R^2}\right) \ge 0$, together with the above inequalities imply 
\begin{equation}\label{approximation_not_finished}
M\ge\left(\ps(P(w))-\frac4{R^2}\right)G \ge \left(\ps(P(w))-\frac4{R^2}\right)A\cdot \OPT \ge \left(A\ps(P(w))-\frac4{R^2}\right) \OPT
\end{equation}
where the last inequallity is by that $A\le 1$. To check $\left(\ps(P(w))-\frac4{R^2}\right) \ge 0$, we notice that since $H(C)\ge1/4$ for any $C\in [0, 1]$ by definition and $P(w) \le 1$, we have $\ps(P(w))\ge1/4$. On the other hand, $K>K_0\ge16000$ by our choice of $K_0$ and $\eps$, so $4/R^2<1/4$. Therefore $\ps(P(w))-\frac{4}{R^2}>0$. The remaining analysis devotes to showing that such a $w$ exists and $A\ps(P(w))$ is not far from $(1-e^{-t})\ps\left(\frac{t^2}{1-e^{-t}}\right)$.

For the chosen horizon $t \in [0.5063, 0.5064]$, we define the following two limiting quantities:
\begin{equation}\label{eq:limit}
 a(W)=1-e^{-t}(1-W),\qquad
 P(W)=\frac{t^2+tW/(1-W)}{a(W)}.
\end{equation}
These two terms are derived by taking $K\to\infty$ so that $\ell/K\to t$, $R/K\to t$, and $q^\ell\to e^{-t}$, which imply $A \to a(W)$ and $P(w^*) \to P(W)$. Notice that at $W=0$, $a(W)p(W)$ is precisely the approximation term in \eqref{eq:constant}.

First, we show that in one iteration of Line~\ref{line_iteration_K} and Line~\ref{line_w_app}, the algorithm picks an $w \in \mathcal{W} $ such that \eqref{approximation_not_finished} holds. In particular, the finite-size and grid estimates in \cref{lem:finite}, stated and proved
in Appendix~\ref{statement_proof_lemmafinite}, ensure that a scale $w\in\mathcal W$
with the required proximity to $w^*$ is enumerated. The next lemma identifies
$W=0$ as the worst case for the limiting coefficient.

\begin{lemma}\label{lem:worst}
For $t\in[0.5063,0.5064]$ and $0\le W\le1/4$, it holds that 
\[
 a(W)\ps(P(W))\ge(1-e^{-t})\ps\left(\frac{t^2}{1-e^{-t}}\right).
\]
\end{lemma}

Combining \cref{lem:worst} with the finite-size and grid estimates in
\cref{lem:finite} and the continuity estimates in \cref{lipschitzcontinuous}
(stated and proved in Appendix~\ref{app_continuity_lemma}) gives
\[
 \frac M\OPT\ge\Gamma-4|t-t_*|-\frac{60}K-\delta
 \ge\Gamma-\varepsilon.
\]
Appendix~\ref{app:submodular-proof-completion} bounds the losses from finite $K$, the share grid, and the rational horizon $t$.
This proves the constant approximation bound.

\textbf{Validity of the returned contract and the running complexity.} We have already shown in \cref{proposition_prunining} that the returned set of agents can be incentivized by contract $\bar{\alpha} z$ for some $z\in [0, 1]$. Moreover, the size of returned set is at most $R$ by \cref{lem:greedy}. Hence, the total contract used to incentivize agents is $Rz\bar{\alpha} \le \frac{R w}{K} =P(w)<1$. These common-share candidates are fair by \cref{prop:non-discriminatory_is_fair}. Small-team candidates have total share at most one and are fair by construction; so are the empty and zero-share candidates. Clearly, once given a fixed $\varepsilon$, $K_0$ and the size of $\mathcal{W}$, i.e., $J$ are constants. The total number of enumerations is thus polynomial. Moreover, our algorithm only requires value queries to oracles. 

Hence, our algorithm runs in polynomial time. This completes the proof.

\section{Conclusions}\label{conclusion_section}

This paper studies fair contracts in team-working settings and characterizes (approximately) optimal fair contracts for submodular and additive reward functions.
A natural extension is to consider reward functions beyond the submodular class. Under our fairness definition, more general functions raise the additional concern of equilibrium selections. As illustrated in \Cref{XOS_notunique}, when $f$ is an XOS function—the next class above submodular functions in the complement-free hierarchy \citep{lehmann2001combinatorial}—a given contract $(S,\alpha)$ may admit multiple equilibria. Even verifying whether there exists an equilibrium in the equilibria set that satisfies the fairness constraints for XOS functions can be computationally challenging.

One way to address the above issue is to focus on either an optimistic or pessimistic equilibrium, maximizing or minimizing the principal's revenue, respectively, thereby reducing equilibrium selection to an optimization problem. Another approach is to consider mixed Nash equilibria, so that agents' expected utilities before swapping are higher on average. We leave the study of fair contracts under more general reward functions as an open direction for future work.

\section*{Disclosure of AI Use}
We used AI tools (GPT-6.1 Sol) for language polishing and manuscript revision.
In particular, the initial version contained a human-derived proof of an (approximately) $0.004$
approximation guarantee. We subsequently used AI to develop the
proof of Theorem~\ref{thm:main}, improving this guarantee to $0.249$. The authors verified the correctness and originality of all content, including references.

\newpage
\bibliographystyle{apalike}
\bibliography{ref}

\newpage
\appendix

\section{Proof of \cref{uniquen_propoistion}}

\begin{proof}
{Note that by \cref{def:fair_contract},} any new equilibrium set of agents $\hat{S}$ for exerting effort must be a subset of $S$, i.e., $\hat{S}\subseteq S$. 

Moreover, there exists at most one agent in $\{i,j\}$ who receives a linear contract that is strictly lower given $\hat{\alpha}$ than $\alpha$. 
Without loss of generality we say this agent is $i$. 
For any agent $k\in S\backslash\{i\}$, we have 
\begin{align*}
\hat{\alpha}_k \geq \alpha_k 
\geq \frac{c_k}{f(k| S\setminus \{k\})} \ge \frac{c_k}{f(k| \hat{S}\setminus \{k\})}
\end{align*}
where the last inequality holds since $f$ is submodular and $\hat{S}\subseteq S$. 
Therefore, agent $k$ has an incentive to exert effort. 
Given our tie breaking rule, we must have $k\in \hat{S}$. 
Finally, whether agent $i$ exerts effort or not depends on whether 
\begin{align*}
\hat{\alpha}_i \geq \frac{c_i}{f(i| S\setminus \{i\})}.
\end{align*}
The unique equilibrium $\hat{S}$ is either $S$ or $S\backslash\{i\}$, depending on the above inequality. 
\end{proof}

\section{Missing Proofs in \cref{characterizationofoptimacontracts}}\label{app_missproffsec3}

\subsection{Proof of \cref{lem:S_ij}}

\begin{proof}
Given the proof in \cref{uniquen_propoistion}, the unique equilibrium $S_{i,j} \in \{S, S\setminus\{j\}\}$. 
In the case that $\alpha_i < \alpha_j$, if $S_{i,j} = S$, 
the team contract $(S, \alpha)$ is not fair for agent $i$ since, after swapping with agent $j$, the set of agents exerting effort remains the same while agent $i$ will receive a strictly higher linear contract $\hat{\alpha}_i = \alpha_j > \alpha_i$. 
Therefore, the unique equilibrium here must satisfy $S_{i,j} = S\setminus\{j\}$.
\end{proof}

\subsection{Proof of \cref{sufficient_necessaryforequi}}
\begin{proof}
For any pair of agents $i,j$, if their linear contracts satisfy $\alpha_i = \alpha_j$, swapping contracts would not affect the equilibrium, and hence the fairness constraints in (\ref{fairnessconstraints_fariteacon}) are satisfied. 
Therefore, we only need to consider the case where $\alpha_i \neq \alpha_j$. We assume $\alpha_i < \alpha_j$ without loss of generality.

\paragraph{Only if.} Suppose $(S, \alpha)$ is a fair contract. By \Cref{lem:S_ij}, we have $S_{i,j} = S\setminus \{j\}$. The fairness constraint (\ref{fairnessconstraints_fariteacon}) for agent $i$ implies $\alpha_i f(S) -c_i \ge \alpha_j f(S_{i,j})-c_i$. Therefore, 
\[
{\alpha_i \ge \alpha_j \cdot \frac{f(S \setminus \{j\})}{f(S)} }.
\]
Moreover, since agent $j\notin S_{i,j}$, we have that after swapping contracts with agent $i$, the utility of agent $j$ satisfies
$\alpha_i f(S_{i,j}) > \alpha_i f(S) - c_j$, which implies 
\[\alpha_i < \frac{c_j}{f(j| S\setminus \{j\})}.\]
Combining the inequalities, the only if direction holds.

\paragraph{If.} We first show that when \Cref{upperlowerboundofcont} holds, $S\setminus j$ is an equilibrium after the swap.  {After swapping the contracts, any agent $k \in S \backslash \{j\}$ continues to exert effort in equilibrium due to submodularity. } 
Moreover, agent $j$ prefers to shirk after swapping the contract since $\frac{c_j}{f(j| S\setminus \{j\})} > \alpha_i$, which further implies that $\alpha_i f(S) -c_j < \alpha_i f(S\setminus \{j\})$.
Therefore, $S_{i,j} = S\setminus j$ is the unique equilibrium after the swap.

Finally, to show that the team contract $(S, \alpha)$ is fair, we only need to show that the fairness constraints (\ref{fairnessconstraints_fariteacon}) hold. 
{Since $\alpha_i \ge \alpha_j \Big( \frac{f(S\setminus \{j\})}{f(S)} \Big)$, which implies that $\alpha_i f(S) -c_i \ge \alpha_j f(S\setminus \{j\})-c_i$, 
the fairness constraint (\ref{fairnessconstraints_fariteacon}) holds for agent $i$. }
Moreover, we have
$\alpha_j f(S) -c_j \ge \alpha_i f(S\setminus \{j\})$. 
This is because 
\[
\alpha_j f(S) - \alpha_i f(S\setminus \{j\}) \ge \alpha_j f(S) - \alpha_j f(S\setminus \{j\}) = \alpha_j f(j|S\setminus j) \ge c_j
\]
where the last inequality holds by the IR constraint.
Therefore, the fairness constraint (\ref{fairnessconstraints_fariteacon}) holds for agent $j$ as well.
\medskip

Combining the above two parts of the analysis, we conclude the proof.
\end{proof}

\subsection{Proof of \cref{lm:1/2}}

\begin{proof}
First, we show that for any $i \in S \setminus \hat{S}$, it holds that $f(i|\hat{S}) \le \frac{1}{2} f(S)$. 
If $\hat{i}$ does not exits, then we must have that $S \setminus \hat{S} = S$ and for any $i\in S \setminus \hat{S}$, it holds that $f(i|\hat{S}) \le f(i) \le \frac{1}{2} f(S)$ by submodularity. If $\hat{i}$ exists, we also have the statement hold; otherwise, if there exists some $i \in S \setminus \hat{S}$ such that $f(i|\hat{S}) > \frac{1}{2} f(S)$, then $f(\hat{S}) + f(i|\hat{S})  > f(S)$ which is a contradiction.

For any other two agents $i,j \in S \setminus \hat{S}$ with contracts $\alpha_j > \alpha_i$, by  \cref{sufficient_necessaryforequi}, we know that
\[\alpha_i \ge \alpha_j (1-\frac{f(j|S\setminus \{j\})}{f(S)}) \ge \alpha_j (1-\frac{f(j|\hat{S})}{f(S)}) \ge \frac{1}{2}\alpha_j\]
where the second inequality follows from submodularity and $\hat{S} \subseteq S\setminus \{j\}$ by definition.
\end{proof}

\section{Missing Proofs in \cref{lossofnondiscrminatory}}

\subsection{Proof of \cref{them_additive_2appr}}

\begin{proof}
    Given any incentivized set of agents $S\subseteq [n]$, the non-discriminatory contract $\alpha$ with minimum expected payment is $\alpha_i = \max_{k \in S}\frac{c_k}{f(k)}$ for agent $i\in S$ and $\alpha_i = 0$ for agent $i\notin S$. Moreover, given some non-discriminatory contract $p \in [0, 1]$, the principal's objective is to find a set of agents $S\subseteq [n]$ maximizing its own revenue
    \[
    \max_{S:~ c_i/f(i) \le p, \forall i \in S}~ (1-|S|\cdot p) f(S)
    \]
    Since the principal pays a constant payment $p$ to every agent $i\in S$, the principal's revenue is thus maximized by greedily adding an agent with the largest $f(i)$ and cut-off wage smaller than $p$, i.e., $\frac{c_i}{f(i)} \le p$. 
    
Since our algorithm finds the optimal non-discriminatory contracts in additive cases, the returned contract achieves $1+\rho$-approximation by combining \Cref{thm_additive_tight_gap}. This concludes the proof.
\end{proof}

\section{ \cref{XOS_notunique} and Its Calculation} \label{app_cal_xos_exa}

\begin{example}[Multiple Equilibria in XOS]\label{XOS_notunique}  Consider an example of $n=4$ agents without normalization for easy exposition. The agents' costs are $c_1 = 0.34, c_2 = 0.25, c_3 =0.5, c_4 = 0.08$. Let $\omega_1, \omega_2, \omega_3$ be three additive functions: $\omega_1 (\{1\}) = \omega_1(\{2\}) =6, \omega_2(\{3\}) = \omega_2(\{4\}) = 6.5, \omega_2 (\{1\}) = \omega_2(\{2\}) =\omega_1(\{3\}) = \omega_1(\{4\}) = 0, \omega_3(\{1\}) = \omega_3(\{2\})=2,  \omega_3(\{3\})= \omega_3(\{4\})=5$ . Define the reward function $f$ using these three additivie functions $\omega_1, \omega_2, \omega_3$. Let the contracts be $\alpha = (0.35,0.27,0.28,0.10)$. Under this contract, the equilibrium set is $S = [n]$. Next, we swap contracts of agent $2$ and $3$, and get $\alpha' = (0.35,0.28,0.27,0.10)$. Under $\alpha'$, we have three equilibria $\mathcal{E}_{S,\alpha'} = \Big\{ \{1, 2\}, \{3, 4\}, \{1, 2, 3, 4\} \Big\}$. If we choose the equilibrium $\{1, 2\}$, we can verify that $(S, \alpha)$ is a fair contract. However, if we choose the equilibrium $\{3, 4\}$, $(S, \alpha)$ is not a fair contract. The calculation is relegated to Appendix \ref{app_cal_xos_exa}.
\end{example}

\paragraph{Calculation of Equilibrium.} The agents' costs are $c_1 = 0.34, c_2 = 0.25, c_3 =0.5, c_4 = 0.08$. Let $\omega_1, \omega_2, \omega_3$ be three additive functions: $\omega_1 (\{1\}) = \omega_1(\{2\}) =6, \omega_2(\{3\}) = \omega_2(\{4\}) = 6.5, \omega_2 (\{1\}) = \omega_2(\{2\}) =\omega_1(\{3\}) = \omega_1(\{4\}) = 0, \omega_3(\{1\}) = \omega_3(\{2\})=2,  \omega_3(\{3\})= \omega_3(\{4\})=5$ . Define the monotone reward function $f$ using these three additivie functions $\omega_1, \omega_2, \omega_3$. Let the contracts be $\alpha = (0.35,0.27,0.28,0.10)$. Under contract $\alpha$, we verify that the equilibrium set is $S = [n]$.
\begin{itemize}
    \item For agent $1$, $\alpha_1 f(S) -c_1 = 4.56 \ge \alpha_1 (S\setminus \{1\}) =  4.55$;
    \item For agent $2$, $\alpha_2 f(S) -c_2 = 3.53 \ge \alpha_2 (S\setminus \{2\}) =  3.51$;
    \item For agent $3$, $\alpha_3 f(S) -c_3 = 3.42 \ge \alpha_3 (S\setminus \{3\}) =  3.36$;
        \item For agent $4$, $\alpha_4 f(S) -c_4 = 1.32 \ge \alpha_4 (S\setminus \{4\}) =  1.2$;
\end{itemize}
Hence, $S = [n]$ is an equilibrium. The agents' utilities are $(4.56, 3.53, 3.42, 1.32)$.

Next, we swap contracts of agent $2$ and $3$, and get $\alpha' = (0.35,0.28,0.27,0.10)$. Under $\alpha'$, there are three equilibria $\mathcal{E}_{S,{\alpha'}} = \Big\{ \{1, 2\}, \{3, 4\}, \{1, 2, 3, 4\} \Big\}$. Consider equilibrium set $ \Big\{ \{1, 2\}, \{3, 4\}\Big\}$.

Under contract $\alpha'$, we verify that the set $\{1, 2\}$ is an equilibrium . 
\begin{itemize}
    \item For agent $1$, $\alpha_1 f(\{1, 2\}) -c_1 = 3.86 \ge \alpha_1 f(\{2\}) =  2.1$;
    \item For agent $2$, $\alpha_2 f(\{1, 2\}) -c_2 = 3.11\ge \alpha_2 f(\{1\}) =  1.68$;
    \item For agent $3$, $\alpha_3 f(\{1, 2, 3\}) -c_3 = 2.74 < \alpha_3 f( \{1, 2\}) =  3.24$;
        \item For agent $4$, $\alpha_4 f(\{1, 2, 4\}) -c_4 = 1.12 < \alpha_4 f(\{1, 2\}) =  1.2$;
\end{itemize}
Hence, $\{1, 2\}$ is an equilibrium. The agents' utilities are $(3.86,3.11,3.24,1.20)$. We can see that when considering this equilibrium, both agents $2$ and $3$ gain more utilities from contracts $(S, \alpha)$, and hence $(S, \alpha)$ is fair.

Under contract $\alpha'$, we verify that the set $\{3, 4\}$ is an equilibrium . 
\begin{itemize}
    \item For agent $1$, $\alpha_1 f(\{1, 3, 4\}) -c_1 = 4.21 < \alpha_1 f(\{3, 4\}) =  4.55$;
    \item For agent $2$, $\alpha_2 f(\{2,3, 4\}) -c_2 = 3.39 < \alpha_2 f(\{3, 4\}) =  3.64$;
    \item For agent $3$, $\alpha_3 f(\{3, 4\}) -c_3 = 3.01 \ge \alpha_3 f( \{4\}) = 1.755$;
        \item For agent $4$, $\alpha_4 f(\{3, 4\}) -c_4 = 1.22 \ge  \alpha_4 f(\{3\}) =  0.65$;
\end{itemize}Hence, $\{3, 4\}$ is an equilibrium. The agents' utilities are $(4.55,3.64,3.01,1.22)$. We can observe that agent $2$ gains more utility from the equilibrium $\{3, 4\}$ than $(S, \alpha)$, i.e., $3.64 > 3.53$. Hence, $(S, \alpha)$ is not fair if considering $\{3, 4\}$.

\section{Missing Proofs in \cref{sec_submodular_approximation}}
\label{apx:computation}

\subsection{Proof of \cref{lemma_computating_psand_t}}
\begin{proof}
    We first show the calculation of \eqref{eq:psi}. For $0<P\le1/2$, every $Pz$ is in the first branch of $H$, and
\[
 \int_0^1\frac{dz}{H(Pz)}
 =\int_0^1\frac{dz}{1-Pz}
 =\left[-\frac1P\log(1-Pz)\right]_{z=0}^{z=1}
 =\frac{-\log(1-P)}P.
\]
For $P>1/2$, split the integral at $z=1/(2P)$:
\begin{align*}
 \int_0^1\frac{dz}{H(Pz)}
 &=\int_0^{1/(2P)}\frac{dz}{1-Pz}
   +\int_{1/(2P)}^1 4Pz\,dz\\
 &=\frac{\log2}{P}+2P-\frac1{2P}
 =2P+\frac cP.
\end{align*}
Taking reciprocals proves \eqref{eq:psi}. At $P=0$, the integral is one. Hence, the continuity of $\ps(P)$ holds at $P=0$. 

Next, we prove the range of the maximizer $t_*$. Let $a(t) = 1-e^{-t}$, and $P(t) = t^2/a(t)$. Then, $P(t)$ is increasing in $t$ as $P'(t) = \frac{2t}{a(t)}-\frac{t^2e^{-t}}{a(t)^2}>0$, which is due to $a(t) > te^{-t}$ for $t>0$.
The limit of $P(t)$ is zero at $t\downarrow0$ and infinity at $t\to\infty$. Hence, there is unique one $\tilde{t}$ such that  $P(\tilde{t}) = 1/2$ where $\tilde{t} \in (0.410,0.411)$.

We first show that $a(t) \ps(P(t))$ is strictly increasing in $t \le \tilde{t} <0.411$. Observe that for $P\le \frac{1}{2}$, we have 
\[
 \ps(P)=\int_0^1(1-P)^s\,ds,
\]
obtained by integrating the exponential $e^{s\log(1-P)}$ in $s$. Note that 
$\ps'(P)=-\int_0^1s(1-P)^{s-1}\,ds <0.$, implying $\ps(P)$ is decreasing in $P$. Hence, for any $t\le \tilde{t}$, we have $\ps(P(t))\ge\ps(1/2)>0.72$. 

By differentiating $\ps'(P)$ once more, we have $
 \ps''(P)=\int_0^1s(s-1)(1-P)^{s-2}\,ds\le0, 
$
implying that $\ps'(P)$ decreases in $P$. Since $\ps'(P) < 0$, so its magnitude is largest at $P=1/2$, and we have 
 $\ps'(P) \ge \ps'(1/2)=-\frac{1-\log2}{(\log2)^2}>-0.65$.

For $t\le0.411$, $e^{-t}>0.662$ for any $t \le \tilde{t}$. Also $a(t)\le t$ for $t>0$, so
\[
 a(t)P(t)'=2t-\frac{t^2e^{-t}}{a(t)}
 \le t(2-e^{-t})<0.411(2-0.662)<0.55.
\]
Therefore, for any $t \le \tilde{t}$, we have the derivative as:
\[
 \frac{d}{dt}[a(t)\ps(P(t))]
 =e^{-t}\ps(P(t))+a(t)\ps'(P(t))P(t)'
 >0.662\cdot0.72-0.65\cdot0.55>0.
\]
implying that $a(t) \ps(P(t))$ is strictly increasing.

Now, consider $t\ge \tilde{t}$. The second branch of \eqref{eq:psi} applies,  giving that 
\[
 a(t)\ps(P(t))=\frac{t^2}{2t^4/a(t)^2+c} > 0,
 \qquad \frac1{a(t)\ps(P(t))}=\frac{2t^2}{a(t)^2}+\frac c{t^2}.
\]
Differentiate the reciprocal explicitly:
\begin{align*}
 \frac{d}{dt}\left[\frac{2t^2}{a(t)^2}+\frac c{t^2}\right]
 &=\frac{4t}{a(t)^2}-\frac{4t^2e^{-t}}{a(t)^3}-\frac{2c}{t^3}\\
 &=\frac2{t^3}\left[\frac{2t^4(a(t)-te^{-t})}{a(t)^3}-c\right].
\end{align*}
whose sign is completely determined by
\[
 \frac{2t^4(a(t)-te^{-t})}{a(t)^3}-c.
\]
The fraction $\frac{2t^4(a(t)-te^{-t})}{a(t)^3}$ is positive, since $a(t)-te^{-t}>0$. The derivative of $a-te^{-t}$ is $te^{-t}$, so the logarithmic derivative of $\frac{2t^4(a(t)-te^{-t})}{a(t)^3}$ is
\[
 \frac4t+\frac{te^{-t}}{a-te^{-t}}-\frac{3e^{-t}}a>\frac1t>0,
\]
The last inequality follows from $e^{-t}/a(t)<1/t$ and the positivity of $\frac{te^{-t}}{a-te^{-t}}$. Hence the $\frac{2t^4(a(t)-te^{-t})}{a(t)^3}$ increases strictly.

Its limit at $t=0$ is zero, as $\frac{2t^4(a(t)-te^{-t})}{a(t)^3} = 2t^^3 (\frac{t^3}{a(t)^3})(\frac{(a(t)-te^{-t})}{t^2})$  and that $a/t\to1$ and $(a-te^{-t})/t^2\to1/2$ when $t\to 0$; its limit $t \to \infty$ is infinity because $a\to1$ and $te^{-t}\to0$. It therefore equals the positive constant $c$ exactly once. This is that for some $t^*$,  the equality \eqref{eq:root} holds.
Note that for any function $f(x)>0$, we have $\left( \frac{1}{f(x)} \right)' = -\frac{f'(x)}{f(x)^2}$. It implies that $ a(t)\ps(P(t))$ increases first till $\eqref{eq:root}$ holds, then decreases.

Finally, we show that in $ 0.5063\le t_* \le 0.5064$. By the above arguments, we know that $t\in (0, t_*]$, $ a(t)\ps(P(t))$ increases; and $ a(t)\ps(P(t))$ decreases for $t\ge t_*$. Hence, $t_*$ is the maximizer.

By substituting $t = 0.5063$, we have that 
\[
 \frac{2t^4(1-e^{-t}-te^{-t})}{(1-e^{-t})^3}-c <0,
\]
Moreover, by substituting $t = 0.5064$, we have that $\frac{2t^4(1-e^{-t}-te^{-t})}{(1-e^{-t})^3}-c >0$. This implies that there must exist $T^* \in [0.5063, 0.5064]$ such that $\frac{2t^4(1-e^{-t}-te^{-t})}{(1-e^{-t})^3}-c =0$. This completes the proof.
\end{proof}

\subsection{Proof of \cref{technical_claim_15}}
\begin{proof}
    Use maximality of $p$ and then subtract the marginal of $h$:
\begin{align*}
 \sum_{i\in S^*\setminus\{h\}}c_i
 &\le p\sum_{i\in S^*\setminus\{h\}}f(i|S^*\setminus i)\\
 &\le p\bigl[f(S^*)-f(h|S^*\setminus h)\bigr]
 =p(1-W)f(S^*).
\end{align*}
\end{proof}

\subsection{Proof of \cref{lem:greedy}}
\begin{proof}
Before pruning, let $S_j$ be the team after $j$ rounds. We analyze the expression $q^{\ell-j}f(S_j)-\pi(S_j)$. Its change in round $j$ is exactly
\begin{align*}
 &\bigl[q^{\ell-j-1}f(S_{j+1})-\pi(S_{j+1})\bigr]
  -\bigl[q^{\ell-j}f(S_j)-\pi(S_j)\bigr]\\
 &=\frac1Kq^{\ell-j-1}f(S_j)
   +q^{\ell-j-1}\bigl[f(S_{j+1})-f(S_j)\bigr]
   -\bigl[\pi(S_{j+1})-\pi(S_j)\bigr].
\end{align*}
If an agent is added, the last two terms are its score. If the round performs no addition, these terms are zero. Thus the change is the best positive score, or zero, plus $q^{\ell-j-1}f(S_j)/K$.

During these rounds, $h$ has not been pruned, so $h\in S^*\cap S_j$. There are at most $K$ agents in $S^*\setminus S_j$, none of which is $h$. 
%The best positive score, or zero, is at least the average of their scores padded by zero entries to a total of $K$ entries. This remains valid even if that average is negative. 
Submodularity and monotonicity give
\begin{align*}
 \sum_{i\in S^*\setminus S_j}\bigl[f(S_j\cup\{i\})-f(S_j)\bigr]
 &\ge f(S_j\cup S^*)-f(S_j)\\
 &\ge f(S^*)-f(S_j).
\end{align*}
%The first inequality follows by adding those agents one at a time: each marginal in the sum is evaluated at the smaller set $S_j$, so it is at least the corresponding marginal in the telescoping addition sequence. 
Also $\pi(S^*\setminus S_j)\le\pi(S^*\setminus\{h\})$, because all prices are nonnegative. Therefore in round $j$, the best positive score, or zero, is at least
\[
 \frac1K\left[q^{\ell-j-1}\bigl(f(S^*)-f(S_j)\bigr)-\pi(S^*\setminus\{h\})\right].
\]
Adding this lower bound to the term $q^{\ell-j-1}f(S_j)/K$ cancels $f(S_j)$. The change in round $j$ is consequently at least
\[
 \frac1K\left[q^{\ell-j-1}f(S^*)-\pi(S^*\setminus\{h\})\right].
\]
Now sum over $j=0,\ldots,\ell-1$. Since $1-q=1/K$,
\[
 \frac1K\sum_{j=0}^{\ell-1}q^{\ell-j-1}=1-q^\ell.
\]
The initial expression is $q^\ell f(\{h\})-\pi_h$, and the final expression is $f(S_\ell)-\pi(S_\ell)$. Rearranging the telescoping sum gives
\[
 f(S_\ell)-\pi(S_\ell)
 \ge(1-q^\ell)f(S^*)+q^\ell f(\{h\})-\pi_h-\frac\ell K\pi(S^*\setminus\{h\}).
\]
The three benchmark substitutions are
\[
 f(\{h\})\ge Wf(S^*),\qquad
 \pi_h=\frac{KpW}{w}f(S^*),\qquad
 \pi(S^*\setminus\{h\})\le\frac{Kp(1-W)}w f(S^*),
\]
where the third follows from \eqref{eq:cost}. They yield
\[
 f(S_\ell)-\pi(S_\ell)
 \ge\left[1-q^\ell+q^\ell W
 -\frac{Kp}{w}\left(W+\frac\ell K(1-W)\right)\right]f(S^*),
\]
which is \eqref{eq:greedy} before pruning. Since pruning only increase the objective, the approximation continues to hold.

The output contains at most the seed and $\ell$ added agents, proving $|U|\le R$, and the stopping rule gives $f(i|U\setminus i)\ge\pi_i$. In particular, \eqref{eq:cost} also implies $\pi(U)\le\sum_i f(i|U\setminus i)\le f(U)$, so $G\ge0$ for every pruned output.
\end{proof}

\subsection{Proof of \cref{monotone_price_to_margianl}}

\begin{proof}
    The conclusion holds mainly due to submodularity. Suppose $i$ survives a deletion of $i_j$ from a current team $S$. Submodularity gives
\[
 f(S\setminus\{i_j\})-f(S\setminus\{i,i_j\})
 \ge f(S)-f(S\setminus\{i\}).
\]
 The price $\pi_i$ does not change, so the ratio $\pi_i/f(i|S\setminus i)$ weakly decreases. This includes zero-price agents under the convention $0/0=0$. Initially $f(i|U\setminus i)\ge\pi_i$, so every ratio is at most one, and a positive price has a positive marginal. Removing a current maximum therefore gives
\[
 1\ge\lambda_1\ge\lambda_2\ge\cdots\ge0.
\]
All the recorded marginals are nonnegative by monotonicity.
\end{proof}

\subsection{Proof of \cref{integration_equality_revenue}}
\begin{proof}
    Since the records are nonincreasing, the deleted records form a prefix of the path. If the remaining set is nonempty, its largest current ratio is the next recorded ratio, which is at most $z$. Thus, for $z>0$ and each survivor,
\[
 f(i|S(z)\setminus i)\ge\frac{\pi_i}{z}
 =\frac{c_i}{z\bar\alpha},
 \qquad (z\bar\alpha)f(i|S(z)\setminus i)\ge c_i.
\]
Hence common share $z\bar\alpha$ induces that suffix. If $z=0$, every current ratio is zero, which implies every surviving price, and therefore every surviving cost, is zero. The zero-share contract induces those agents under the effort tie-breaking rule. Empty suffixes are feasible as well.

 Let $v_j = f(i_j|S_{j-1}\setminus i_j)$ at step $j$. 
The value of a suffix telescopes along the rest of the same path:
\[
 V(z)=\sum_{j:\lambda_j\le z}v_j.
\]
Indeed, the value removed at each later step is precisely its recorded $v_j$, and the terminal empty set has value zero. On the full path,
\[
 f(U)=\sum_jv_j,\qquad
 \pi(U)=\sum_j\pi_{i_j}=\sum_j\lambda_jv_j.
\]
The equality $\pi_{i_j}=\lambda_jv_j$ also holds when both the price and marginal are zero. Because the sum is finite, it can be interchanged with the integral. Each layer $v_j$ is present precisely for $z\in[\lambda_j,1]$, an interval of length $1-\lambda_j$. Consequently,
\begin{equation}
 \int_0^1V(z)\,dz
 =\sum_j(1-\lambda_j)v_j=f(U)-\pi(U)=G.
\end{equation}
More explicitly, the contribution of that layer is
\[
 \int_0^1 v_j\,\mathbf1\{\lambda_j\le z\}\,dz=(1-\lambda_j)v_j.
\]
Thus $G$ is exactly the area under the retained-value function $V$. No loss has occurred in this identity, including when several critical ratios are equal.
\end{proof}

\subsection{Completion of the proof of \cref{lemma_z_pruning_lowerbound}}\label{app:ratio-thinning-completion}
\begin{proof}
Recall that $U$ is the pruned greedy output, $|U|\le R$,
$\bar\alpha=P/R$, $0\le P\le1$, and $R\ge4$.
For $z\in[0,1]$, let $S(z)$ be the suffix remaining after all records
$\lambda_j>z$ have been deleted, and let $V(z)=f(S(z))$.
If $S(z)$ is empty, the empty contract attains the zero bound.
Otherwise put $k=|S(z)|$ and $C=Pz$. For a remaining cardinality
$1\le j\le k$, write $y=j/R$. The main-text argument proves feasibility
at common share $z\bar\alpha$ and the revenue lower bound
\eqref{eq:subset}. Here $0\le C\le1$ and $j\le k\le R$, so
$1-Cy\ge0$. We lower bound its coefficient $\frac{y}{k/R}(1-Cy)$ by the following cases.

    \textbf{Case 1: $C\le1/2$.} Let $j=k$, i.e., keeping all the agents. The revenue coefficient is
\[
 1-Ck/R\ge1-C=H(C),
\]
because $k/R\le1$.

\textbf{Case 2: $C>1/2$ and $k/R\le1/(2C)$.} Again let $j=k$. Then $Ck/R\le1/2$, so revenue is at least $V(z)/2$. Since $C>1/2$, we have $1/(4C)\le1/2$, giving revenue at least $H(C)V(z)$. 

\textbf{Case 3: $C>1/2$ and $k/R>1/(2C)$.} Choose
\[
 j=\left\lfloor\frac{R}{2C}\right\rfloor,\qquad y=j/R.
\]
This is a valid cardinality: $C\le1$ and $R\ge4$ imply $R/(2C)\ge R/2\ge2$, so $j\ge1$; the case assumption implies $j<k$. Moreover,
\[
 0\le\frac1{2C}-y<\frac1R,
 \qquad Cy\le\frac12.
\]
In \eqref{eq:subset}, the factor $1/(k/R)$ is at least one, and the remaining factors are nonnegative. Thus its right side is at least $y(1-Cy)V(z)$. Completing the square gives the exact identity
\[
 y(1-Cy)=\frac1{4C}-C\left(y-\frac1{2C}\right)^2
\]
The difference from the continuous maximum is quadratic in the cardinality rounding error. Substituting the bound below $1/R$ gives
\[
 y(1-Cy)\ge\frac1{4C}-\frac C{R^2}=H(C)-\frac C{R^2}.
\]
This prove the second part of the lemma.

Finally, note that while we consider $z$ chosen as in Line~\ref{minimum_delectionline}, the analysis is the similar if we choose any $z$ such that the set $S$ is the same.  Hence, the conclusion holds for any $z\in [0, 1]$.
\end{proof}

\subsection{Statement and Proof of \cref{lem:finite}}\label{statement_proof_lemmafinite}
For $t\in[0.5063,0.5064]$, $K\ge1000$, and $0\le W\le1/4$, recall
\[
 \ell=\lfloor tK\rfloor,\qquad R=\ell+1,\qquad q=1-1/K,
 \qquad A=1-q^\ell(1-W).
\]
Let $w^*$ be the quantity in \eqref{eq:cap}, let $P(w)=Rw/K$,
and let $a(W),P(W)$ be the limiting quantities in \eqref{eq:limit}.
For the grid assertion, use the capped accuracy $0<\varepsilon\le1/16$,
$K_0=\lceil1000/\varepsilon\rceil$, $\delta=\varepsilon/100$,
$J=\lceil\log6/\log(1+\delta)\rceil$, and
$\mathcal W=\{\tfrac12(1+\delta)^j:j=0,\ldots,J\}$.

\begin{lemma}\label{lem:finite}
For $K\ge1000$, $t\in[0.5063,0.5064]$, and $0\le W\le1/4$, it holds that $P(W)<0.78$ and the finite quantities in \eqref{eq:cap} obey
\begin{equation}\label{eq:finite}
 |A-a(W)|\le\frac2K,\qquad
 \left|\frac RK w^*-P(W)\right|\le\frac{40}K,
 \qquad \frac12<w^*<3.
\end{equation}
For $K>K_0=\lceil 1000/\eps \rceil$, $\frac12<w^*<3$ implies the existence of $w \in \mathcal{W}$ such that $w\in [w^*, (1+\delta)w^*]$ and $P(w)<0.9$, so the algorithm picks such a $w$ during enumeration.
\end{lemma}

\begin{proof}
\textbf{Proving $P(W)<0.78$.} Notice that $P(0) = \frac{t^2}{1-e^{-t}}$ is increasing in $t \in [0, 1]$. 
We have the following bounds
\[
 0.602<e^{-t}<0.603,\quad 0.397<1-e^{-t}<0.398,
 \quad 0.645<P(0)<0.646
\]
hold on this horizon interval; rational certificates are given in the appendix. Differentiate \eqref{eq:limit} using $a'(W)=e^{-t}$:
\begin{align*}
 P'(W)
 &=\frac{\dfrac{t}{(1-W)^2}a(W)
     -e^{-t}\left(t^2+\dfrac{tW}{1-W}\right)}{a(W)^2}\\
 &=\frac{t\bigl[1-e^{-t}+e^{-t}W^2-te^{-t}(1-W)^2\bigr]}
         {(1-W)^2a(W)^2}.
\end{align*}
Further notice that 
\[
 1-e^{-t}+e^{-t}W^2-te^{-t}(1-W)^2> 1-e^{-t} - te^{-t} >0.
\]
where the second inequality is by $e^t>1+t$ for $t>0$. Therefore $P'(W)>0$.

Since $P$ increases in $W$, its maximum on this interval is at $W=1/4$. At that endpoint,
\[
 t^2+\frac{tW}{1-W}\le0.5064^2+0.5064/3=0.42524096,
\]
while $a(W)=1-e^{-t}+e^{-t}/4>0.397+0.602/4=0.5475$. Consequently $P(W)<0.78$ throughout the interval.

\textbf{Proving $|A-a(W)|\le\frac2K$.} From $\ell=\lfloor tK\rfloor$ and $R=\ell+1$,
\[
 0\le t-\frac\ell K<\frac1K,
 \qquad 0<\frac RK-t\le\frac1K.
\]
Since $A-a(W)=(e^{-t}-q^\ell)(1-W) = (e^{-t}-e^{\ell \log q})(1-W)$, it can be seen that $e^{\ell \log q} = e^{\lfloor tK\rfloor \log (1-1/K)} \approx e^{-t}$ if $K$ is sufficiently large. Our next analysis is to prove this intuition. 

Expand $-\log(1-1/K)$ into its positive power series,
\[-\log\left(1-\frac1K\right)=\frac1K+\sum_{j=2}^{\infty}\frac{1}{jK^j} \le \frac{1}{K} + \frac{1}{2}\sum_{j=2}^{\infty}\frac{1}{K^j}, \]
implying that 
\[
 0\le-\log(1-1/K)-1/K \le \frac{1}{2}\sum_{j=2}^{\infty}\frac{1}{K^j}=\frac1{2K^2(1-1/K)},
\]
Multiplying this estimate by $\ell$ and comparing $\ell/K$ with $t$ gives
\[
 |{-\ell\log q}-t|
 \le\frac1K+\frac{\ell}{2K^2(1-1/K)}
 \le\frac1K+\frac{t}{2(K-1)}\le\frac2K.
\]
The function $e^{-s}$ has derivative magnitude at most one for $s\ge0$. Applying this bound to $s=-\ell\log q$ and $s=t$ yields $|q^\ell-e^{-t}|\le2/K$. Hence, $|A-a(W)|=|(e^{-t}-q^\ell)||(1-W)| \le 2/K$|.

\textbf{Proving $\left|\frac RK w^*-P(W)\right|\le\frac{40}K$.} Notice that 
\[
 0<A\le1,\qquad A\ge a(W)-2/K>1-e^{-t} -2/K > 0.397-0.002>0.39.
\]
The finite load $P(w^*) = (R/K)w^*$ and its limit $P(W)$ are
\[
 \frac{(R/K)[W+(\ell/K)(1-W)]}{A(1-W+W/K)},
 \qquad
 \frac{t[W+t(1-W)]}{a(W)(1-W)}.
\]
Their denominators have the lower bounds
\[
 A(1-W+W/K)>0.39\cdot\frac34>0.29,
 \qquad a(W)(1-W)>0.397\cdot\frac34>0.29.
\]
The difference between these denominators is bounded by
\begin{align*}
 &|A(1-W+W/K)-a(W)(1-W)|\\
 &\le|A-a(W)|(1-W+W/K)+a(W)W/K\\
 &\le\frac2K+\frac1{4K}<\frac3K.
\end{align*}
where the inequalities are by $1-W+W/K \le 1$ and $a(W)\le 1$. The two numerators satisfy
\begin{equation}\label{two_numerations}
 W+(\ell/K)(1-W)\le W+t(1-W)
 \le\frac14+0.5064\cdot\frac34<0.63.
\end{equation}
Therefore, using this bound together with the floor errors gives the numerator comparison
\begin{align*}
 &\left|\frac RK\left[W+\frac\ell K(1-W)\right]
       -t[W+t(1-W)]\right|\\
 &\le\left|\frac RK-t\right|\left[W+\frac\ell K(1-W)\right]
       +t\left|\frac\ell K-t\right|(1-W)\\
 &<\frac{0.63+0.5064}{K}<\frac2K.
\end{align*}
We thus have that 
\begin{align*}
     &|\frac{(R/K)[W+(\ell/K)(1-W)]}{A(1-W+W/K)}-
 \frac{t[W+t(1-W)]}{a(W)(1-W)}| \\
 &\le |\frac{(R/K)[W+(\ell/K)(1-W)]}{A(1-W+W/K)} - \frac{t[W+t(1-W)]}{A(1-W+W/K)}| + |\frac{t[W+t(1-W)]}{A(1-W+W/K)} - \frac{t[W+t(1-W)]}{a(W)(1-W)}| \\
 &\le |\frac{2}{K}\frac{1}{0.29}|+ |t[W+t(1-W)]| \frac{|A(1-W+W/K) - a(W)(1-W)|}{|A(1-W+W/K)|\cdot |a(W)(1-W)|} \\
 & \le |\frac{2}{K}\frac{1}{0.29}| + |0.5064 \cdot 0.63| |\frac{3}{K}| |\frac{1}{0.29^2}| \\
 & < \frac{40}{K}
\end{align*}

\textbf{Proving $\frac{1}{2}<w^*<3$.} By \eqref{two_numerations}, the numerator of  $w^*$ is less than $0.63$. Its lower bound is
\[
 W+(\ell/K)(1-W)\ge\ell/K
 >0.5063-1/K\ge0.5053>1/2,
\]
The denominator is at most one, and is greater than $0.39\cdot(3/4)$. Therefore
\[
 \frac12<w^*<\frac{0.63}{0.39\cdot(3/4)}<3.
\]

\medskip

By construction, the grid $\mathcal{W}$ covers the interval $\frac{1}{2}<w^*<3$. Therefore, it implies $w\in [w^*, (1+\delta)w^*]$ exists.

For $K>K_0$, $40/K<0.04\varepsilon$. Thus, $P(w^*) \le P(W) + \frac{40}{K} <0.78+0.04\eps$. By construction and the choice of $\eps$, we have 
\[
 P(w)\le(1+\delta)(0.78+40/K)
 <(1+\varepsilon/100)(0.78+0.04\varepsilon)<0.9.
\]
\end{proof}

\subsection{Proof of \cref{lem:worst}}

\begin{proof}
Notice that $a(0)\ps(P(0))=(1-e^{-t})\ps\left(\frac{t^2}{1-e^{-t}}\right)$. If we show that $a(W)\ps(P(W))$ is increasing in $W\in [0, 1/4]$, we prove the claim.

Recall that in the proof of \cref{lem:finite}, $P(W)$ is increasing in $W$. Hence, $P(W)\ge P(0)>1/2$, and we always use the second branch of \eqref{eq:psi}. Since that 
\[
 \ps'(P)=\frac{c-2P^2}{(2P^2+c)^2},\qquad
 a(W)P'(W)=\frac{t}{(1-W)^2}-e^{-t}P(W).
\]
By the product and chain rules,
\begin{align*}
 \frac{d}{dW}[a(W)\ps(P(W))]
 &=e^{-t}\frac{P(W)}{2P(W)^2+c}
   +a(W)\frac{c-2P(W)^2}{(2P(W)^2+c)^2}P'(W)\\
 &=\frac{4e^{-t}P(W)^3+\dfrac{t}{(1-W)^2}(c-2P(W)^2)}
         {(2P(W)^2+c)^2}.
\end{align*}

By \cref{lem:finite}, $P(W)>P(0)>0.645$ and $0.193<c<0.194$, $2P(W)^2-c>0$. Dividing the derivative numerator by the positive quantity $e^{-t}(2P(W)^2-c)$ shows that positivity is equivalent to
\[
 \frac{4P(W)^3}{2P(W)^2-c}>
 \frac{te^t}{(1-W)^2}.
\]
For the right-hand side,
\[
 \frac{te^t}{(1-W)^2}
 <\frac{0.5064}{(3/4)^2\cdot0.602}<1.50.
\]
For the left-hand side, differentiating the fraction with respect to $P$ gives
\[
 \frac{d}{dP}\left(\frac{4P^3}{2P^2-c}\right)
 =\frac{4P^2(2P^2-3c)}{(2P^2-c)^2}>0
 \qquad(P\ge0.645),
\]
The fraction also increases with $c$. Therefore
\[
 \frac{4P(W)^3}{2P(W)^2-c}
 >\frac{4(0.645)^3}{2(0.645)^2-0.193}>1.67.
\]
The derivative of $a(W)\ps(P(W))$ is thus positive throughout $[0,1/4]$, proving the claim.
\end{proof}

\subsection{Statement and Proof of Continuity Lemma}\label{app_continuity_lemma}

\begin{lemma}\label{lipschitzcontinuous}
    $\ps(P)$ is $1$-Lipschitz, and $(1-e^{-t})\ps(\frac{t^2}{1-e^{-t}})$ is $4$-Lipschitz for $t\in[0.5063,0.5064]$.
\end{lemma}

\begin{proof}
On $0<P\le1/2$, the first branch of \eqref{eq:psi} has the integral representation
\[
 \ps(P)=\int_0^1(1-P)^s\,ds,
\]
obtained by integrating the exponential $e^{s\log(1-P)}$ in $s$. The representation extends to $P=0$ by continuity. Differentiating under this integral gives
\[
 \ps'(P)=-\int_0^1s(1-P)^{s-1}\,ds.
\]
For $0\le P\le1/2$ and $s\in[0,1]$, the factor $(1-P)^{s-1} \le 2$. Consequently $|\ps'(P)|\le\int_0^12s\,ds=1$.

On $P\ge1/2$, \eqref{eq:psi} gives
\[
 \ps'(P)=\frac{c-2P^2}{(2P^2+c)^2},\qquad
 |\ps'(P)|=\frac{2P^2-c}{(2P^2+c)^2}\le\frac1{8c}<1.
\]
The inequality follows from
\[
 (2P^2+c)^2-8c(2P^2-c)=(2P^2-3c)^2\ge0;
\]
the strict upper bound one uses $c>1/8$. Both values at $P=1/2$ equal $1/(2\log2)$, and both first derivatives equal $(\log2-1)/(\log2)^2$. Thus the branches join continuously with matching derivatives. The derivative bound on each branch proves that $\ps$ is $1$-Lipschitz on $[0,1]$. 
%It is positive and decreasing, and $\ps\le1$, either by its integral definition or by comparison with $\ps(0)=1$.

Denote $a=1-e^{-t}$ and $P=t^2/a$. We have
\[
 a'=e^{-t},\qquad
 P'=\frac{2t}{a}-\frac{t^2e^{-t}}{a^2}>0,
 \qquad aP'=2t-\frac{t^2e^{-t}}a\le2t.
\]
Positivity follows from $a>te^{-t}$, which is equivalent to $e^t>1+t$. The product and chain rules, together with the bounds just proved, give
\[
 \left|\frac{d}{dt}[a\ps(P)]\right|
 \le e^{-t}\ps(P)+a|\ps'(P)|P'\le1+2t<4
 \quad\text{for }0.5063\le t\le0.5064.
\]

%Both arguments between $t$ and $t_*$ are in the horizon interval, so this derivative estimate justifies the mean value bound used in the theorem.
\end{proof}

\subsection{Completion of the proof of \cref{thm:main}}\label{app:submodular-proof-completion}
\begin{proof} 
Continue the large-team case of the main-text proof.
The accuracy and horizon satisfy $0<\varepsilon\le1/16$,
$K_0=\lceil1000/\varepsilon\rceil$, $\delta=\varepsilon/100$,
$t\in[0.5063,0.5064]$, and $|t-t_*|\le\varepsilon/1000$.
Let $S^*$ be an optimal team, with $\OPT>0$, $K=|S^*|>K_0$,
$h\in\operatorname*{arg\,max}_{i\in S^*}r_i(S^*)$,
$p=r_h(S^*)>0$, and
$W=f(h\mid S^*\setminus\{h\})/f(S^*)\in[0,1/4]$.
Recall $\ell=\lfloor tK\rfloor$, $R=\ell+1$, $q=1-1/K$, and
$A=1-q^\ell(1-W)$. Let $w^*$ be defined by \eqref{eq:cap}.
By \cref{lem:finite}, the grid enumerates a scale
$w\in[w^*,(1+\delta)w^*]$ with $P(w)=Rw/K<0.9$.
In that iteration, $\bar\alpha=w/K$, $\pi_i=c_i/\bar\alpha$,
$G=f(U)-\pi(U)$, and $M$ is the best candidate revenue, including
the empty contract. The main proof established \eqref{approximation_not_finished}.
Use the limiting quantities $a(W),P(W)$ from \eqref{eq:limit}.
    
Now, we are ready to show $A\ps(P(w))$ is not far from $(1-e^{-t})\ps\left(\frac{t^2}{1-e^{-t}}\right)$.
Recall $P(w^*) \le P(w) \le (1+\delta) P(w^*)$ and $P(w^*) \le P(w) < 1$ by \cref{lem:finite}. Therefore, we have 
\[
 0\le P(w)-\frac RK w^*
 \le\delta\frac RK w^*\le\delta.
\]
Combining this with \cref{lem:finite} gives
\[
 |P(w)-P(W)|\le\frac{40}K+\delta.
\]
Note that $0\le P(w), P(W) \le 1$ by \cref{lem:finite} . The $1$-Lipschitz estimate for $\ps$ (\cref{lipschitzcontinuous}) yields
\[
 |\ps(P^+)-\ps(P(W))|\le\frac{40}K+\delta.
\]
To compare the complete coefficients, write their difference explicitly:
\begin{align*}
 &A\ps(P(w))-a(W)\ps(P(W))\\
 &=(A-a(W))\ps(P(w))
   +a(W)[\ps(P(w))-\ps(P(W))]\\
 &\ge-\frac2K-\left(\frac{40}K+\delta\right)
 =-\frac{42}K-\delta.
\end{align*}
where the inequality is by  \cref{lem:finite} and  $\ps\le1$, $a(W)\le1$. Thus
\[
 \frac M\OPT\ge a(W)\ps(P(W))-\frac{42}K-\delta-\frac4{R^2}.
\]

By Lemma~\ref{lem:worst}, the leading coefficient is at least its value at $W=0$. Also $4/R^2\le16/K^2$ due to $R = \lfloor tK\rfloor +1 \ge tK \ge \frac{K}{2}$. The $4$-Lipschitz holds for $t \in [0.5063,0.5064]$, which gives
\[
 (1-e^{-t})\ps\left(\frac{t^2}{1-e^{-t}}\right)
 \ge\Gamma-4|t-t_*|.
\]
where $\Gamma$ is defined in \eqref{eq:constant}. Putting these estimates together,
\begin{align*}
 \frac M\OPT
 &\ge(1-e^{-t})\ps\left(\frac{t^2}{1-e^{-t}}\right)-\frac{42}K-\delta-\frac{16}{K^2}\\
 &\ge\Gamma-4|t-t_*|-\frac{60}K-\delta\\
 &\ge\Gamma-\frac{4\varepsilon}{1000}-\frac{60\varepsilon}{1000}-\frac{\varepsilon}{100}
 \ge\Gamma-\varepsilon.
\end{align*}
This proves the constant approximation bound.
\end{proof}

\section{Missing Proofs in \cref{section_additive_tpfas}}\label{app_additive_fptsa_proofs}

\subsection{Proof of \cref{lemma:approxlarge}}

\begin{proof} Let $i \in S^*$  be the agent such that $f(i) \ge (1-\gamma) f(S^*)$. The revenue from incentivizing agent $i$ is 
    \[(1-\frac{c_i}{f(i)})f(i) \ge (1-\sum_{k \in S^*}\frac{c_k}{f(k)})f(i) \ge (1- \sum_{k\in S^*} \alpha_k^*)f(i) \ge (1-\gamma) \opt\]
    where the second inequality is by the optimal minimum-share structure that $\alpha^*_k = \max\Big\{\mathcal{L}_{S^*}^*, \frac{c_k}{f(k)}\Big\}$.
\end{proof}

\subsection{Proof of \cref{lemma_upper_lower}}

\begin{proof}
    First, $\tilde{f}(S) \le f(S)$ holds by our discretization step. The second  inequality follows by observing that 
    \[
\tilde{f}(S) = \sum_{i \in S} \tilde{f}(i) \ge \sum_{i \in S} ({f}(i) -  \delta) \ge  \sum_{i \in S} {f}(i) -  n\delta = \sum_{i \in S} {f}(i) - \frac{\eps}{2n} f(i^*) \ge (1-\frac{\eps}{2n} )f(S),
\]
where the first inequality holds by observing that the discretization step is $\delta$. 
\end{proof}

\subsection{Proof of \cref{dynamicprogramingreturensafaircontract}}
\begin{proof}
    By Lemma \ref{lemma_upper_lower} and $\bar{i} = \bar{i}(S, \tilde x /(1-\frac{\eps}{2n}))$, we know that $\mathcal{L}(S, f(S)) \le \tilde{\mathcal{L}} (\bar i, \tilde x/(1-\frac{\eps}{2n}))$ holds. It implies that by Corollary \ref{lprimgretearthanlfaircontractcoro}, the solution forms a fair contract.
\end{proof}

\subsection{Proof of \cref{1minussumgreatereps}}\label{app_1minussumgreatereps}

We prove the lemma by contradiction.
Suppose we have $(1-\sum_{i \in S^*} \alpha_i^* ) -\frac{\eps}{\gamma} < 0$, implying
$
1 -\frac{\eps}{\gamma}<\sum_{i \in S^*} \alpha_i^* 
$.
Hence, there must exist one agent $k \in S^*$ such that 
\[
\frac{1-\frac{\eps}{\gamma}}{n} < \alpha_k^* 
\]
To prove the lemma, we show that by removing the agent $k$, the principal gains more revenue. This leads to a contradiction to the optimality of $S^*$.

To this end, we first observe that 
\[
(1-\sum_{i\in S^*\setminus k} \alpha_i (S^*\setminus k)) f(S^*\setminus k) \ge(1-\sum_{i \in S^*\setminus k} \alpha_i^* )  f(S^*\setminus k)
\]
where $\alpha_i(S)\triangleq \max\{ \frac{c_i}{f(i)}, \mathcal{L}(S, f(S))\}$ is the fair contract with minimum payment for a given set $S$. 
Then, the inequality follows by observing that $\mathcal{L}(S^*\setminus k, f(S^*\setminus k)) \le \mathcal{L}(S^* \setminus k, f(S^*)) \le \mathcal{L}(S^*, f(S^*))$.

Our main analysis is devoted to showing that 
\begin{equation}\label{eq_ineqaulityremovingagent}
(1-\sum_{i \in S^*\setminus k} \alpha_i^* )  f(S^*\setminus k) > (1-\sum_{i \in S^*} \alpha_i^* )f(S^*) 
\end{equation}
The left-hand side is
\begin{align*}
    (1-\sum_{i \in S^*\setminus k} \alpha_i^* )  f(S^*\setminus k) 
    &= (1-\sum_{i \in S^*} \alpha_i^*  + \alpha_k^*)[f(S^*) -f(k)]\\
    &=(1-\sum_{i \in S^*} \alpha_i^* ) f(S^*) -  (1-\sum_{i \in S^*} \alpha_i^* ) f(k) + \alpha_k^* f(S^*) - \alpha_k^* f(k)
\end{align*}
Hence, to prove  \Cref{eq_ineqaulityremovingagent}, we only need to show $-  (1-\sum_{i \in S^*} \alpha_i^* ) f(k) + \alpha_k^* f(S^*) - \alpha_k^* f(k) > 0$, which is equivalent to
\[
\alpha_k^* f(S^*) - \alpha_k^* f(k) > (1-\sum_{i \in S^*} \alpha_i^* ) f(k).
\]
Note that by the contradiction assumption, we have 
\[
(1-\sum_{i \in S^*} \alpha_i^* ) f(k) < \frac{\eps}{\gamma}f(k)
\]
and, recalling that $1-\frac{\epsilon}{\gamma}< \sum_{i \in S^*} \alpha^*_i$, we get 
\[
\alpha_k^* f(S^*) - \alpha_k^* f(k) > \frac{1-\frac{\eps}{\gamma}}{n} (f(S^*) -f(k)).
\]

Hence, if the following equality holds
\begin{align}\label{eq:equivalentContraddiction}
\frac{1-\frac{\eps}{\gamma}}{n} (f(S^*) -f(k)) \ge \frac{\eps}{\gamma}f(k),
\end{align}
then also \Cref{eq_ineqaulityremovingagent} holds.
Moreover, it holds 
\[
\frac{1-\frac{\eps}{\gamma}}{n} (f(S^*) -f(k)) \ge \frac{\eps}{\gamma}f(k) \Longleftrightarrow \frac{f(S^*) -f(k)}{f(S^*) -f(k) + nf(k)} \ge \frac{\eps}{\gamma} \Longleftrightarrow \frac{\frac{f(S^*)}{f(k)} -1}{\frac{f(S^*)}{f(k)} -1 + n} \ge \frac{\eps}{\gamma}
\]
Now, notice that $\frac{x -1}{x -1 + n}$ is monotone increasing in $x$ for $x\ge 1$.
Furthermore, since we are in Case 2, we have $\frac{f(S^*)}{f(k)} > \frac{1}{1-\gamma}$ as $k\in S^*$. Hence, we have 
 \[
 \frac{\frac{f(S^*)}{f(i)} -1}{\frac{f(S^*)}{f(i)} -1 + n} > \frac{\frac{1}{1-\gamma} -1}{\frac{1}{1-\gamma} -1 + n}= \frac{\gamma}{\gamma +n(1-\gamma)}
>  \frac{\eps}{\gamma}
 \]
 where the last inequality holds due to $\eps = \frac{1}{2} \frac{\gamma^2}{\gamma +n(1-\gamma)}$. 

This means that $S^*\setminus k$ gives the principal a larger revenue, implying that 
$S^*$ is not an optimal set, which is a contradiction. This concludes the proof.

\subsection{Proof of \cref{technicalclaimreserveparotionintermideate}}\label{app_technicalclaimreserveparotionintermideate}

Recall that the optimal fair contract $\alpha^*$ is the least $S^*$-incentive contract with minimum share $\mathcal{L}(S^*, f(S^*))$. The second part of the claim follows by observing that $\tilde f(S^*) \le f(S^*)$ implies $\frac{c_{\bar{i}}}{f(\bar i)} (1-\frac{f(\bar i)}{\tilde f(S^*)})  \le {\frac{c_{\bar{i}}}{f(\bar i)} (1-\frac{f(\bar i)}{ f(S^*)})} \le\mathcal{L}(S^*, f(S^*))$ where the second inequality is by $\bar{i} \in S^*$.

   The first part of the claim is by observing that  
\begin{align}
    1-\sum_{i\in S^* }\tilde\alpha_{i} 
    &= 1- \sum_{i \in S'} \tilde{\mathcal{L}} (\bar i, \tilde{f}(S^*)/ (1-\frac{\eps}{2n})) - \sum_{i \in S^* \setminus S'} \frac{c_i}{f(i)} \nonumber\\
    & = 1- \sum_{i \in S'} \frac{c_{\bar{i}}}{f(\bar i)} (1-\frac{f(\bar i)}{\tilde f(S^*)/ (1-\eps/(2n))})  -\sum_{i \in S^*\setminus S'} \frac{c_i}{f(i)} \nonumber\\
    & = 1- \sum_{i \in S'} \frac{c_{\bar{i}}}{f(\bar i)} (1-\frac{f(\bar i)}{\tilde f(S^*)})  -\sum_{i \in S^*\setminus S'} \frac{c_i}{f(i)} - \sum_{i \in S'}  {\frac{c_{\bar{i}}}{f(\bar{i})}}\frac{f(\bar i)}{\tilde{f}(S^*) }\frac{\eps}{2n} \nonumber\\
    &\ge \Big( 1-  \sum_{i \in S'} \frac{c_{\bar{i}}}{f(\bar i)} (1-\frac{f(\bar i)}{\tilde f(S^*)})   -\sum_{i \in S^*\setminus S'} \frac{c_i}{f(i)} \Big) - \eps \nonumber\\
    & = \Big( 1- \sum_{i \in S^*} \hat{\alpha}_i  \Big) - \eps \label{hatalpahgrateralpah} \\
    &\ge (1-\sum_{i \in S^*} \alpha_i^*) -\eps  \nonumber\
\end{align}
where the first inequality is by that $\frac{f(\bar i)}{\tilde{f}(S^*) } \le \frac{f(i^*)}{\tilde{f}(i^*)} \le \frac{\tilde{f}(i^*) + \delta} {\tilde{f}(i^*)} \le 2$ and $\frac{c_{\bar{i}}}{f(\bar{i})} \le 1$ since $\bar{i}, i^* \in S^*$. Due to Lemma \ref{1minussumgreatereps}, we have 
\[
(1-\sum_{i \in S^*} \alpha_i^*) -\eps \ge \frac{\eps}{\gamma} -\eps > 0.
\]
Additionally, (\ref{hatalpahgrateralpah}) proves the second inequality of the first part of the claim. This concludes the proof.

\subsection{Proof of \cref{fptas_compexity_reslut}}

\begin{proof}
Our algorithm enumerates agent pairs $(i^*, \bar{i})$, which takes time $O(n^2)$. Given an agent pair $(i^*, \bar{i})$ and an estimate $\tilde x$, one needs to construct the agent set $S_{\bar{i}, \tilde x, i^*}$, which takes time $O(n)$. 

Given a pair of agents $(i^*, \bar{i})$, we define the discretization step $\delta$. Since agent $i^*$ gives the largest contribution in this iteration,  the number of different values $\tilde{x} = k\delta$ that we need to consider is at most 
\[k \le \frac{nf(i^*)}{\delta} = O(\frac{2n^3}{\eps}).\]
Hence, the iteration over $\tilde{x}$ is $O(\frac{2n^3}{\eps})$. 

Next, we examine the running time of dynamic programming (\ref{dynamic_pro}). For each estimate $x = k\delta$, we have a feasible set $|S_{\bar{i}, x, i^*}| \le n$ and solve the dynamic programming in time $O(\frac{2n^4}{\eps})$. 

Since we define $\eps = \frac{1}{2} \frac{\gamma^2}{\gamma +n(1-\gamma)}$ and all the operations here only need polynomial time, our algorithm requires a running time of $\poly(\frac{1}{\gamma}, n)$. 
\end{proof}

\subsection{Dynamic Programming for Problem (\ref{dynamic_pro})}\label{app_endymica_proendom}

In this section, we show the dynamic programming to solve Problem (\ref{dynamic_pro}). We assume $i^*$, $\bar{i}$ are two different agents. The analysis is similar if they are the same agent.  Given $\tilde{x}$, Problem (\ref{dynamic_pro}) is equivalent to 
\begin{equation}\label{dynamic_pro_app}
\begin{aligned}
    \min_{\{i^*, \bar{i}\}\subseteq S \subseteq S_{\bar{i}, \tilde x, i^*}} & \quad \sum_{i\in S}\tilde\alpha_{i} \\
    \text{subject. to}  &  \quad \tilde{f}(S) = \tilde x
    \end{aligned}
\end{equation}

Let $\mathcal{D}_{k, y}$ be the optimal set of agents when considering first $k$ agents in $S_{\bar{i}, \tilde x, i^*}$ and size of constraints must have $\tilde{f}(\mathcal{D}_{k, y}) = y$, where $y = k\delta$ are for $k = 0,1, 2, \dots, \lfloor \frac{2n^3} {\eps} \rfloor $. 
Moreover, 
let $\mathcal{V}_{k,y}$ be the corresponding minimum objective value in Problem (\ref{dynamic_pro_app}). Therefore, when considering the $(k+1)^{\text{th}}$ agent in $S_{\bar{i}, \tilde x, i^*}$, denoted as $a$, the state-transition equations are
\begin{equation}\label{value_state_trans}
    \mathcal{V}_{k+1,y} = 
    \begin{cases} 
\mathcal{V}_{k,y-\tilde f(a)} + \tilde\alpha_{a} ,  & \text{if~} \mathcal{V}_{k,y-\tilde f(a)} + \tilde\alpha_{a} < \mathcal{V}_{k,y}, \\
\mathcal{V}_{k,y}, & \text{otherwise}, 
\end{cases}
\end{equation}
Correspondingly, the state-transition equations for set $\mathcal{D}$ are
\begin{equation}\label{set_state_trans}
    \mathcal{D}_{k+1,y} = 
    \begin{cases} 
\mathcal{D}_{k,y-\tilde f(a)} \cup \{a\} ,  & \text{if~} \mathcal{V}_{k,y-\tilde f(a)} + \tilde\alpha_{a} <\mathcal{V}_{k,y}, \\
\mathcal{D}_{k,y}, & \text{otherwise}, 
\end{cases}
\end{equation}
Let the new domain set $\tilde{S} = S_{\bar{i}, \tilde x, i^*} \setminus \{i^*,\bar{i}\}$. The algorithm is depicted in Algorithm \ref{syanmicnprogram28}. Note that Liner \ref{line3intitionalization} in Algorithm \ref{syanmicnprogram28} ensures that the agents $i^*, \bar{i}$ are included in the returned solution set if non-empty.

\begin{algorithm}[t]
\caption{Dynamic Programming for Problem (\ref{dynamic_pro}) \\
\textbf{Input:} $\{\tilde\alpha_i\}_{i \in \mathcal{N}}$, an additive reward functions $f$, an estimate $\tilde x$, $n, \eps, \delta$, the new domain set $\tilde{S}$ and agents $ i^*, \bar{i}$   \\
\textbf{Output:} an incentivized set of agents and the objective value for Problem (\ref{dynamic_pro}).
} \label{syanmicnprogram28} 
Let $\mathcal{D}_{k, y} = \emptyset$ for all $y = \ell\delta$ with $ \ell\in \{0,1, 2, \dots, \lfloor \frac{2n^3} {\eps} \rfloor\}$ and all $k = \{0, 1, \dots, |\tilde{S}| \}$\;
Let $\mathcal{V}_{k, y} = \infty$ for all $y = \ell\delta$ with $ \ell\in \{0,1, 2, \dots, \lfloor \frac{2n^3} {\eps} \rfloor\}$ and all $k = \{0, 1, \dots, |\tilde{S}| \}$\;
$\mathcal{D}_{0, \tilde{f}(i^*) + \tilde{f}(\bar{i})} = \{i^*, \bar{i}\}$, $\mathcal{V}_{0, \tilde{f}(i^*) + \tilde{f}(\bar{i})} = \tilde{\alpha}_{i^*} + \tilde{\alpha}_{\bar{i}}$, $k=1$ \label{line3intitionalization}\;
\For{$a \in \tilde{S}$}{
  \For{$y= \ell \delta,  ~\ell \in \{0,1, 2, \dots, \lfloor \frac{2n^3} {\eps} \rfloor\}$}{
  Update $\mathcal{V}_{k, y}$ and $\mathcal{D}_{k, y}$  according to (\ref{value_state_trans}) and (\ref{set_state_trans}) \;
  }
  $k = k+1$\;
}
\Return the set of agents $\mathcal{D}_{|\tilde{S}|, \tilde{x}}$ and the corresponding objective value $\mathcal{V}_{|\tilde{S}|, \tilde{x}}$.
\end{algorithm}

\section{NP-Hardness: Proof of \cref{hardness_additive}}\label{app_nphardpproof}

Our proof is inspired by \cite{dutting2023multi} and by a reduction from the subset sum problem with the size of the set being $k \in \mathbb{N}$\footnote{Notice that the standard version of the subset sum problem does not require the set to be of a given size $k$. However, it is straightforward to see that even this restricted version is \np-hard. }.  The key idea of constructing our instance is to ensure that if there exists such a subset $I$ of size $k$, we have a fair contract that would incentivize the set $I$ of agents.

\begin{proofof}{\Cref{hardness_additive}}
    We reduce from the following subset sum problem with a size constraint $k \in \mathbb{N}$: Given a set of positive integers $\{w_1,\dots, w_m\}$ with $\sum_{i \in [m]} w_i=W$, we are required to find a set of indexes $I$ such that $\sum_{in \in I} w_i=\frac{W}{2}$ and $|I| = k$.

    We build an instance of the fair contract problem as follows. Let $\delta=  \frac{1}{4Wmk}$. Then,  we add an agent $i$ for each integer $w_i$, and set
    \begin{itemize}
    \item $f(i)= w_i \delta +\frac{1}{2m}$,
    \item $c_i=\frac{(w_i \delta+\frac{1}{2m})^2}{W\delta+\frac{k}{m}}$,
    \end{itemize}
    We show that if the subset sum instance is satisfiable, then there exists a fair contract with principal's utility $\frac{W\delta+\frac{k}{m}}{4}$; 
    Otherwise, all the possible fair contracts have utility strictly smaller than $\frac{W\delta+\frac{k}{m}}{4}$. This is sufficient to prove the statement.

    \paragraph{Sufficiency.} Suppose a set $I\subseteq [m]$ of size $k$ such that $\sum_{i \in I} w_i=\frac{W}{2}$. We set  the contracts for agents as follows: Let $\alpha_i=\frac{c_i}{f(i)} = \frac{w_i \delta+\frac{1}{2m}}{W\delta+\frac{k}{m}}$ for each agent $i\in I$, and $\alpha_i=0$ for all $i \notin I$. Clearly, the set of agents $I$ are incentivized to exert efforts.

    Then, we can see that the principal's revenue under the contract $\alpha$ is 
    \begin{align*}
    \left(1-\sum_{i \in I} \alpha_i\right) f(I)&= \left(1-\sum_{i \in I}\frac{(w_i \delta+\frac{1}{2m})}{W\delta+\frac{k}{m}}\right) \sum_{i \in I} (w_i \delta +\frac{1}{2m}) \\
    &= (1-\frac{\frac{W \delta}{2}+ \frac{k}{2m}}{W\delta+\frac{k}{m}}) (\frac{W \delta}{2}+ \frac{k}{2m}) \\
    & = \frac{W \delta+ \frac{k}{m}}{4}.
    \end{align*}

    Now, we show that the contract $\alpha$ is fair. {By \Cref{sufficient_necessaryforequi}, we only need to show that for two agents $i,j \in I$ such that $\alpha_i<\alpha_j$, it should hold that $\frac{c_j}{f(j)} > \alpha_i \ge \alpha_j(1-\frac{f(j)}{f(I)})$.}
    
    Consider any two agents $i,j \in I$ such that $\alpha_i<\alpha_j$. Clearly, as $\alpha_j = \frac{c_j}{f(j)}$, we know that $\frac{c_j}{f(j)} > \alpha_i $ is satisfied.
Hence, we want to show $\alpha_i \ge \alpha_j(1-\frac{f(j)}{f(I)})$, which is equivalent to
    \begin{align*}
        \alpha_i f(I)  \ge \alpha_j f(I \setminus j)
    \end{align*}
    Equivalently, we need to show:
    \begin{align*}
       \frac{(w_i \delta+\frac{1}{2m})}{W\delta+\frac{k}{m}}  (\frac{W \delta}{2}+ \frac{k}{2m})\ge \frac{(w_j \delta+\frac{1}{2m})}{W\delta+\frac{k}{m}} \sum_{i \in I \setminus j} (w_i \delta +\frac{1}{2m}).  
    \end{align*}
The above inequality is proved to hold by observing that 
    \begin{align*}
        \frac{(w_i \delta+\frac{1}{2m})}{W\delta+\frac{k}{m}}  (\frac{W \delta}{2}+ \frac{k}{2m}) & \ge  \frac{\frac{1}{2m}}{W\delta+\frac{k}{m}}  \frac{k}{2m} \\
        & \ge \frac{\frac{1}{2m}+\frac{1}{4mk}}{W\delta+\frac{k}{m}}  \frac{k-\frac{1}{2}}{2m} \\
        & \ge \frac{(w_j \delta+\frac{1}{2m})}{W\delta+\frac{k}{m}} \left(\frac{k-1}{2m}+\sum_{i \in I \setminus j}w_i \delta\right)  \\
        & \ge \frac{(w_j \delta+\frac{1}{2m})}{W\delta+\frac{k}{m}} \sum_{i \in I \setminus j} (w_i \delta +\frac{1}{2m})
    \end{align*}
    where the second inequality follows from $\frac{k}{2m}\ge \left(\frac{1}{2m}+\frac{1}{4mk}\right) \left(k-\frac{1}{2}\right) = \frac{k}{2m} - \frac{1}{8mk}$, and the third one by $\delta W = \frac{1}{4mk}$, which implies that 
    \[\frac{1}{4mk} = \delta W \ge \delta w_j, \quad \text{and }\quad \frac{k-\frac{1}{2}}{2m} = \frac{k-1}{2m} + \frac{1}{4m} \ge \frac{k-1}{2m} + \frac{1}{4mk} \ge  \frac{k-1}{2m}+\sum_{i \in I \setminus j}w_i \delta\]
    This concludes the first part of the proof.

    \paragraph{Necessity.} 
    We show an even stronger result. In particular, we show that if there does not exist a satisfying subset $I$ of size $k$, then any contract, even not fair, would give the principal a revenue strictly smaller than $\frac{W\delta+\frac{k}{m}}{4}$.
    
    Indeed, using the results in  \citep{dutting2023multi}, the minimum payments to incentivize any set $I$ of agents to exert efforts are by paying agent $i\in I$ the cut-off wage $\alpha_i = \frac{c_i}{f(i)}$. 
   Hence,  the maximum possible revenue that the principal can get by incentivizing set $I$  is
     \[\left(1-\sum_{i \in I}\frac{(w_i \delta+\frac{1}{2m})}{W\delta+\frac{k}{m}}\right) \sum_{i \in I} (w_i \delta +\frac{1}{2m}).\]
    
    Now, consider any set $I\subseteq [m]$ of size $k$. It should hold that $\sum_{i \in I} w_i\neq W/2$, and hence  $\sum_{i \in I} (w_i \delta +\frac{1}{2m})\neq \frac{W \delta + k/m}{2}$.
Then, we consider any set $I\subseteq [m]$ of size $k'\neq k$. We show that $\sum_{i \in I} (w_i \delta +\frac{1}{2m})\neq \frac{W \delta + k/m}{2}$ holds. By contradiction, if it holds as equality, we have
\[
\sum_{i \in I} (w_i \delta +\frac{1}{2m}) = \frac{W \delta + k/m}{2}
\]
By substituting $\delta=  \frac{1}{4Wmk}$, it leads to 
\[
\sum_{i \in I} w_i -\frac{W}{2} = (k-k')2k\cdot W
\]
If $k> k'$, then it must imply that $
\sum_{i \in I} w_i -\frac{W}{2} > 2k\cdot W
$, which is impossible. Otherwise, if $k < k'$, it must imply that $
\frac{W}{2} - \sum_{i \in I} w_i > 2k\cdot W
$, which is impossible either. Hence, we get that $\sum_{i \in I} (w_i \delta +\frac{1}{2m})\neq \frac{W \delta + k/m}{2}$ .

    Therefore, we can upper bound the principal's revenue with 
     \[\max_{x \in \mathbb{R}\setminus \{ \frac{W \delta + k/m}{2}\}} \left(1-\frac{x}{W\delta+\frac{k}{m}}\right) x<\frac{W\delta+\frac{k}{m}}{4}.\]
    This concludes the proof.
\end{proofof}

\section{Missing Algorithms in the Main Content}\label{app_missing_algo}

\begin{algorithm}[ht]
\caption{Non-discriminatory contracts for additive functions \\
\textbf{Input:} costs of efforts $\{c_i\}_{i \in [n]}$, reward functions $\{f(i)\}_{i \in [n]}$, number of agents $n$,   \\
\textbf{Output:} set of agent $S$ and non-discriminatory contract $\alpha$.
} \label{non_discriminatory_add}
Sort all the agents in decreasing order according to $f(i)$ \;
$S \gets \emptyset$, $\alpha = 0$ \;
\For{$i = 1, \dots, n$}{
  $\hat{S} = \{i\}$, $\hat{\alpha} = \frac{c_i}{f(i)}$ \;
  \For{$j = 1, 2, \dots, n$}{
    \If{$\frac{c_j}{f(j)} \le \hat{\alpha}$ and $j\neq i$}{
        \If{$(1-(|\hat S|+1)\hat{\alpha}) f(\hat S\cup\{j\}) > (1-(|\hat S|)\hat{\alpha}) f(\hat S) $}{
         $\hat{S} = \hat{S} \cup \{j\}$\;
        }
    }
    }
    \If{$(1-|S|{\alpha}) f(S) < (1-|\hat S|\hat{\alpha}) f(\hat S) $}{
      $S=\hat{S}$, $\alpha = \hat{\alpha}$ \;
    }
}
\Return $S$ and $\alpha$.
\end{algorithm}

\begin{algorithm}[ht]
\caption{FPTAS algorithm \\
\textbf{Input:} costs of efforts $\{c_i\}_{i \in [n]}$, an additive reward functions $f$, number of agents $n$, a constant $\gamma >0$   \\
\textbf{Output:} an incentivized set of agent $\hat{S}$.
} \label{fptasforadditive} 
$S_1 \gets \arg\max_{i \in \mathcal{N}}~ (1-\frac{c_i}{f(i)}) f(i)$ \label{line_enumeraset_algo}\;
$S_2 \gets$ Return a set of agents from Algorithm \ref{case2algoadditive}\;
Construct the optimal fair contracts $\alpha_{S_1}$ (resp. $\alpha_{S_2}$) for $S_1$ (resp. $S_2$) by \cref{theorem_optimal_share} \;
\Return $S = \arg\max_{S \in \{S_1, S_2\}} \rev(S,\alpha_S) $.
\end{algorithm}

%\newpage

%\input{constant_old}

\end{document}